\documentclass[%
 reprint,
groupedaddress,
 amsmath,amssymb,
 aps,  prx,
floatfix,
]{revtex4-2}
\usepackage{graphicx}
\usepackage{dcolumn}
\usepackage{bm}
\usepackage{multirow}
\usepackage{xcolor}

\begin{document}

\preprint{APS/123-QED}

\title{How criticality meets bifurcation in compressive failure of disordered solids}

\author{Ashwij Mayya}
\author{Estelle Berthier}%
\altaffiliation[Current address : ]{Arnold-Sommerfeld-Center for Theoretical Physics and Center for NanoScience, Ludwig-Maximilians-Universit\"at M\"unchen, D-80333 M\"unchen, Germany}
\author{Laurent Ponson}%
\email{laurent.ponson@upmc.fr}
\affiliation{Institut Jean Le Rond D’Alembert UMR 7190, Sorbonne Université, CNRS, Paris, France}%

\date{\today}

\begin{abstract}
Continuum mechanics describes compressive failure as a standard bifurcation in the response of a material to an increasing load: damage, which initially grows uniformly in the material, localizes within a thin band at failure. Yet, experiments recording the acoustic activity preceding localization evidence power-law distributed failure precursors of increasing size, suggesting that compressive failure is a critical phenomenon. We examine here this apparent contradiction by probing the spatial organization of the damage activity and its evolution until localization during compression experiments of 2D cellular solids. The intermittent damage evolution measured in our experiments is adequately described by a non-stationary depinning equation derived from damage mechanics and reminiscent of critical phenomena. In this description, precursors are damage cascades emerging from the interplay between the material's disorder and the long-range stress redistributions following individual damage events. Yet, the divergence of their characteristic size close to failure, which we observe in our experiments, is not the signature of a transition towards criticality. Instead, the system remains at a fixed distance to the critical point at all stages of the damage evolution. The divergence results from the progressive loss of stability of the material as it is driven towards localization. Thus, our study shows that compressive failure is a standard bifurcation for which the material disorder plays a marginal role. It also shows that precursory activity constitute by-products of the evolution towards localization and can therefore be used to assess  the residual lifetime of structures.
\end{abstract}


\maketitle
\section{Introduction}
Damage localization is the standard mode of failure of materials under compression. Decoding this degradation process is therefore the cornerstone of the design of reliable and safe structures such as buildings, bridges, tunnels and a countless number of mechanical parts under compressive loading conditions. During their life in service, these structures may progressively lose their mechanical integrity. Comprehending damage evolution to predict their remaining lifetime is an essential component of modern tools of structural design and predictive maintenance. Yet, the appropriate theoretical concepts for describing damage spreading and ultimately localization are still vigorously debated and  constitute an active topic of research~\cite{koivisto2016,kadar2020,biswas2020,debski2021}.

Continuum damage mechanics is a powerful approach for describing the compressive failure of materials such as rocks, ceramics or mortar~\cite{ashby1990,lockner1991,kachanov1987,mazars1989}. In this framework, discrete damage mechanisms like microcrack growth are described at a continuum scale through the degradation of the local elastic stiffness of the material~\cite{kachanov1993,fortin2006, fortin2009, manzato2012,manzato2014,tal2016,thilakarathna2020}. Beyond some critical load level, this softening leads to a bifurcation from the homogeneous damage field to a localized damage that only grows within a thin band and leading to material failure~\citep{rudnicki1975,olsson1999}. 

In parallel, and almost independently to the development of damage mechanics, the intermittent dynamics of damage growth preceding compressive failure has attracted a lot of attention. Acoustic emissions have been used as a preferential means of experimental investigation. Experimental measurements reveal that damage grows through bursts that display robust scale-free statistics~\citep{lockner1991,lockner1993,petri1994,fortin2009,deschanel2006,davidsen2007,rosti2009,baro2013,castillo2013,baro2018,weiss2019}. Accounting for material disorder, various theoretical works~\citep{sornette1994, zapperi1997, zapperi1997b, sornette2002,truskinovsky2020} have proposed to describe failure as a discontinuous (first-order) phase transition where the precursory damage events emerge from the sweeping of an instability. These ideas were primarily discussed in the context of toy models of failure, using e.g. random fuse models. As a result, a direct comparison with the statistical properties of precursors measured experimentally is not possible, leaving unresolved the applicability of these concepts to real materials. Motivated by the observation of an increase of the precursors' size close to localization~\cite{guarino1998,girard2010,baro2013,kun2014,kandula2019,weiss2019}, an alternative scenario in which compressive failure is described as a continuous (second-order) phase transition was also proposed~\cite{roux1988,delaplace1996,garcimartin1997,moreno2000,girard2010,weiss2014,renard2018,weiss2019}: similarly to a large range of driven disordered elastic systems~\citep{barabasi,sethna,wiese2}, the bursts of activity characterizing the response of damaging materials are interpreted as critical fluctuations, or avalanches, that are reminiscent of the so-called depinning transition, a critical phenomenon emerging from the competition between disorder and elastic interactions. Above a depinning threshold, damage is thus expected to grow at some finite speed, eventually leading to failure. When approaching this critical point, the material should then display scale-free fluctuations with diverging length and time scales, a feature that has been observed in some compression experiments~\cite{renard2018,weiss2019}.

Despite the appeal of such a scenario, it comes in direct contradiction with continuum damage models which describe compressive failure as a standard bifurcation taking place in homogeneous solids, for which material disorder and hence precursors play a minor role. The objective of this work is to reconcile these two seemingly incompatible scenarios: \textit {Is compressive failure a standard bifurcation or a depinning transition ? Are its precursors the result of a transition towards a critical point ?} These interrogations have crucial engineering implications: \textit{ Do precursory damage events foretell impending failure?} 
Recently, we examined this issue in a 1D toy model~\citep{berthier2021}. However, the generalization of our results to real materials was limited by the short-range interactions and the system dimension considered in our model. Here, we follow a different approach: we start from the in-depth characterization of the damage precursors in a model experimental system. Subsequently, experimental observations are confronted with the competing scenarios, yielding our two main results :
\begin{itemize}
    \item The scale-free statistics of the precursory activity is reminiscent of the critical avalanche dynamics during depinning of a driven disordered elastic interface. Yet, in stark contrast with the critical transition scenario, the divergence of the length and the time scales of precursors close to failure results from the on-coming localization, a standard bifurcation taking place in homogeneous solids. 
    \item The approach towards the bifurcation does not drive the damaging solid towards a critical point. Instead, it remains at some finite distance from criticality during the whole process of damage accumulation.
\end{itemize}

Our article is structured as follows. First, we present our experimental setup (section~\ref{sec2}). We carry compression test of cohesionless soft cellular solids. We show that this material behave like an elasto-damageable medium, ensuring the applicability of our findings to other brittle materials. The characteristic features of the precursory damage activity are then analyzed at both the global scale (using the force-displacement response of the specimen) and the local scale (using full-field measurements of the mechanical quantities). This multi-scale characterization provides the complete (non-stationary) statistical structure of precursors that follows robust scaling laws. To rationalize these experimental observations, we use damage mechanics that we extend to disordered solids (section~\ref{sec3}). We derive the evolution equation of the damage field that reveals the complex connection with driven disordered elastic interfaces and depinning transition. Our approach also captures damage localization that is described as a standard instability. In section~\ref{sec4}, we come back to our experimental data and validate several non-standard aspects of the proposed depinning model. The compatibility of both competing scenarios with our experimental observations is discussed in section~\ref{sec5}.  We argue that damage spreading  is reminiscent of the non-stationary depinning dynamics of a driven disordered elastic interface that culminates in a standard bifurcation at localization. Section~\ref{sec6}  provides a direct engineering application of our work. We bring the experimental proof of concept that precursors can be harnessed for predicting the residual lifetime of structures. The analysis of the acoustic emissions recorded during our experiments is presented in section~\ref{sec7}. Their statistical similarity with the acoustic precursors recorded during the failure of standard brittle solids ensures the generality of our results. The numerical resolution of the proposed damage model is carried in section~\ref{num_model}. It provides a comprehensive interpretation of the statistics of failure precursors observed in our experiments. Finally, the broader implications of the proposed non-stationary depinning scenario o other phenomena including amorphous plasticity are discussed in section~\ref{sec9}. The methods employed in our experiments and our model are briefly presented in the Annexes A and B. A thorough description is provided in the Supplementary Information~\cite{supp_info}.

\section{Experimental investigation of failure precursors}
\label{sec2}
\subsection{Damage localization}
Taking inspiration from Poirier \textit{et al.}'s experiments~\cite{poirier1992}, we perform compression tests of 2D cohesionless soft cellular solids as shown in Fig.~\ref{fig:fig1}(a). The results presented below are based on ten different experiments during which the evolution of the material microstructure and the associated damage is tracked using full-field measurements. (see Appendix \ref{app_a}1 for a detailed description of our experimental set-up). Following Karimi \textit{et al.}~\cite{karimi2019}, the effect of friction between cells is described at a mesoscopic continuum scale by introducing an equivalent elasto-damageable medium. As shows in the following sections and justified in detail in Appendix~\ref{app_a}2, such a cohesionless cellular solid is a model system  that mimics the jerky dissipative response of brittle disordered solids under compression. Dissipation taking place at the scale of the individual cells is tracked in space and time using a high speed camera (see Appendix \ref{app_a}4-5 and SI Sec.~1 and~2 for details). We thus circumvent the drawbacks inherent to X-ray tomography that provides the detailed spatial structure of damage events in 3D materials but at the cost of temporal resolution~\cite{renard2019,cartwright2020}. We also overcome the limitations of acoustic emissions that provide highly resolved time series but with a rather poor spatial resolution~\cite{lennartz2014,davidsen2017}. 

First, we focus on the average mechanical response of the specimen. A typical force-displacement curve recorded during a compression test under displacement control conditions is shown in Fig.~\ref{fig:fig1}(b). As evidenced from the snapshots of the experiments taken at different load levels (top inset of Fig.~\ref{fig:fig1}(b) and Supplementary Video S1), the specimen initially deforms rather uniformly, even beyond the linear elastic regime. We consider the deviation to cell circularity beyond the elastic limit as a measure of the damage level, $d_\circ$. As shown in the inset of Fig.~\ref{fig:fig1}(c), the larger $d_\circ$, the lower the Young's modulus $E(d_\circ)$ of the material, as expected for elasto-damageable media. Together with its Poisson's ratio $\nu(d_\circ) \simeq \nu_\circ$ that remains nearly constant, the function $E(d_\circ)$ describes the impact of damage on the mechanical response of our 2D elasto-damageable solid (see Appendix \ref{app_a}2). Damage is observed to grow homogeneously from the elastic limit $F = F_\mathrm{el}$ until the peak load $F = F_\mathrm{c}$ (refer to Table. S1 for a list of notations), except for the cells close to the boundary where friction with the lateral wall prevails. Tracking the collapsing of cells from their deviation to circularity $d_\circ > 0.5$, we observe that at peak load, no cell has collapsed yet, except at the boundaries. After peak load, however, a band consisting of progressively collapsing cells appears, initiating from the top corners of the structure. The localization band is clearly visible further away from peak load, as shown on  the snapshot in the upper right corner of Fig.~\ref{fig:fig1}(b). The onset loading $\Delta_\mathrm{c}$ of localization is inferred  from the evolution of the vertical strain $\langle \epsilon_\mathrm{yy} \rangle_\mathrm{\mathcal{R}_2}$ averaged over the bottom region $\mathcal{R}_2$ of the specimen (indicated in Fig.~\ref{fig:fig1}A): as displayed in the lower inset of Fig.~\ref{fig:fig1}(b), for $\Delta> \Delta_\mathrm{c}$, $\langle \epsilon_\mathrm{yy} \rangle_\mathrm{\mathcal{R}_2}$ saturates and departs from the strain  $\epsilon_\mathrm{yy}^\mathrm{ext}$ imposed by the loading machine. On the contrary, the strain $\langle \epsilon_\mathrm{yy} \rangle_\mathrm{\mathcal{R}_1}$ measured in the upper region $\mathcal{R}_1$ follows the imposed strain. This confirms that damage localization starts at peak load. To confirm this important result, we investigate the spatial distribution of damage growth rate near peak load (see Appendix \ref{app_a}5). Before peak load, it is rather homogeneously distributed, while righter after peak load, its value is several times larger in the thin band where cells start to progressively collapse.

This  behavior is consistent with the observations made  in a wide range of brittle solids \cite{baud2004,lenoir2007,lennartz2014,mcbeck2020,kandula2022}. It is also captured by the non-local damage models recently proposed in Berthier \textit{et al.}~\cite{berthier2017} and Dansereau \textit{et al.}~\cite{dansereau2019} (see SI Sec.~S3F~\cite{supp_info} for the analytical prediction). If the experiments were under force control conditions, as in most real-life structural applications, a sudden collapse of the cells resulting in  the catastrophic failure of the specimen would also occur at peak load. 
\subsection{Precursors as cascades of damage events}
\begin{figure}[t]
\centering
\includegraphics[width=\linewidth]{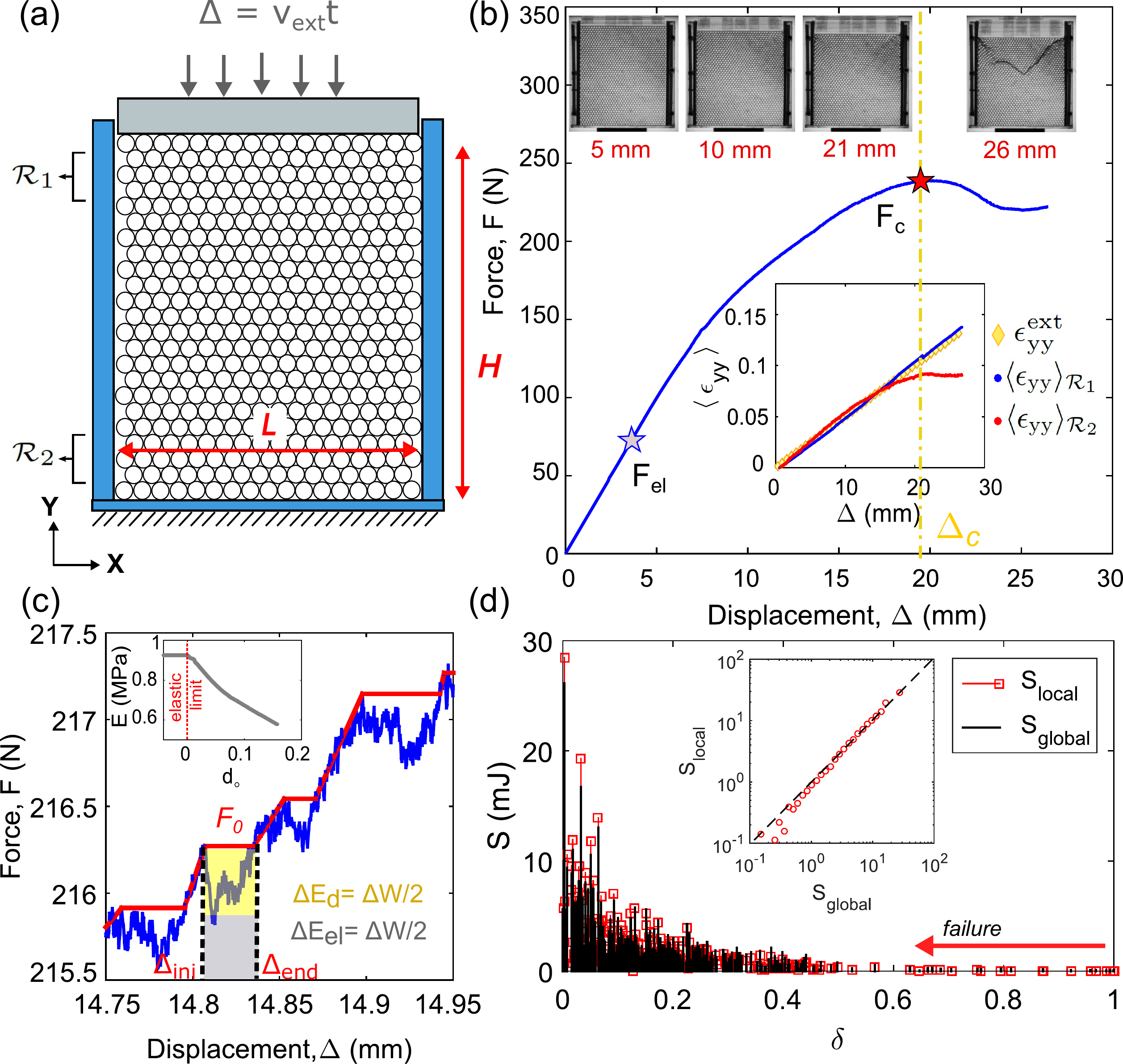}
\caption{(a) Schematic of the compression experiment depicting the front view of the hexagonal packing of soft cellular solids.  The box dimensions are $205\mathrm{mm}\times170\mathrm{mm}\times30\mathrm{mm}$. Owing to the displacement $\Delta$  applied to the specimen through a piston moving at a rate $v_{\mathrm{ext}}$, the cells undergo a deformation that is recorded using a high speed camera. The following figures describe the typical mechanical response of the specimens, as observed in one of our experiments. (b) Force-displacement response of the specimen. Top insets: The emergence of a localization band of collapsed cells, corresponding to highly localized deformations, is visible on the snapshots of the specimen taken at different load levels. Bottom inset : The strain level averaged over the bottom region $\mathcal{R}_2$ indicated in panel (a) saturates after localization for $\Delta > \Delta_\mathrm{c}$ while the one measured in the upper region $\mathcal{R}_1$ follows the linear trend $\epsilon_\mathrm{yy}^\mathrm{ext} = \Delta /H = v_\mathrm{ext} \, t /L$ imposed by the loading machine. (c) Construction of an equivalent force control experiment from the mechanical response of the specimen measured under displacement control. The start and the end of a damage precursor taking place at a constant force $F_0$ are denoted by $\Delta_\mathrm{ini}$ and $\Delta_\mathrm{end}$. The precursor size $S_\mathrm{global} = \Delta E_\mathrm{d}$ defined as the dissipated energy during the event corresponds to half the work of the external force $\Delta W = F_0 \, (\Delta_\mathrm{end} - \Delta_\mathrm{ini})$. Inset : The stiffness degradation of the effective medium with increasing damage. (d) Variations of the precursor size $S$, in terms of dissipated energy, with the distance to failure $\delta$. Inset : Comparison of precursor sizes computed from the global analysis and the local analysis.}
\label{fig:fig1}
\end{figure}

\begin{figure*}
\centering
\includegraphics[width=0.95\linewidth]{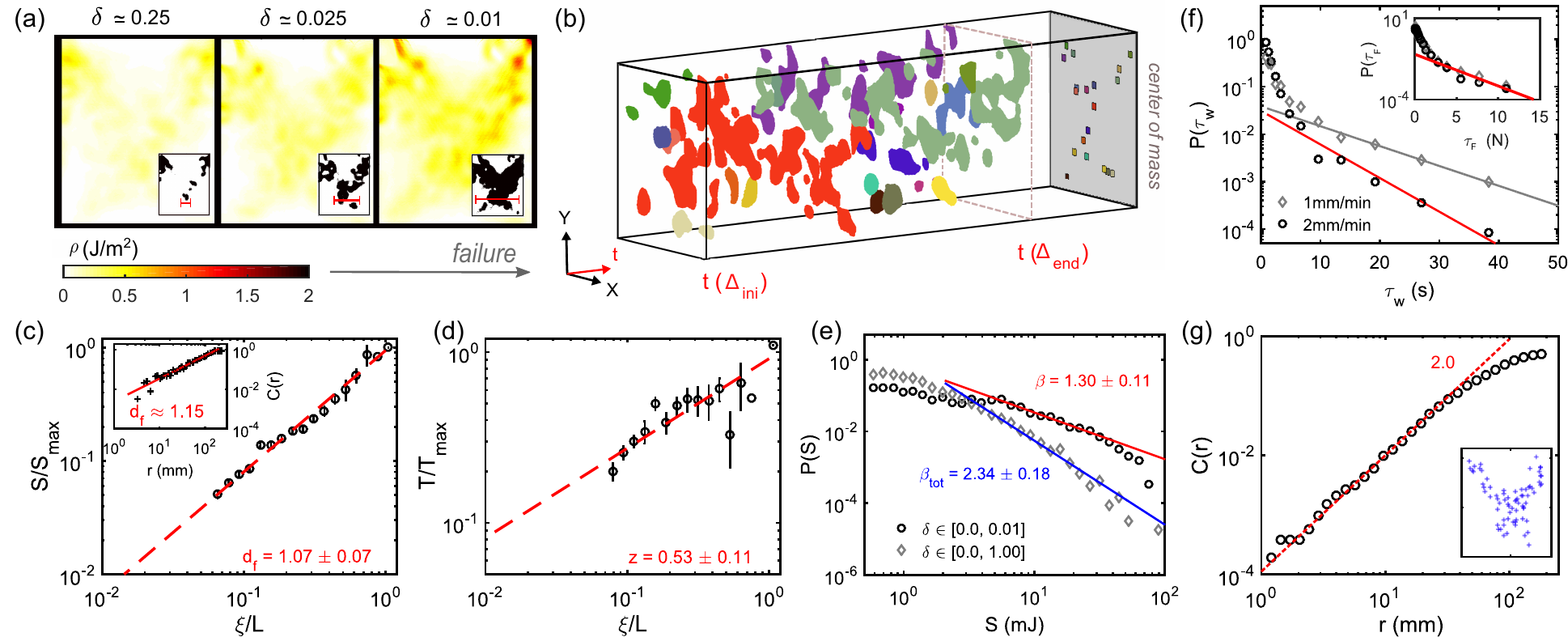}
\caption{(a) Energy density maps, $\rho(\vec{x})$, and their corresponding binary thresholded formats (inset) of energy dissipation at various distances to failure in a typical experiment. The common color bar depicting the scale for the dissipation energy density $\rho$ ($\mathrm{J/m^2}$) is presented under the first panel. The spatial extent of the cascades extracted using the auto-correlation of the thresholded maps are denoted by the red bracket in the inset. (b) Evolution of clusters within a cascade (second panel in (a), at $\delta \simeq 0.025$) with colors representing unique cluster IDs. The center of mass of each cluster is shown in the background with a marker of same color as the cluster. Here, the third axis is time.  The largest clusters appear in multiple slices of time. Scaling of the size (in terms of dissipated energy) (c) and duration (d) of cascades, normalized by their maximum values, as a function of their characteristic spatial extent $\xi$, normalized by the system size $L$. Inset in (c) : Pair correlation function of the centers of mass of the clusters within a cascade permitting to extract the fractal dimension $d_\mathrm{f} \simeq 1.15$.  (e) Distribution of the cascade sizes obtained during the whole duration of experiments (diamonds) and in the vicinity of final failure (circles). (f) Distribution of waiting time between cascades. The waiting time is defined as the difference in time-stamps of the arrival of two successive avalanches $\tau_w \mathrm{(s)} = t_i(\Delta_{ini}) - t_{i-1}(\Delta_{ini})$. Inset:  distribution of waiting times considered as the difference in value of force, $\tau_\mathrm{F} \mathrm{(N)} = F_i - F_{i-1}$. (g) Fractal analysis of the spatial distribution of the seeds of the precursors in a typical experiment (inset), showing a behavior reminiscent of a spatial Poisson process ($d_\mathrm{f} \simeq 2$). The statistical analyses is based on data recorded during ten experiments.}
\label{fig:fig2}
\end{figure*}

We now analyze the precursory damage activity taking place before peak load. A closer examination of the force-displacement curve in Fig.~\ref{fig:fig1}(c) reveals sudden force drops of amplitude much larger than the precision $\pm 0.05~\mathrm{N}$ of our load cell. These drops are followed by a linear increase of the force, recovering the force drop albeit with a degraded macroscopic stiffness (lower slope). The investigation of the specimen response under different loading rates shows similar behaviors. This alternating sequence of damage growth and elastic re-loading is reminiscent of the avalanche dynamics observed in driven disordered elastic systems~\cite{barabasi,sethna,wiese2}. We construct the mechanical response of the specimen (in red) in an equivalent force control experiment where displacement jumps (from $\Delta_\mathrm{ini}$ to $\Delta_\mathrm{end}$) at constant force correspond to a cascade of damage growth, also called an avalanche (Fig.~\ref{fig:fig1}(c)). The precursors defined this way can be shown to be statistically similar to those that would be measured during an actual force control experiment (see SI, Sec.~S4 for a numerical validation). As under force control conditions damage cascades take place at constant force, the work of the loading machine $\Delta W$ during the event can be shown to contribute equally to the increase $\Delta E_\mathrm{el}$ of the elastic energy and to the dissipation $\Delta E_\mathrm{d}$ by damage (see Appendix A3). Hence, $\Delta E_\mathrm{d} = \Delta W/2 = (\Delta_\mathrm{end} - \Delta_\mathrm{ini})\, F_0/2$ and we define this quantity as the precursor size $S_\mathrm{global}$.
The evolution of $S_\mathrm{global}$ with the distance to failure $\delta = (F_\mathrm{c} - F)/(F_\mathrm{c} - F_\mathrm{el})$ is shown in Fig.~\ref{fig:fig1}(d). The introduction of $F_\mathrm{el}$ ensures that $\delta = 1$ corresponds to the beginning of the damage accumulation regime. However, choosing another definition, as e.g., $(F_\mathrm{c} - F)/F_\mathrm{c}$, does not modify our conclusions. We note that the cascading dynamics and its amplification on approaching failure $(\delta \rightarrow 0)$ observed in our 2D cellular material under compression are reminiscent of the intermittent damage activity evidenced by acoustic emissions in standard brittle materials (see Fig.~\ref{fig:fig1}(d)).

Alternatively, we can also identify and characterize the precursors at the local scale using our (time-resolved) full-field measurement of the displacement and the damage field (see SI Sec.~1C and Figs. S1(c)-(f)~\cite{supp_info} for details on the local analysis). From these quantities, we compute the field of stored elastic energy in the effective elasto-damageable medium. Considering energy balance at the local scale, we can thus determine the dissipation energy density $\rho(\vec{x},\Delta)$ that we integrate over an avalanche $\rho(\vec{x}) = \int_{\Delta_\mathrm{ini}}^{\Delta_\mathrm{end}} \rho(\vec{x},\Delta) \, d\Delta$. 

Maps of dissipation energy density $\rho(\vec{x})$ depicting the complex spatial structure of precursors are presented in Fig.~\ref{fig:fig2}(a). We observe a diffuse pattern, yet containing locally well defined regions of varying intensity and size. These  clusters are reminiscent of the time and space correlated structure of incremental damage events. In practice, highly correlated individual damage events can be grouped together by implementing a spatio-temporal clustering algorithm on the fields $\rho(\vec{x},\Delta)$ recorded during $\Delta_\mathrm{ini}< \Delta < \Delta_\mathrm{end}$. This reveals the cluster-like structure of precursors, illustrated in Fig.~\ref{fig:fig2}(b) (see also Video S2) for the precursor shown in the second panel of Fig.~\ref{fig:fig2}(a).
 
The energy $S_\mathrm{local} = \int \rho(\vec{x}) d\vec{x}$ dissipated during the cascade compares well with the precursor size $S_\mathrm{global}$ inferred from the force-displacement response, see Fig.~\ref{fig:fig1}(d). It is also in good agreement with the precursor size computed using the field of dissipated energy inferred from the incremental damage field, thus validating the assumption of local energy balance (see SI Sec. S2, Figs. S2(d) and S2(e)~\cite{supp_info}). Finally, this agreement comes in support to the description of our cohesionless cellular solid as an elasto-damageable medium.

\subsection{Statistical characterization of precursors}
We now explore the properties of the damage cascades observed in our experiments. First, their spatial extent is determined from (thresholded) maps of dissipation energy density (insets of Fig.~\ref{fig:fig2}(a)). The employed threshold value $\rho^\ast$ is inferred from the distribution of local dissipation densities that follows an exponential decay $P(\rho) \propto \mathrm{e}^{-\rho / \rho^\ast}$   (see SI Sec.~S1(c) and Fig. S1(g)~\cite{supp_info}). We extract the characteristic length $\xi$ from the 2D auto-correlation of thresholded dissipation map (SI Sec.~S1(d) and Fig. S1(h)~\cite{supp_info}). The length grows with the avalanche size as $S \propto \xi^{d_\mathrm{f}}$, where $d_\mathrm{f} \simeq 1.07$ is the fractal dimension, see Fig.~\ref{fig:fig2}(c). While cascades spread over the whole specimen (see Fig.~\ref{fig:fig2}(a)), $\xi$ represents the spatial extent of the largest clusters constituting the cascade. Interestingly, $\xi$ reaches the specimen size $L$ on approaching failure, which implies an upper limit on the size of the precursors.
An independent estimate of the fractal dimension of the precursors is obtained from the spatial distributions of the clusters. We identify the location of their center of mass, as illustrated on the right-end of Fig.~\ref{fig:fig2}(b), and then compute the correlation function $C(r) \propto r^{d_\mathrm{f}}$ defined as the fraction of pairs of points whose separation is less than $r$~\cite{hentschel1983}. This provides a fractal dimension $d_\mathrm{f} \simeq 1.15$ (see inset of~Fig.~\ref{fig:fig2}(c)) compatible with the one obtained from the spatial distribution of the individual damage events. Note that the spatial extent $\xi$ of a damage cascade is different from its size $S$ that corresponds to the energy dissipated by damage during the cascade.

We now seek to determine the characteristic duration of damage cascades. We come back to the force-displacement response and explore the sequence of load drops observed within an avalanche. The precursor duration $T$ is defined as the number of load drops (also see SI Sec.~S1B and Figs.~S1(a) and S1(b)~\cite{supp_info}). It scales with the characteristic length $\xi$ of the precursor as $T \propto \xi^{z}$, see Fig.~\ref{fig:fig2}(d), with dynamic exponent $z \simeq 0.53$.

Thus, a damage cascade is characterized by its size, its spatial extent and its duration. All three quantities are related to each other by scaling laws. The probability distribution of these quantities is studied in~Fig.~\ref{fig:fig2}(e), where we focus on the distribution $P(S)$ of precursor sizes, the other distributions $P(\xi)$ and $P(T)$ being inferred from the previous scaling laws. Considering all the precursors ($\delta \in [0, 1]$) or only the ones close to localization $(\delta \in [0, 0.01])$, both distributions follow a power-law statistics but with two different exponents $\beta_\mathrm{tot} \simeq 2.34$ and $\beta \simeq 1.30$, respectively. This difference results from the 
increase of the size of the largest precursors on approaching failure (see SI Sec.~S4A~\cite{supp_info,amitrano2012}). The exponent $\beta$ will be connected later with the marginal stability of the material elements~\cite{lin2014,lin2015} that will be investigated in Sec.~\ref{sec4}A.

Finally, we characterize the correlations in the sequence of damage cascades.  The distribution  $P(\tau_\mathrm{w})$ of waiting times separating two successive damage cascades is shown in Fig.~\ref{fig:fig2}(f). It follows an exponential law $P(\tau_\mathrm{w}) \propto \mathrm{e}^{-\tau_\mathrm{w}/\tau_\mathrm{w}^\star}$, defining a characteristic waiting time $\tau_\mathrm{w}^\star$. This result is at odd with the power-law distribution of waiting times separating acoustic events in compression experiments~\cite{davidsen2007,rosti2009,baro2013}. To confirm our observation, we perform additional experiments with a loading rate $v_\mathrm{ext}$ twice smaller. Interestingly, we also measure an exponential distribution, but with a characteristic waiting time about twice larger (see Fig.~\ref{fig:fig2}(f)). We thus replace the waiting time $\tau_\mathrm{w}$ by the force increment $\tau_\mathrm{F}$ separating two successive precursors, so that distributions corresponding to different loading rates collapse on a single curve, as shown on the inset of Fig.~\ref{fig:fig2}(f). The exponential distribution of waiting time, characteristic of uncorrelated events described by a Poisson process, suggests that precursors are triggered independently from each others. This is further confirmed by the spatial distribution of the seeds (first damage event) of precursors that we define as the center of mass of the cluster appearing at $t(\Delta_{ini})$.  The fractal analysis of the precursor seeds provides $C(r) \propto r^{2}$ a behavior reminiscent of spatially uncorrelated events (see Fig.~\ref{fig:fig2}(g)).

\section{Theoretical modeling of compressive failure}
\label{sec3}
The statistical features of the precursors measured in our experiments strikingly remind the avalanche dynamics of elastic interfaces driven in disordered media. In these models, an elastic interface responds to a continuously increasing  drive and exhibits scale free avalanches or crackling noise~\cite{barabasi,sethna,wiese2}. The size, spatial extent and duration of the avalanches are related by scaling laws with universal exponents that depend on the interface elasticity and its dimension. For interfaces with long-range elasticity, avalanches are formed by a set of correlated clusters that are spatially disconnected similar to damage clusters within a cascade observed in our experiments (see Fig.~\ref{fig:fig2}(b)). Taking inspiration from Weiss~{\it et al.}~\cite{weiss2014} and using the non-local theory proposed in Dansereau~{\it et al.}~\cite{dansereau2019}, we derive below an evolution equation of the damage field in the specimen for a compression test under force control conditions. This theoretical formulation sheds light on the connection with models of driven disordered elastic interfaces. We provide here the main ingredients of the derivation of the damage evolution law, the detailed calculations being presented in SI Sec.~S3A-D~\cite{supp_info}

First, we assume that the material behaves as an elasto-damageable solid. We thus introduce a damage field $d(\vec{x},t)$ that describes the level of damage accumulated in the specimen at the location $\vec{x}$ and time $t$. Damage growth is inferred from a balance of energy, by comparing two quantities: the local driving force $Y[d(\vec{x},t),t]$ which provides the rate of elastic energy released for an incremental growth of damage, and the damage resistance $Y_\mathrm{c}[d(\vec{x},t)]$ which provides the material resistance to damage and corresponds to the rate of energy dissipated for an incremental growth of damage. The first quantity is similar to the elastic energy release rate introduced in fracture mechanics, which drives crack propagation and is a quadratic function $Y \propto \sigma_0^2$ of the nominal compressive stress $\sigma_\circ = F/(L \, W)$ applied by the test machine~\cite{lawn,berthier_thesis}. The damage resistance  is equivalent to the fracture energy introduced in fracture mechanics.  Note that in our model, $Y_\mathrm{c}$ depends not only on $\vec{x}$ as precursors emerge from the material inhomogeneities, but also on $d$ as a damage event in $\vec{x}$ may change the subsequent failure resistance in the same material element. The damage then increases in the material element $\vec{x}$ if the local value of the driving force $Y(\vec{x})$ reaches the material resistance $Y_\mathrm{c}(\vec{x})$.

To describe the damage field fluctuations resulting from the material heterogeneities, we introduce a reference damage level $d_\circ = \langle d(\vec{x},t_\circ)\rangle$ and the damage field perturbations $\Delta d(\vec{x},t) = d(\vec{x},t)  - d_\circ$ over the time $\delta t  = t - t_\circ  \ll t_\circ$ to ensure that $\langle \Delta d(\vec{x},t) \rangle_{\mathrm{\vec{x}}} \ll d_\circ$. The driving force and the damage resistance  can then be linearized as $Y[d(\vec{x},t),\sigma_\circ] = Y_\circ(d_\circ,\sigma_\circ) + \Delta Y[\Delta d(\vec{x},t), \sigma_\circ]$ and $Y_\mathrm{c}[d(\vec{x},t)] = Y_\mathrm{c \circ}(d_\circ) + \Delta Y_\mathrm{c}[\Delta d(\vec{x},t)]$. The zero-order equation $Y_\circ(d_\circ,\sigma_\circ) = Y_\mathrm{c \circ}(d_\circ)$ provides the relationship between the reference damage level $d_\circ$ and the reference applied load $\sigma_\circ = \sigma(t_\circ)$. In the following, we investigate how the damage field perturbations $\Delta \dot{d}(\vec{x},t) \propto \Delta Y[\Delta d(\vec{x},t),\sigma_\circ] - \Delta Y_\mathrm{c}[\Delta d(\vec{x},t)]$ evolves over time. We write the total driving force as the sum of three terms~\citep{weiss2014,dansereau2019}, 
\begin{equation}
\begin{split}
  \Delta \dot{d}(\vec{x},t) \propto & \, \mathcal{K}(\sigma_\circ) \left[v_\mathrm{m}(\sigma_\circ) \, t - \Delta d(\vec{x},t) \right] +  \\
       &\psi(\sigma_\circ) \ast  [\Delta d(\vec{x},t) - \langle \Delta d\rangle_\mathrm{\vec{x}}] - y_\mathrm{c}[\vec{x}, d(\vec{x},t)].   
\end{split}
\label{eq1}
 \end{equation}
The first (local) term comprises the effect of the driving by the test machine, where the driving speed $v_\mathrm{m} \propto v_\mathrm{ext}$ sets the damage growth rate. Considering a pseudo-interface of position $\Delta d(\vec{x},t)$, this term acts as a rigid plate moving at a speed $v_\mathrm{m}$ and pulling on the interface with springs of stiffness $\mathcal{K}$ (see SI Fig.~S3~\cite{supp_info} for a schematic representation). The second term is non-local. It describes the interactions within the specimen, its values in $\vec{x}$ depends on the damage level $\Delta d(\vec{x},t)$ everywhere in the specimen. In practice, the kernel $\psi(\sigma_\circ)$ (provided in Eq.~\eqref{kernel}) describes the spatial structure of the redistribution of driving force taking place in the aftermath of an individual damage event. It decays as $\psi \propto 1/r^{2}$~\cite{weiss2014,dansereau2019}. Its angular dependence is shown in Fig.~\ref{fig:fig3}(a) for the particular case of uni-axial compression. It exhibits a quadrupolar symmetry with non-positive regions (in blue).
Hence only a fraction of the neighboring elements are reloaded in the aftermath of a damage event while the others (located above and below the damaged element) are actually unloaded. The third term represents the effect of material disorder. Its spatial average is close to zero as the contribution of the hardening $\langle \Delta Y_\mathrm{c} \rangle_\mathrm{\vec{x}}= \eta \, \langle \Delta d \rangle_\mathrm{\vec{x}}$ (where $\eta$ is a hardening parameter and observed in our experiments) is taken into account in the first term (see SI Sec.~S2 and Figs.~S2(b) and S2(c)~\cite{supp_info}). The presence of $d(\vec{x},t)$ as an argument of the disorder term implies that the evolution equation~\eqref{eq1} is strongly non-linear, leading to the rich phenomenology that we now discuss.

Equation~(\ref{eq1}) provides a clear connection between damage evolution and disordered elastic interfaces: the accumulated damage field is analogous to a 2D elastic interface $\Delta d(\vec{x},t)$ driven at the speed $v_\mathrm{m}$ through a 3D disordered medium (see SI Sec.~S3E, Fig.~S3~\cite{supp_info}). As a result, damage is expected to grow through bursts characterized by scaling laws involving critical exponents reminiscent of the so-called depinning transition. As a first test of our model, we compare the theoretically predicted exponents  with the one measured experimentally. Investigating the avalanches dynamics of 2D interfaces with non-positive interactions in the context of amorphous plasticity, Lin~{\it et al.}~\cite{lin2014,lin2016} predicted the exponent values $\beta = 1.51$, $d_\mathrm{f} = 1.10$ and $z = 0.57$ that agree reasonably well with the ones measured in our experiments. Corrections to these predictions from the numerical solution of the evolution equation~(\ref{eq1}) are provided at the end of our manuscript, in section \ref{num_model}. They improve further the agreement with the experimentally measured exponents.

Despite the ability of this approach to describe the scaling behavior of precursors, we note that two important features of the damage evolution equation (\ref{eq1}) differ from standard models of driven elastic interfaces. First, the long-range elastic interactions result in both reloading and unloading of material elements in the aftermath of a damage event. Second, a subtler but more important feature  is that the evolution equation~\eqref{eq1} describes a non-stationary depinning scenario culminating in a bifurcation at the localization threshold. In the following, we examine these aspects in details and discuss their implications on the damage accumulation process preceding failure.

\section{Atypical aspects of damage evolution as a driven disordered elastic interface}
\label{sec4}
\subsection{Non-positive elastic interactions} 
\begin{figure}[t]
\includegraphics[width=\linewidth]{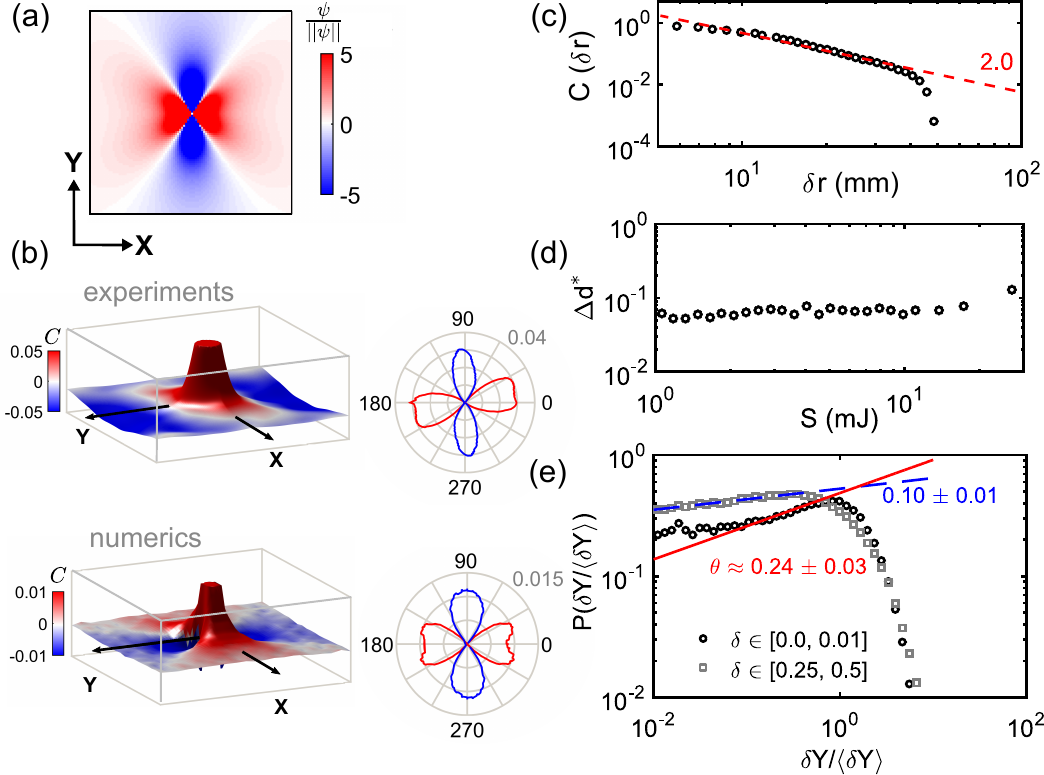}
\caption{(a) Angular distribution of the long-range interaction kernel derived theoretically for the case of uni-axial compression. (b) 2D auto-correlation map of the incremental damage field of precursors and angular distribution of the correlations at a fixed distance as obtained from the 50 largest avalanches measured in one of our experiments and our numerical simulations. (c) Variations of the correlation function of the incremental damage field with distance along the horizontal $x$-axis and comparison with the scaling $\psi \propto 1/\delta r^2$ of the theoretical interaction kernel (dashed line). (d) Variations of the depth $\Delta d^{\ast}$ of the damage cascade with the cascade size $S$. (e) Distribution of the local distance to failure $\delta Y(\vec{x})$ close and far from failure.}
\label{fig:fig3}
\end{figure}

We describe below a methodology that we employ to characterize the elastic interactions driving the cooperative dynamics of damage. This method takes advantage of our full field characterization of damage precursors. Indeed, it turns out that the spatial distribution of events within an avalanche encrypts the range and the anisotropy of the elastic interactions, as we will see below. The first step consists in computing the incremental damage field $\delta d(\vec{r})$ during an avalanche. Its 2D auto-correlation function  $C(\vec{\delta r}) = \langle \delta d(\vec{r}) \cdot \delta d(\vec{r} + \vec{\delta r}) \rangle_\mathrm{\vec{r}}$ averaged over several avalanches is presented in Fig.~\ref{fig:fig3}(b). The angular distribution (at a fixed distance) shows a clear quadrupolar symmetry similar to the re-distribution pattern of the theoretical interaction kernel (Fig.~\ref{fig:fig3}(a)).
Remarkably, the correlation along the horizontal axis where the reloading is maximal decays as $C(\delta r) \propto 1 / \delta r^2$ (see Fig.~\ref{fig:fig3}(c)), a behavior also in line with the theoretical predictions of the elastic kernel, Eq.~\eqref{kernel}. These observations support further the applicability of the non-local damage mechanics to our experiments.

The presence of an unloading region in the interaction kernel has several important implications.
First at the scale of a material element, the approach to failure is non-monotonic as the local driving force for damage can both increase and decrease over time. As a result, the probability that an elements damages more than an once during a cascade is low. In practice, the depth of avalanches $\Delta d^\ast$ defined as the average damage increment of the elements involved in the cascade is constant and does not vary with the precursor size $S$, as shown in Fig.~\ref{fig:fig3}(d) (see also SI Sec. S4A~\cite{supp_info}). This  is at odds with the behavior of driven elastic interfaces with positive interactions for which the depth of avalanches scales with their size. Another crucial, yet more subtle difference relates to the distribution of net driving force $\delta Y(\vec{x}) = Y_\mathrm{c}(\vec{x}) - Y(\vec{x})$ that controls the (marginal) stability of the specimen. $\delta Y(\vec{x}) > 0$ provides the increment of driving force required for triggering damage. Its distribution, computed over all the material elements, is expected to scale as $P(\delta Y) \propto \delta Y^\theta$~\cite{lin2014,lin2016,lin2015,liu2016, budrikis2017}. Positive interactions lead to $\theta = 0$, pointing out the presence of a finite number of material elements close to failure. On the contrary, the number of elements close to failure vanishes for sign-changing interactions, leading to $\theta > 0$. The experimental determination of the exponent $\theta$ is quite challenging, as it requires a priori the knowledge of the material disorder. In practice, we circumvent this difficulty by computing  the driving force $Y(\vec{x},t)$ (according to SI Eq.S4~\cite{supp_info}) at each time step for each material element and determine $Y_\mathrm{c}[d(\vec{x},t)]$ retrospectively from the value of $Y(\vec{x},t)$ when the material element damages (see  Appendix \ref{app_a}4, SI Sec.~2 and Fig.~S2(a)~\cite{supp_info} for details on the method). Figure~\ref{fig:fig3}(e) shows the distribution $P(\delta Y/\langle \delta Y \rangle)$ close and far from localization. In both cases, we measure a positive exponent $\theta > 0$, a particularly non-trivial property that comes in support of the proposed model. Interestingly, $\theta$ increases as the specimen approaches failure, a feature that possibly arises from the non-stationary nature of the evolution equation~(\ref{eq1}). The exponent $\theta$ has been shown to be related to $\beta = 2 - \frac{\theta}{\theta+1}\frac{d}{d_\mathrm{f}}$~\cite{lin2014} that characterizes the distribution of precursor size $S$. This scaling relation predicts a slight decrease of $\beta$ on approaching failure, a subtle effect that is consistent with our experimental observations. A similar trend of an increasing $\theta$ has been reported in direct simulations of sheared amorphous solids that are also characterized by sign-changing interactions~\cite{lin2015, ozawa2018}.

\subsection{Divergence of precursors}
\begin{figure}[bp]
\includegraphics[width=\linewidth]{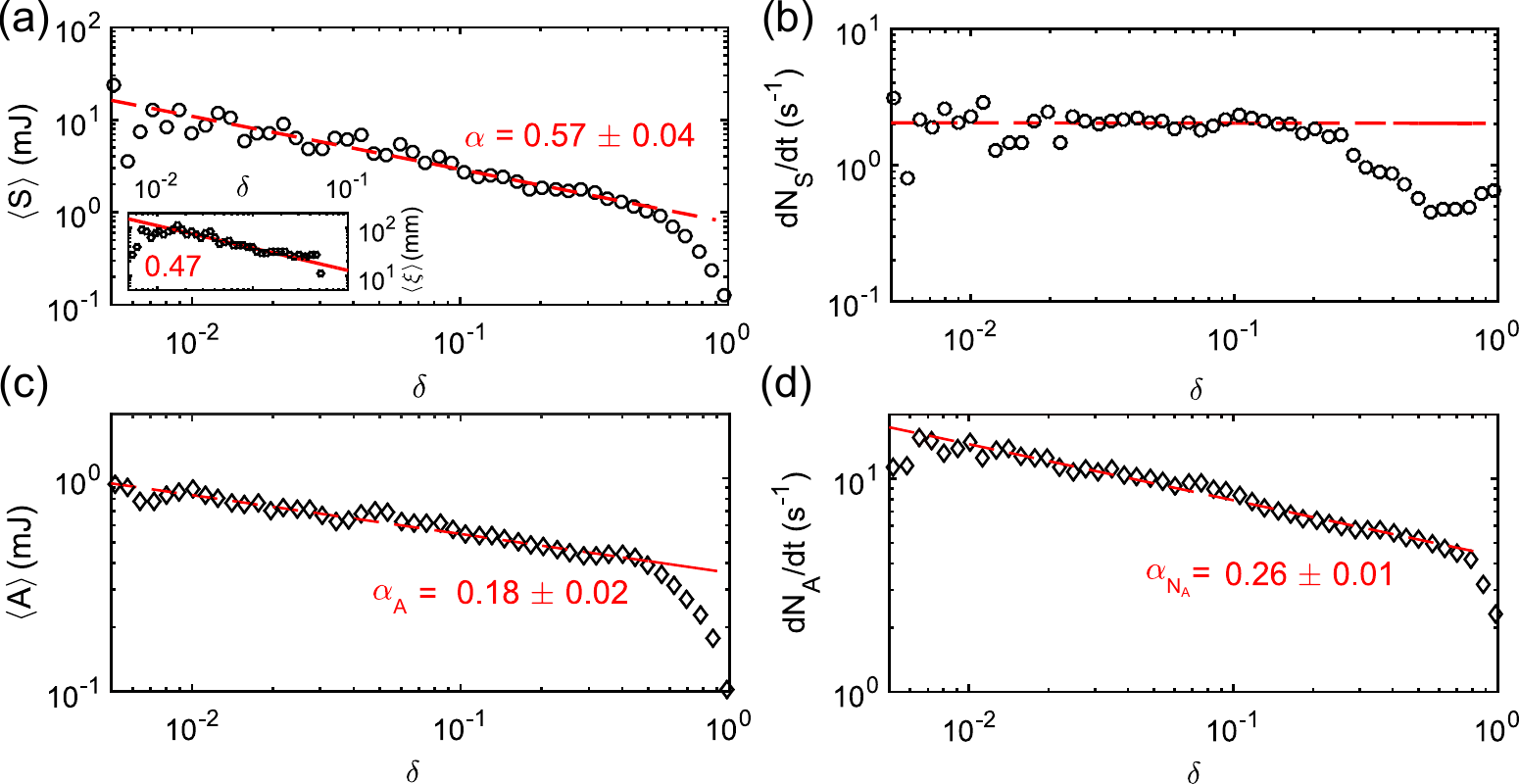}
\caption{Variation with distance to failure of (a) the average precursor size $\langle S \rangle$, the average spatial extent of precursors $\langle \xi \rangle$ in the inset, (b) the activity rate  $dN_\mathrm{S}/dt$, (c) the average size of load drop events $\langle A \rangle$ and (d) the event activity rate $dN_\mathrm{A}/dt$. While $S$ and $dN_\mathrm{S}/dt$ characterize the intermittent damage activity under force controlled conditions, $A$ and $dN_\mathrm{A}/dt$ are relevant for displacement imposed conditions. The product of both quantities $dE_d/dt = \langle S \rangle\,dN_\mathrm{S}/dt = \langle A \rangle \,dN_\mathrm{A}/dt $ provides the dissipation rate that also diverges as $dE_d/dt \sim \delta^{-\alpha}$ with $\alpha = 1/2$.}
\label{fig:fig4}
\end{figure}

We now come back on the observations made in Fig.~\ref{fig:fig1}(d) and Fig.~\ref{fig:fig2}(a)  of an increase of the size and the spatial extent of precursors close to failure. As shown in Fig.~\ref{fig:fig4}(a), the average precursor size increases as a power-law with the distance to failure, $\langle S \rangle \sim 1/\delta^{-\alpha}$  where $\alpha \simeq 0.57$. Following the scaling relations $S \propto \xi^{d_{\mathrm{f}}}$ and $T \propto \xi^z$, the associated length and time scales then also diverge on approaching failure. This is confirmed by the variations of the precursor spatial extent directly measured from our local analysis (see SI Sec.1D~\cite{supp_info})  in the inset of Fig.~\ref{fig:fig4}(a). 

Notably, the activity rate $dN_S/dt$ i.e., the number of cascades per interval of time is rather constant during damage accumulation, see Fig.~\ref{fig:fig4}(b). This is in line with our previous observation of an exponential distribution of waiting times, supporting further that precursors emerge from a (random) Poisson process. As the dissipation rate $dE_\mathrm{d}/dt$ during the intermittent damage evolution writes as the product of the average precursor size with the precursor rate, $dE_\mathrm{d}/dt = \langle S \rangle \, dN_\mathrm{S}/dt$, we  obtain $dE_\mathrm{d}/dt \propto 1/\delta^{-\alpha}$. As argued in the next section, the divergence of the dissipation rate on approaching peak load is reminiscent of damage localization, a feature that results from the loss of stability of the specimen. To further test this idea, we reanalyze our data considering the actual displacement imposed conditions that also gives rise to damage localization at peak load. The dissipation rate then writes as the product of the average load drop size $\langle A \rangle$  with their activity rate $dN_A/dt$ (refer to SI Sec.1B and Fig.S1(b)~\cite{supp_info} for the definition of precursors under displacement imposed loading conditions). We observe that both load drops $\langle A \rangle \propto \delta^{-\alpha_{\mathrm{A}}}$ and precursor rate $dN_A/dt \propto\delta^{-\alpha_{N_A}}$ diverge on approaching localization, see Fig.~\ref{fig:fig4}(c) and (d). The exponent $\alpha = \alpha_A + \alpha_{N_A} \simeq 0.44$ characterizing the divergence of the dissipation rate under displacement imposed conditions is close to $1/2$, as expected and accounts for the numerical observations of Girard~\textit{et al.}~\cite{girard2010} who reported $dE_\mathrm{d}/dt \propto \delta^{-0.4}$. 

\section{Damage localization : depinning transition or standard bifurcation ?} 
\label{sec5}

The divergence of the precursor size, and hence of the characteristic timescale and length scale of precursors on approaching failure supports \textit{a priori} the interpretation of compressive failure as a critical point~\cite{weiss2014,weiss2019}. However, the comparison between the scaling exponents predicted by the critical point scenario and the exponents measured in our experiments tells a different story (see Table~\ref{tab:tab0}). 
\begin{table}[h]
\centering
\caption{Comparison between the exponents predicted by the critical point scenario that interprets failure as a depinning transition and the exponent measured in our experiments. The three exponents provided below describes the divergence $\sim \delta^{-\gamma}$ of the avalanche size $S$, its spatial extent $\xi$ and its duration $T$ on approaching failure. The theoretical exponents describe the behavior of a driven 2D interface with long-range non-positive elastic interactions approaching the depinning transition~\cite{lin2014,weiss2019}.}
\begin{ruledtabular} 
\begin{tabular}{ccccc}
   & \multicolumn{2}{c}{Critical point scenario} & \multicolumn{2}{c}{Experiments} \\ 
\hline
 & $\gamma$ &  prediction - 2D & $\gamma$ &  \\ \hline
 $S$ & $\nu d_\mathrm{f}$ &  1.27 & $\alpha$ & 0.57 \\
$\xi$ & $\nu$ &  1.16 & $\alpha/d_\mathrm{f}$ & 0.53 \\
$T$ & $\nu z$ &  0.66 & $z\alpha/d_\mathrm{f}$ & 0.28 \\
\end{tabular}
\end{ruledtabular}
\label{tab:tab0}
\end{table}

To explain the difference between the theoretical predictions and our experimental observations, we further develop the model proposed in Sec.~\ref{sec3}. We focus on the non-stationary aspects of the evolution equation~\eqref{eq1}, namely the stiffness $\mathcal{K}$ and the driving speed $v_\mathrm{m}$ that are given by the following expressions:
\begin{equation}
    \begin{split}
        \mathcal{K}(\sigma_\circ) &= \left.\frac{\partial (Y_\mathrm{c\circ}- Y_\circ)}{\partial d_\circ}\right|_{\sigma_\circ} 
        \\
        v_\mathrm{m}(\sigma_\circ) &= v_\mathrm{ext}\left(\frac{\left.\partial Y_\circ/ \partial \sigma_\circ\right|_{d_\circ}}{\mathcal{K}(\sigma_\circ)}\right).
    \end{split}
    \label{eq2}
\end{equation}
$\mathcal{K}$ controls the stability of the damage evolution. Indeed, a negative value of $\mathcal{K}$ implies that the net driving force $Y - Y_\mathrm{c}$ increases with the damage level, leading to its unstable growth and thus failure. It turns out that $\mathcal{K}$ goes to zero on approaching peak load (see SI Sec.~S3F~\cite{supp_info}), in line with the stability condition under force controlled conditions. The driving speed $v_\mathrm{m}$ is inversely proportional to $\mathcal{K}$ and hence goes to infinity. A linear expansion of the damage evolution equation close to peak load $\sigma_\circ < \sigma_\mathrm{c}$ (see SI Sec.~S3G~\cite{supp_info}),  provides the  asymptotic behavior of the damage growth rate $\Dot{\Delta d} \sim 1/\sqrt{\delta}$. As $\Delta \dot{d} \propto v_\mathrm{m}$, owing to~\eqref{eq2} we obtain $v_{\mathrm{m}} \sim 1/\sqrt{\delta}$ and thus $\mathcal{K} \sim \sqrt{\delta}$.

What are then the consequences of the divergence of the speed of the pseudo-interface at peak load ? 
As the rate of dissipated energy is controlled by the damage growth rate, one expects $dE_\mathrm{d}/dt \propto 1/\sqrt{\delta}$. Considering the intermittency of damage evolution, the dissipation rate writes as the product of the precursors' size with the precursors' rate, $dE_\mathrm{d}/dt = \langle S \rangle \, dN_\mathrm{S}/dt$. As $dN_\mathrm{S}/dt$ remains constant during the experiment (Fig.~\ref{fig:fig4}(b)), a feature expected for disordered elastic interfaces, it follows that $\langle S \rangle \sim 1/\sqrt{\delta}$, a prediction that accounts for our experimental observations.

The implication of our model is clear: the divergence of the size of the precursors close to failure, and hence the divergence of the length scale and time scale of the fluctuations, result from the presence of a standard bifurcation at peak load. The progressive loss of stability of the specimen on approaching peak load is accompanied by a divergence of the damage growth rate (and thus a divergence of the precursor size), a behavior that has nothing to do with the presence of disorder. This mechanism is in stark contrast with the divergence of fluctuations near a critical point, a feature that vanishes if the disorder is shut down. 
\begin{figure}[t]
\centering
\includegraphics[width=0.9\linewidth]{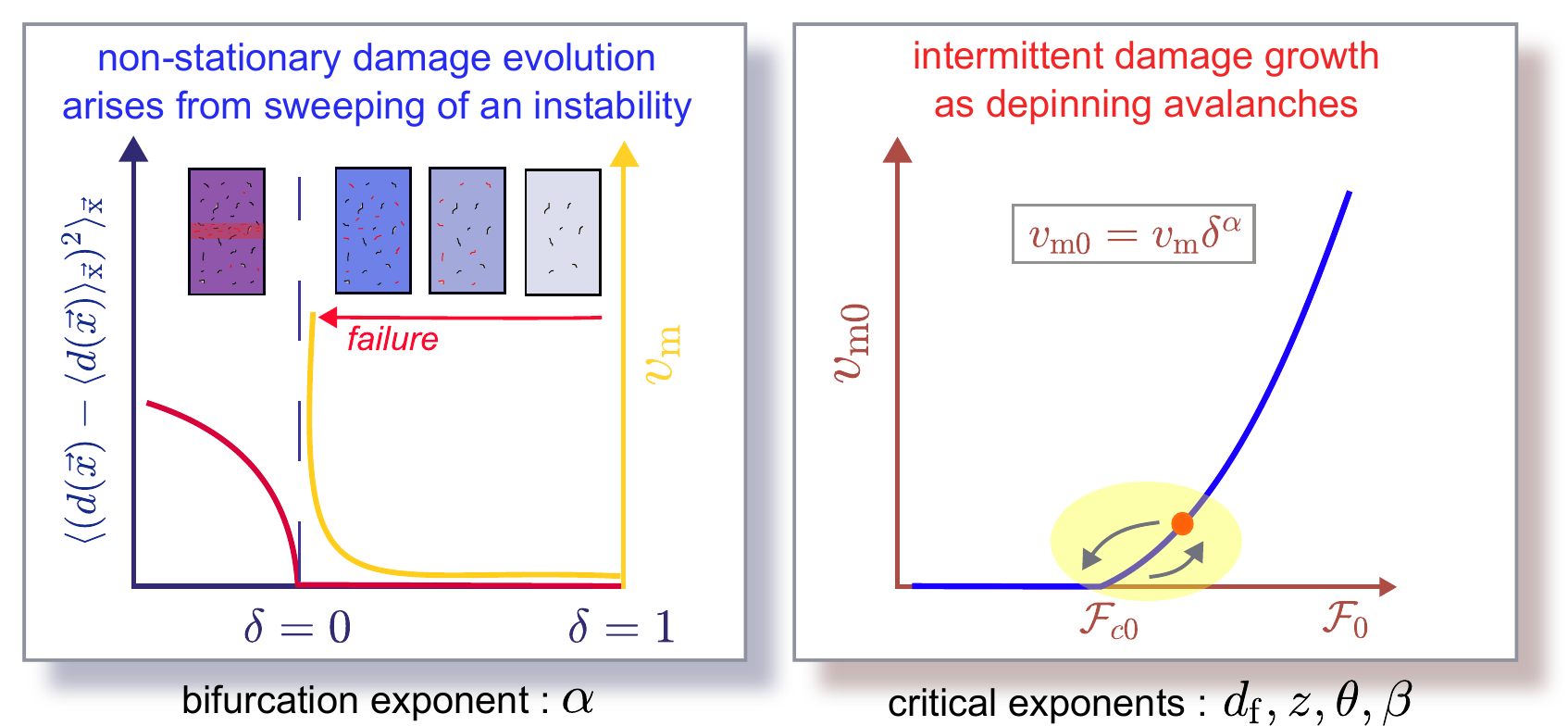}
\caption{Interpretation of the two phenomena underlying compressive failure of disordered solids. The left panel describes the behavior of the specimen as it is driven to failure. The amplitude of the fluctuations of the damage field (in red) and the dissipation rate (yellow) increase near failure ($\delta = 0$). The spatial variations $\langle (d(\vec{x}) - \langle d(\vec{x})\rangle_{\vec{\mathrm{x}}})^2\rangle_{\vec{\mathrm{x}}}$ of the accumulated damage are essentially zero before peak load $(\delta < 0)$, but rapidly increase after localization $(\delta > 0)$. The state of the specimen bifurcates from a homogeneously growing damage to a localized damage growth. This bifurcation is preceded by an increase $v_\mathrm{m} \sim \delta^{-\alpha}$ of the dissipation rate, and thus an increase of $\langle S\rangle \sim \delta^{-\alpha}$ of the fluctuation amplitude. The right panel describes the process of damage growth at a fixed distance to failure. By renormalizing the non-stationary features of the evolution equation (see Eq.\eqref{eq3}), we realize that the damage field behaves as an elastic interface driven in a disordered medium at constant speed. $v_{\mathrm{m}0} \simeq v_\mathrm{m}\delta^\alpha$ at a fixed distance above the critical point (here, the depinning transition) even when we reach peak load ($\delta = 0$) and thus bifurcation.}
\label{fig:fig5}
\end{figure}

To further support this claim, we rewrite the damage evolution law~(\ref{eq1}) using the new variable $\underset{\widetilde{}}{d}(\vec{x})  = d(\vec{x}) \, \sqrt{\delta}$. The obtained expression ensures a straightforward connection with standard (stationary) depinning models~\cite{barabasi, wiese2}:
\begin{equation}
\begin{split}
  \underset{\widetilde{}}{\Delta \dot{d}} &\propto \mathcal{K}_0 \left[v_{\mathrm{m}0} \, t - \Delta \underset{\widetilde{}}{d}(\vec{x},t) \right] +  \\
       &~~~~ {\psi} (d_\circ \sqrt{\delta})/\sqrt{\delta} \ast  [\Delta \underset{\widetilde{}}{d}(\vec{x},t) - \langle \Delta \underset{\widetilde{}}{d}\rangle_\mathrm{\vec{x}}] - y_c[\vec{x}, \underset{\widetilde{}}{d}(\vec{x},t)].   
\end{split}
\label{eq3}
 \end{equation}
Under this form, the evolution equation displays both a constant stiffness $\mathcal{K}_0 = \mathcal{K}/\sqrt{\delta}$ and a constant driving speed $v_{\mathrm{m}0} = v_\mathrm{m}\sqrt{\delta} $. The normalized damage field $\underset{\widetilde{}}{d}$ behaves as an elastic interface driven at constant finite speed during the whole damage accumulation regime. As a result, the specimen remains at a fixed distance to the depinning transition that corresponds to the limit $v_\mathrm{m0} \rightarrow 0$ and $\mathcal{K}_0 \rightarrow 0$. Our conclusions are in stark contrast with a depinning scenario where the divergence of the precursory activity results from the interpretation of compressive failure as a critical transition and the evolution of the specimen towards this critical point. In other words, the increasing applied load drives the specimen towards instability without driving it towards more criticality. The intricate connection between damage precursors and localization is illustrated in  Fig.~\ref{fig:fig5}. 

We now would like to  highlight the strategic value of our findings to structural health monitoring. In our theoretical description of compressive failure, the evolution of  precursors is described by robust scaling laws that are independent of the material properties and the loading conditions. In particular, as they apply for both force and displacement driven experiments, they serve as early warning signals of impending failure.

\section{Failure prediction from precursory activity}
\label{sec6}
\begin{figure}[bp]
\centering
\includegraphics[width=0.75\linewidth]{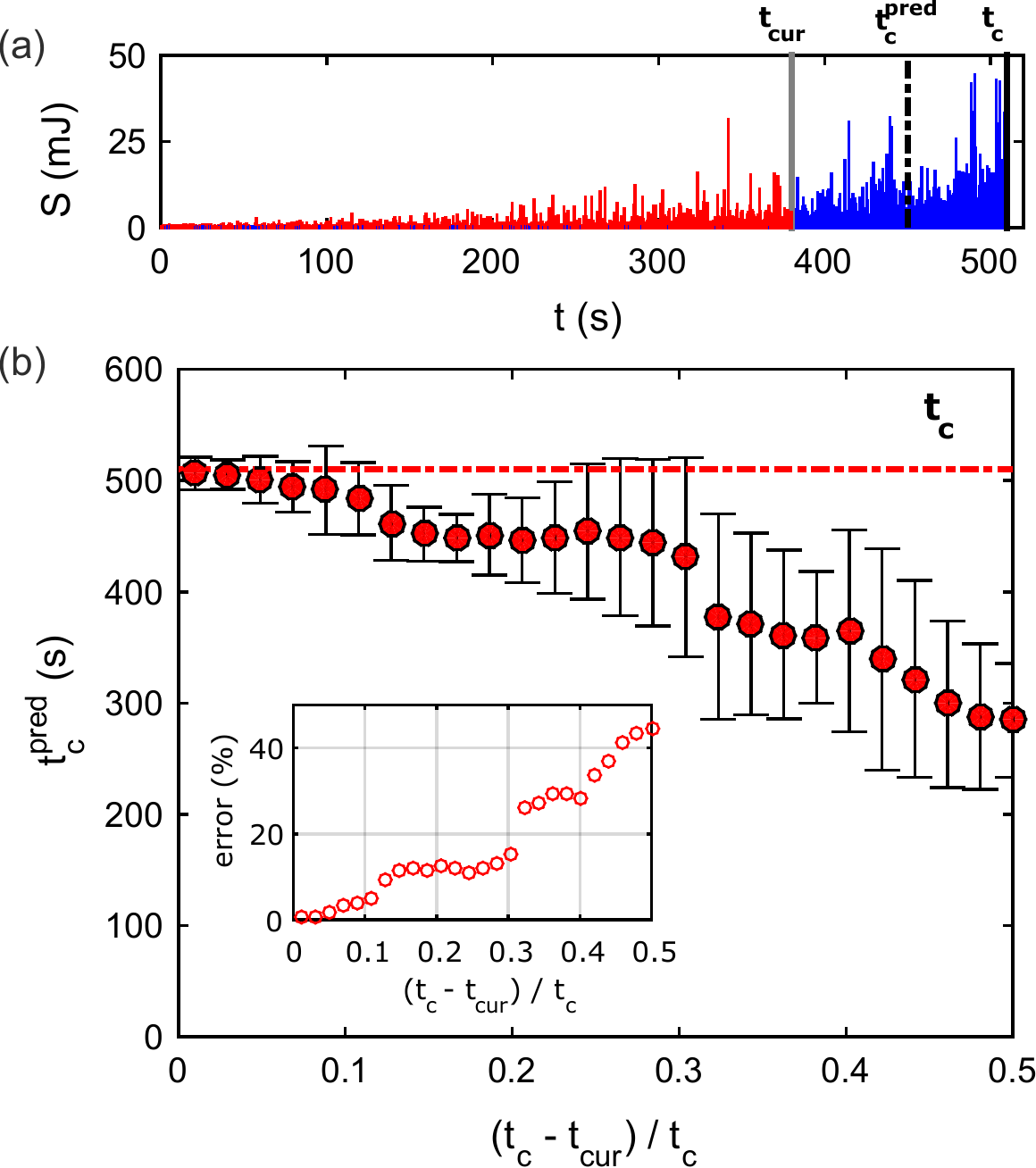}
\caption{(a) Time series data of failure precursors obtained during the compression test and data available for prediction (red) at $t_\mathrm{cur}$. (b) Variation of the predicted remaining lifetime at different instances $t_\mathrm{cur}$ represented as a fraction of the time to failure. The error bars provide intervals with 90$\%$ confidence levels. Inset : Error on the predicted remaining lifetime.}
\label{fig:fig6}
\end{figure}
We now harness the scaling behavior of the precursors and bring an experimental proof of concept of their predictive power by inferring the residual life-time of our specimen. We perform a retrospective failure prediction using the cascade size $S$ measured during the damage accumulation regime. To do so, we  follow the idea of Anifrani~\textit{et al.}~\cite{anifrani1995} and use the methodology proposed by Mayya~\textit{et al.}~\cite{Mayya}. Considering a time-series of measured precursory activity for the equivalent force-controlled experiment, the normalized distance to failure here writes as $\delta = (t_\mathrm{c} - t)/t_\mathrm{c}$ where  $t = 0$ corresponds to $F = F_\mathrm{el}$ and $t = t_\mathrm{c}$ corresponds to $F = F_\mathrm{c}$. We rewrite the scaling law for the cascade size variations as 
\begin{equation}
\centering
\langle S \rangle  = S_0/\sqrt{t_\mathrm{c} - t}
\label{eq.predict}
\end{equation}
\noindent where $S_0$ is a constant. Re-arranging the terms, we obtain $\langle S \rangle^2 \, t =   \langle S \rangle^2 \, t_\mathrm{c} + S_0$, an expression that can be used for performing a linear-regression of our experimental data set $(t, S)$ shown in Fig.~\ref{fig:fig6}(a). The average size of the precursors is obtained over a non-overlapping time window of $10~\mathrm{s}$. The prediction is made at time $t_\mathrm{cur}$ so that only the precursors recorded at time $t < t_\mathrm{cur}$ can be used for the prediction. Note, however,  that we only use a short period (here $100~\mathrm{s}$) before $t_\mathrm{cur}$ to make the prediction.

The linear regression provides $t_\mathrm{c}^\mathrm{pred}$ that is shown in Fig.~\ref{fig:fig6}(b) as a function of $t_\mathrm{cur}$. As shown in inset, the error on the predicted failure time reduces as the prediction is made closer to the actual failure time $t_\mathrm{c}$. The prediction lies within $10~\%$ error when the prediction is made in the last $25~\%$ of the total lifetime. Note that the same methodology can be implemented using the duration or the rate of events (under displacement controlled conditions only), thus providing several independent measurements to forecast final failure. Interestingly, the predictions are conservative by providing shorter residual lifetime than the actual one.  Importantly,  the proposed methodology does not require monitoring from the beginning of the damage accumulation phase. The robustness of our method strongly argues that the statistical analysis of failure precursors based on the proposed scenario is an efficient quantitative tool of damage monitoring and lifetime prediction of structures.

\section{Acoustic emissions}
\label{sec7}
\begin{figure}[t]
\centering
\includegraphics[width=\linewidth]{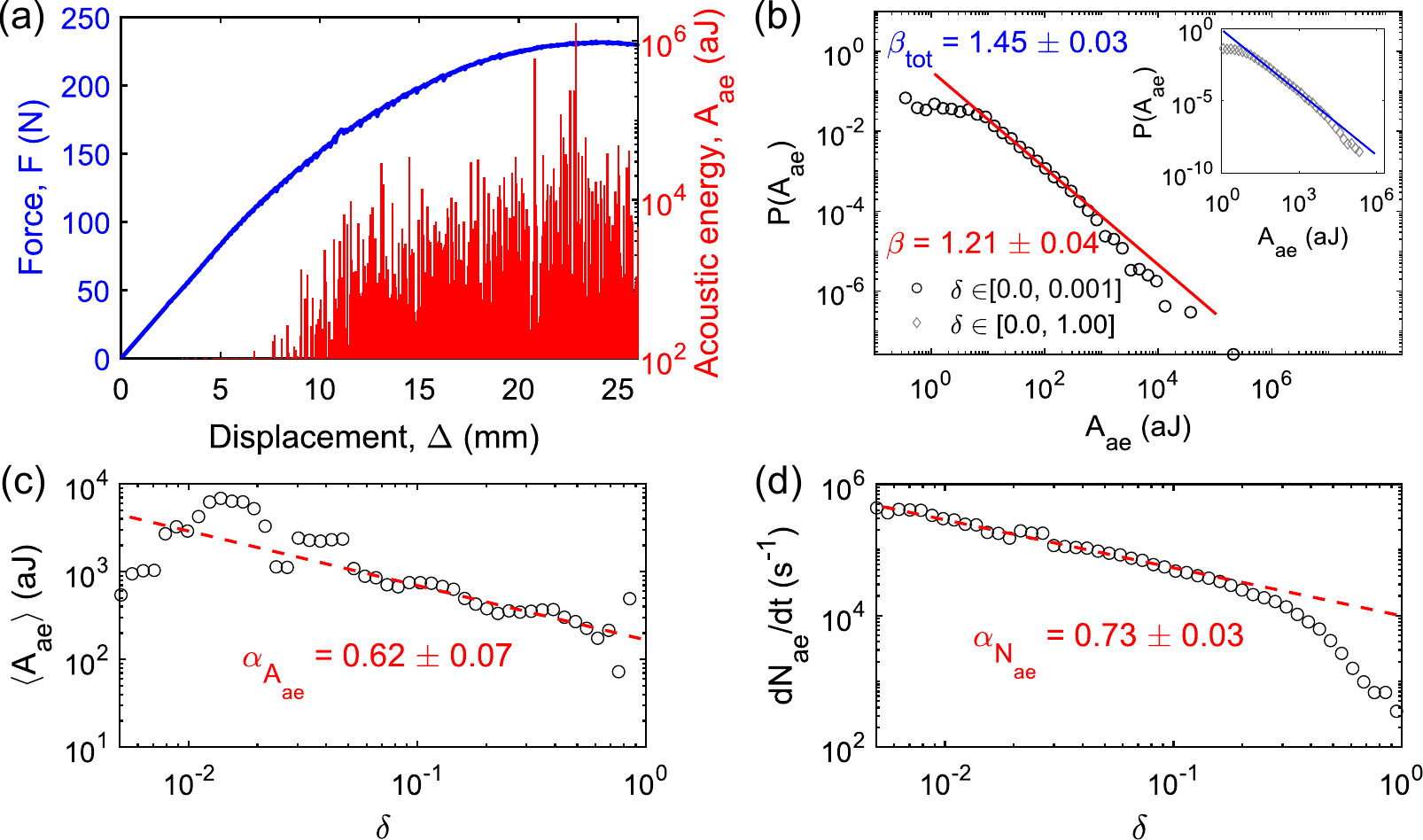}
\caption{(a)  Acoustic emissions recorded during a typical experiment of compressive failure. (b) Distribution of acoustic event energy recorded during the whole duration of experiments (inset) and close to failure (main panel). Variations of the average value of (c) the acoustic energy $A_{\mathrm{ae}}$ and (d) their activity rate $dN_\mathrm{ae}/dt$ with distance to failure $\delta$.}
\label{fig:fig7}
\end{figure}

We now verify the generality of our results and their applications to standard brittle solids. To this effect, we analyze the acoustic emissions accompanying damage growth  in our experiments (see Appendix \ref{app_a}1 for experimental details). A typical acoustic emission time-series record is shown in Fig.~\ref{fig:fig7}(a). First, we observe bursts of acoustic emissions separated by silent periods, a behavior typical of compressive failure of brittle disordered materials. Then, we see that the acoustic activity intensifies closer to peak load, near localization, a behavior that is also observed close to failure in standard brittle solids, and that we explore in more details below. The distribution of acoustic event energy is shown in Fig.~\ref{fig:fig7}(b) and in the inset. It follows a power law with an exponent $ \beta_{\mathrm{tot}_\mathrm{ae}} \simeq 1.45$ that is slightly smaller when we consider only the acoustic emissions close to failure $\beta_\mathrm{ae} \simeq 1.21$. Tracking now the evolution of the acoustic activity, we find  that the energy $\langle A_\mathrm{ae}\rangle$ of the acoustic events increases as  $\langle A_\mathrm{ae}\rangle \propto \delta^{-\alpha_{A_{\mathrm{ae}}}}$ on approaching failure, see Fig.~\ref{fig:fig7}(c). 
The acoustic activity rate $dN_\mathrm{ae}/dt \propto \delta^{-\alpha_{N_\mathrm{ae}} }$ also increases as a power law on approaching failure, see Fig.~\ref{fig:fig7} (d). As a result, the rate of acoustic energy $dE_\mathrm{ae}/dt$  defined as the product of the average acoustic event energy with the activity rate $dE_\mathrm{ae}/dt \propto \langle A_\mathrm{ae}\rangle\cdot dN_\mathrm{ae}/dt$  increases as $dE_\mathrm{ae} \propto \delta^{-\alpha_\mathrm{ae}}$ where $\alpha_\mathrm{ae} = \alpha_{A_\mathrm{ae}} + \alpha_{N_\mathrm{ae}} \simeq 1.35$.  A comparison of the exponents measured in our experiments with the one reported in the literature shows that damage spreading in our experimental system shares strong similarities with damage spreading in standard brittle solids. In particular, the exponents $\beta_{\mathrm{tot}_\mathrm{ae}} \simeq 1.4~\mathrm{and}~\alpha_{N_\mathrm{ae}}\simeq 0.7$ measured in our experiments are similar to the one measured in rocks~\cite{baro2013, nataf2014b, davidsen2021}, coal~\cite{xu2019} and concrete~\cite{weiss2019}. Beyond confirming  the applicability of our experimental and theoretical findings to a broad range of brittle materials, these results confirm versatility of acoustic emissions as a damage monitoring tool. \\
We would like to discuss the intricate connection between the acoustic emissions ~\cite{zreihan2019} and the mechanical energy cascades that have been thoroughly characterized in our experiments (see Secs.~\ref{sec2}B and ~\ref{sec2}C). We clearly observe here that acoustic events do not correspond to the damage cascades characterizing the mechanical response of the specimen. First, we find that their number (typically $10^3$) during an experiment is much larger than the number of mechanical precursors (typically $10^2$). Then, we observe that the exponent characterizing the increase of emitted acoustic energy on approaching failure $\alpha_\mathrm{AE}$ is much larger than the exponent $\alpha$ describing the increase of dissipated rate of mechanical energy. Finally, we observe that the rate of acoustic events increase on approaching failure, an observation in stark contrast with the constant activity rate observed for damage cascades, see Fig.~\ref{fig:fig4}(b). \\
We infer that the acoustic emissions may rather be reminiscent of the dissipative processes taking place at much lower spatio-temporal scales within damage cascades. Such a scenario  also agrees with the power-law distribution of inter-event waiting time of  acoustic events~\cite{baro2013, davidsen2017} that contrasts with the exponential distribution observed here for damage cascades, see Fig.~\ref{fig:fig2}(f). A deeper analysis of our experimental data may be needed to understand the connection between mechanical precursors and acoustic events. Nevertheless, their increase on approaching failure may be harnessed for tracking damage growth and to predict the residual lifetime of structures, after calibration of material specific exponents~\cite{Mayya}.

\section{Numerical solution of the damage evolution equation} \label{num_model}
To conclusively validate our interpretation of the non-stationary avalanche dynamics preceding compressive failure, we numerically solve the damage evolution equation~(\ref{eq1}) using a 2D cellular automata (see Appendix \ref{app_b}) for both force (SI Sec.~S4A~\cite{supp_info}) and displacement (SI Sec.~S4B~\cite{supp_info}) control conditions. We recover that intermittent damage accumulation culminates in a bifurcation that manifests as the emergence of a localization band at peak load (SI Figs. S4(a)-(b) \& S5(a)-(b)~\cite{supp_info}). The exponents characterizing the damage cascades are measured numerically (SI Figs. S4(c)-(j) \& S5(c)-(j)~\cite{supp_info}) using the methods employed for analyzing the experimental data, thus allowing for a systematic comparison with the statistical features of the precursors measured experimentally. Numerical and experimental exponents are  provided in Table~\ref{tab:tab1}. We also proceed to a comparison with exponent values reported in the literature (see SI Table S2~\cite{supp_info}). The good agreement supports our theoretical framework as an adequate description. 
Importantly, the similarities between the statistics in force and equivalent-force control  validate our method of reconstruction of the precursors from our experimental data set and interpretation of failure built from the analyses. We also verify numerically the method employed to characterize the elastic interactions from the incremental damage field in Fig.\ref{fig:fig3}(b). 
\begin{table}[h]
\centering
\caption{Exponents from experiments and numerical model.}
\begin{ruledtabular} 
\begin{tabular}{llcc}
   & Definition & Experiments & Simulations${}^{\dagger}$ \\  
\hline 
$d_\mathrm{f}$ &  $S\propto \xi^{d_\mathrm{f}}$  & 1.07 $\pm$~0.07  & 1.15 \\
$z$ & $T\propto  \xi^{z}$    & 0.53 $\pm$~0.11  & 0.62      \\
$\theta$& $P(\delta Y) \propto  \delta Y^\theta$& 0.24 $\pm$~0.03  & 0.35~(0.18) \\
$\beta$& $P(S) \propto  S^{-\beta}$  & 1.30 $\pm$~0.11  & 1.36~(1.34) \\
$\alpha$ & $S \propto  \delta^{-\alpha}$ & 0.57 $\pm$~0.04  & 0.48~(0.60) \\
$z/d_\mathrm{f}$ & $T \propto  S^{z/d_\mathrm{f}}$  & 0.49 $\pm$~0.14  & 0.53~(0.64) \\
$\beta_{tot}$& $\beta_{tot} = \beta + \frac{2-\beta}{\alpha}$  & 2.32 $\pm$~0.18 & 2.2~(2.13)  \\
\label{Table2}
\end{tabular}
\end{ruledtabular} 
\\
${}^{\dagger}$Values in brackets are from equivalent force control scenario.
\label{tab:tab1}
\end{table} 
\section{Implications and conclusion}
\label{sec9}
We would like to conclude by highlighting the implications of our findings. First, our results may be relevant for any brittle solids such as rocks, concrete and mortar, the generality of our approach being bonded to damage mechanics that we used to describe our experiments (see Annexe~2). By showing that compressive failure is not a critical phase transition, but instead a standard bifurcation, we pave the way for reliable predictions of the failure load of brittle solids and the use of precursory events to anticipate their forthcoming failure. Indeed, this finding implies that the stability analysis of a {\it homogeneous} specimen is sufficient to predict the failure load of disordered solids. This approach, used in this study, captures the emergence of the localization band at peak load and, more generally, the stress-strain response of the specimen. Further experimental validation of the application of these concepts to more complex materials are a part of on-going studies. Another direct implication of our finding is the prediction of the forthcoming failure from the statistics of precursors. The interpretation of compressive failure as a standard bifurcation provides robust scaling laws that can be harnessed to predict failure.

We also would like to highlight the implications of our work for a larger class of phenomena. Compressive failure is an archetypical example of seemingly critical phenomena, as it is preceded by precursors with diverging size and duration on approaching failure. We show that this apparent criticality is not bounded to the presence of a critical point. Instead, we show that a standard bifurcation in presence of disorder and long-range interactions results in an intermittent response with fluctuations that diverge on approaching the bifurcation point.  We expect these ideas to be relevant for understanding the relationship between intermittent plastic flow and shear banding in amorphous solids, where the nature of the yielding transition and the localization has been vigorously debated these recent years~\cite{lin2014,parisi2017,ozawa2018}. Our interpretation of the specimen being at a finite distance above the critical point as it evolves towards failure aligns well with the conclusion drawn by Lin {\it et al.}~\cite{lin2015}. Using a discrete model of amorphous plasticity, they show that avalanche size diverges on approaching failure. Yet, this divergence is not reminiscent of a critical behavior. Instead, they show that the avalanche size is controlled by the slope $\frac{d\sigma}{d\epsilon}$ of the stress-strain response. The scaling law $\langle S\rangle \propto \frac{1}{d\sigma/d\epsilon}$ is compatible with the one obtained in our experiments $\langle S \rangle \propto \frac{1}{\sqrt{\delta}}$ (see Fig.\ref{fig:fig4}(a)). This further supports the relevance of the scenario proposed here for sheared amorphous media.

In summary, we investigated experimentally the precursors to localization during compressive failure of 2D cohesionless soft cellular solids. This model system was shown to behave like a wide range of standard (and somehow more complex) brittle materials under compression. We used this simplified brittle disordered material to characterize in depth (i) the highly correlated sequence of spatially coherent clusters of failure events that compose precursors, (ii) the divergence of the size, the spatial extent and the duration of these precursors on approaching failure, (iii) the presence of the long-range elastic interactions decaying as $\sim 1/r^2$ that drives this cooperative dynamics, (iv) the power-law distributed acoustic bursts accompanying these damage cascades and finally (v) the emergence of a localization band at peak load. We characterized the non-stationary statistics of the damage cascades observed during the damage accumulation regime prior localization. We then derived from continuum damage mechanics an evolution equation of the damage field that was shown to account quantitatively for all the scaling properties measured experimentally. The avalanche-dynamics of damage growth was thus shown to be reminiscent of a non-stationary depinning scenario that reconciles the two competing approaches proposed to describe compressive failure, namely standard bifurcation and critical transition.

Ultimately, failure precursors have been shown to be merely the by-products of the progressive loss of stability of the specimen as it approaches localization. Contrary to the critical transition scenario, specimens driven towards failure stay at a fixed distance to criticality. Nevertheless, the evolution of the statistical features of precursors can be harnessed to anticipate and even predict the forthcoming failure. 

\appendix

\section{Experimental characterization of precursors}
\label{app_a}
\subsection{Experimental set-up } Our specimen consists of a hexagonal packing of about 1500~soft cylinders made of polypropylene placed in a transparent Plexiglas box of dimensions $205\mathrm{mm}\times170\mathrm{mm}\times30\mathrm{mm}$, as shown in Fig. \ref{fig:fig1}(a). The cylinders are $25~\mathrm{mm}$ long with a $5\mathrm{mm}$ diameter. Displacement is applied to the upper layer through a Plexiglas beam using an AG-X Shimadzu test machine. The force experienced by the specimen is measured using a $10\mathrm{kN}$ load cell and sampled at a rate of $100~\mathrm{Hz}$. During the test, images are recorded every $0.1$ seconds from the lateral side of the box, using a Baumer HXC20 camera with a resolution of $2048\times1088$ pixels and binarized using the open source Fiji software~\cite{fiji}. Precursory activity recorded during ten experiments with loading rate of $2~\mathrm{mm/min}$ and two experiments with loading rate of $1~\mathrm{mm/min}$ were analyzed to determine the statistical structure of precursors. The peak load measured in our  experiments is $ \mathrm{F}_c = 228 \pm 4~N$. Acoustic emissions were recorded during four of the ten experiments using two low frequency sensors (type R3$\alpha$ from Mistras group, Europhysical Acoustics) that were placed on the compression platens. The signals crossing the fixed threshold 27 dB from the transducers are preamplified (40 dB) and transferred to a PC1-2 acquisition system.  A detailed analysis in a study of acoustic emissions during dislocations showed that the scaling exponents are robust even for the range of the timing parameters that may result in superposition of events~\cite{lebyodkin2013}. Still, we account for the cellular nature of the packing and set the event detection time post the first crossing of threshold as  well as the  lockout time for signals after the event as $500 \mu s$. Calibration of the timing parameters was performed by recording signals from pencil lead-breaks on metallic rods placed within the cylindrical cells.

\subsection{Cohesionless soft cellular solids under compression as model brittle solids}
We justify here why packings of cohesionless soft cells employed in our experiments can be considered as model materials to investigate the intermittent response of {\it brittle} disordered solids under compression. First, we show that our packings of soft cells behave like standard elasto-damageable media, a description that has been shown to accurately describe the compressive response of brittle solids~\cite{ashby1990}. Then, we explain how friction between cells that control the non-elastic response of our packings leads to a mechanical response that is similar to the ones of brittle solids under compression.

First, we validate the applicability of damage mechanics to describe the average mechanical response of our packings of soft cells. We remind that elasto-damageable solids behave elastically until some critical load level is reached. Then, damage (the level of which is described by an internal variable noted $d_\circ$) increases. The impact of damage on the mechanical response of the material is described by the decrease of the material Young's modulus $E(d_\circ)$ and the variations of its Poisson's ratio $\nu(d_\circ)$ where $d_\circ$ is the accumulated damage. In our experiments, $d_\circ$ is obtained from the deviation of the cell geometry to cell circularity, a definition that is justified below.
 
\begin{figure}[bp]
\centering
\includegraphics[width=\linewidth]{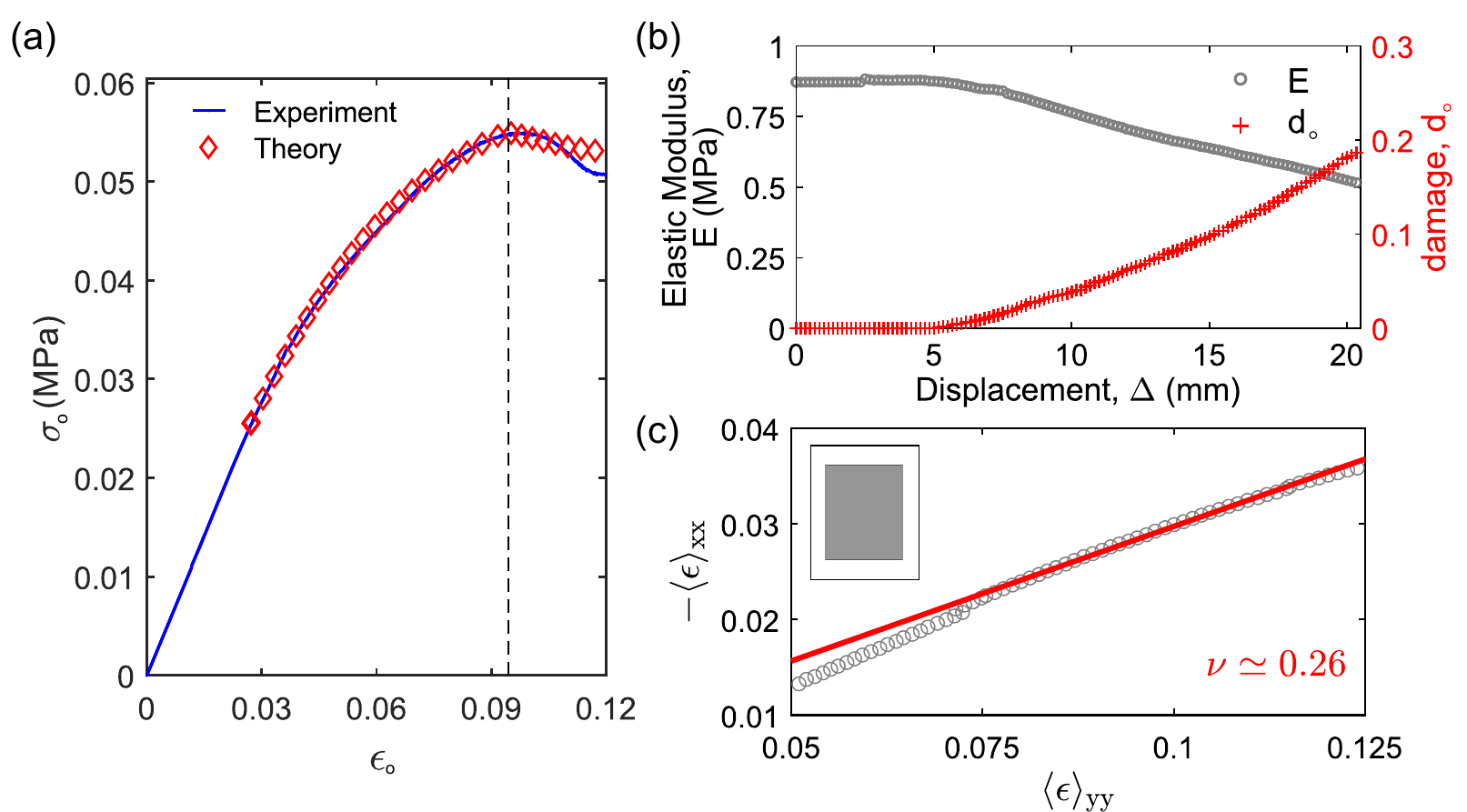}
\caption{Suitability of damage mechanics to describe the average mechanical response of the cohesionless frictional cellular solids used in our experiments. (a) Comparison of the mechanical response predicted by damage mechanics with the experimental stress-strain curve. (b) Variations of the elastic modulus and the average damage level during a representative experiment as inferred from the global mechanical response and our tracking of the cells position. (c) Determination of the Poisson's ratio $\nu$ inferred from the linear fit of the lateral strain vs the normal one after averaging over the central gray region of the specimen shown in inset. $\nu \simeq \nu_\circ(d_\circ) = 0.26$ is found to be nearly constant and independent of $d_\circ$}
\label{fig:fig8}
\end{figure}

The Young's modulus of our packings is obtained from the instantaneous specimen stiffness $E = (1 - \nu^2) \sigma_\circ/\epsilon_\circ$ leading to the following  quadratic decrease $E \simeq E_\circ - a_\circ d_\circ^2$. Its Poisson's ratio $\nu \simeq 0.26$, obtained from the ratio $-\langle \epsilon_\mathrm{xx}\rangle/\langle \epsilon_\mathrm{yy}\rangle$ computed in the central region of the specimen, see Fig.~\ref{fig:fig8}(c), is found to be constant throughout our experiments and nearly independent of the damage level $d_\circ$. 

The force-displacement response predicted by elasto-damageable theory is shown in Fig.~\ref{fig:fig8}(a). It results from the balance $Y[\sigma_\circ(d_\circ),E(d_\circ),\nu] = Y_\mathrm{c}(d_\circ)$ between the elastic energy release rate $Y$ and the damage energy $Y_\mathrm{c}$ (see SI Section~3.A for the detailed calculation).The predicted behavior captures very well our experimental data. The macroscopic stress $\sigma_\circ$ and strain $\epsilon_\circ$ are computed here as parametric functions of the average damage level $d_\circ$\\
\begin{equation}
\left \{
\begin{array}{lcl} 
\vspace{3pt} \sigma_\circ(d_\circ) & = & \displaystyle \frac{E(d_\circ)}{\sqrt{1-\nu^2}} \, \sqrt{\frac{-2Y_{c\circ}(d_\circ)}{E'(d_\circ)}},\\ 
\epsilon_\circ(d_\circ) & = & \displaystyle \sqrt{1 - \nu^2} \sqrt{\frac{-2Y_{c\circ}(d_\circ}{E'(d_\circ)}}.
\label{stress-strain}
\end{array}
\right .
\end{equation}
The average damage resistance $Y_{c\circ}(d_\circ)$ is inferred from the displacement fields measured by tracking the cell positions, following the procedure described in Appendix \ref{app_a}4. Note that beyond capturing the {\it average} mechanical response of the packings of soft cells used in our experiments, elasto-damageable theory captures the {\it fluctuations} around this average response, and in particular the statistics of failure precursors in its most subtle details. The application of the elasto-damageable theory to disordered brittle solids is described in Section~3 of SI and the comparison of the predictions derived from this model with our experimental observations is summarized in Table~\ref{Table2}. Note also that the long-range interactions $\sim 1/r^2$, a unique feature of elasto-damageable solids, is retrieved in our experiments (see Fig.~3(a)-(c)). Last but not least, elasto-damageable theory predicts (see SI Section 3F) the emergence of localization at peak load, a prediction in perfect agreement with our experimental observations (see Appendix~A.5. for the experimental determination of the onset of damage localization).

The applicability of damage mechanics to our packings of soft cells support the generality of our findings and their applicability to a broad range of brittle disordered solids. We now would like to provide a microscopic interpretation to this seemingly surprising behavior. The inelastic behavior of our packings of soft cells is governed by the friction between neighboring cells. This rearrangement of the cells explains the local softening of the material that is appropriately descried by a local decrease of the Young's modulus $E(d_\circ)$ within the theoretical framework of damage mechanics. Meanwhile, cells rearrangements alter the geometry of the initially circular cell. As a result, the deviation to cell circularity is an appropriate of these rearrangements and the associated material softening, justifying the definition of the damage level $d_\circ$ employed in our study.

Last but not least, friction is very often considered as the central damage mechanism in standard brittle solids under compression. This is clearly evidenced by the large success of the Mohr-Coulomb theory (a model that directly derives from the assumption that the different regions of the material can slide on each other) to describe the inelastic response of brittle materials under compression.

Interestingly, Karimi \textit{et al.}~\cite{karimi2019} previously noticed in their numerical work that soft granular solids can be described by damage mechanics. They also observed in their simulations that such systems failed through the emergence of a localization band at peak load and even noticed that the size of the failure precursors diverge as a power-law, two observations in agreement with our experimental observations.

Overall, these various observations come in support to our central assumption: packings of frictional soft particles behaves as elasto-damageable solids, a conclusion already drawn by~\citet{houdoux2018,houdoux2021} from compression experiments carried on standard granular solids.
\\

\subsection{Precursor size in an equivalent force control scenario} 
We explain here why the energy dissipated by damage during a precursory damage event follows the expression $ S_\mathrm{global} = (\Delta_\mathrm{end} - \Delta_\mathrm{ini}) F_0/2 $. We compute first the elastic energy stored in the specimen right before (for $\Delta = \Delta_\mathrm{ini} $ ) and right after (for $\Delta = \Delta_\mathrm{end} $) the precursory damage event. They follow $E_\mathrm{el, ini} = \Delta_\mathrm{ini} F_0/2$ and $ E_\mathrm{el, end} = \Delta_\mathrm{end} F_0/2$, respectively. This leads to the following expression $\Delta E_\mathrm{el} = E_\mathrm{el, end} - E_\mathrm{el, ini} =  (\Delta_\mathrm{end} - \Delta_\mathrm{ini}) F_0/2$ for the variation of the total elastic energy stored in the specimen during one precursory event. Note that this quantity is positive, meaning that the stored elastic energy does increase during a precursory event.
We now compute the work of the external force during such an event. As the force is constant during the considered force-controlled scenario, we obtain $\Delta W =  (\Delta_\mathrm{end} - \Delta_\mathrm{ini}) F_0$. This quantity is defined positive and corresponds to the energy injected by the loading machine within the specimen during a precursory event.
The determination of the energy dissipated by damage during a precursor is now in order. Observing that the energy injected by the loading machine during one precursor compensates exactly the new elastic energy stored in the specimen and the one dissipated by damage, one obtains $\Delta W = \Delta E_\mathrm{el} + S_\mathrm{global} \Rightarrow S_\mathrm{global} = \Delta W -  \Delta E_\mathrm{el} = (\Delta_\mathrm{end} - \Delta_\mathrm{ini}) F_0/2 $.
Interestingly, this balance of energy during one precursor can be understood graphically from the force-displacement response of the specimen during the event. As shown in Fig.~\ref{fig:fig1}(c), the work $\displaystyle \Delta W = \int_{\Delta_\mathrm{ini}}^{\Delta_\mathrm{end}} F(\Delta) \mathrm{d}\Delta = F_0 \int_{\Delta_\mathrm{ini}}^{\Delta_\mathrm{end}}  \mathrm{d}\Delta = (\Delta_\mathrm{end} - \Delta_\mathrm{ini}) F_0$ of the external force during one precursor corresponds to the area of the rectangle located below the force plateau. This rectangle can be divided in two other rectangles of similar size $(\Delta_\mathrm{end} - \Delta_\mathrm{ini}) F_0/2$, one of them corresponding to the additional stored energy $\Delta E_\mathrm{el}$ (represented in grey in Fig.~\ref{fig:fig1}(c)) while the other one corresponding to the dissipated energy $\Delta E_\mathrm{d} = S_\mathrm{global} $ (represented in yellow). This equipartition of energy between the new stored elastic energy and the dissipated one provides a simple way to track the size of failure precursors during the phase of damage accumulation.
\\

\subsection{Tracking damage evolution }The continuous image acquisition allows for the tracking of both position and circularity
(used here to define damage) of individual cells. We then compute the displacement fields $u_x (\vec{x},t)~\mathrm{and}~u_y(\vec{x},t)$ and the damage  field $d(\vec{x},t)$ of the effective elasto-damageable medium following the coarse-graining procedure described in~\citet{glasser2001}. To resolve the damage cascades from the image stack, we refer to the macroscopic response where  an equivalent force control scenario is constructed from the sequence of load drops. We thus obtain the start and end of each damage cascade that are noted  $t(\Delta_{\mathrm{ini}})$ and $t(\Delta_{\mathrm{end}})$,  respectively. The damage events belonging to the image stack constituting the cascades are then clustered based on their connectivity (26-connected neighborhood). This procedure provides a space-time dissipation map of failure precursors composed of a series of highly correlated clusters. We derive the strain field as $\epsilon_{xx}(\vec{x}) = \frac{du_x}{dx},~ \epsilon_{yy}(\vec{x}) = \frac{du_y}{dy}~\mathrm{and}~ \epsilon_{xy}(\vec{x}) = \frac{1}{2}\left(\frac{du_x}{dy} + \frac{du_x}{dy}\right)$. The field of elastic energy stored in the specimen per unit volume is then obtained from the relation,
\begin{equation}\label{eq:Eel}
\begin{split}
    E_\mathrm{el}^\mathrm{local}(\vec{x},t) =& \frac{E_\mathrm{local}(\vec{x},t)}{2(1 - \nu^2)} \left[\epsilon_\mathrm{xx}^2(\vec{x},t) +\right. \\
    &\epsilon_\mathrm{yy}^2(\vec{x},t) + 2 \, \nu \epsilon_\mathrm{xx}(\vec{x},t) \epsilon_\mathrm{yy}(\vec{x},t)  + \\
    & \left. 2 (1 - \nu)\epsilon_\mathrm{xy}^2(\vec{x},t) \right].
 \end{split}   
\end{equation}
The field of damage driving force is obtained as $Y[d(\vec{x},t)] = \frac{dE_\mathrm{el}^\mathrm{local}}{dd}$. The field of damage resistance $Y_{c}[d(\vec{x},t)]$ is inferred from the field of damage driving force $Y[d(\vec{x},t)]$ computed in a retrospective manner. If damage grows locally at $(\vec{x}_0,t_i)$, we assign $Y_c[d(\vec{x}_0),t_i] = Y[d(\vec{x}_0),t_i]$ as well as for the preceding time-steps $(t_\mathrm{i-})$. We find that the average damage resistance $Y_\mathrm{c\circ}(d_\circ)$ {\it increases} with the damage level $d_\circ$. A linear approximation of the hardening behavior $Y_\mathrm{c\circ}(d_\circ) \sim \eta d_\circ$ with a hardening coefficient $\eta \simeq 45$ provides a good description of our experimental data.

\subsection{Onset of damage localization} 
\begin{figure}[h]
\centering
\includegraphics[width=\linewidth]{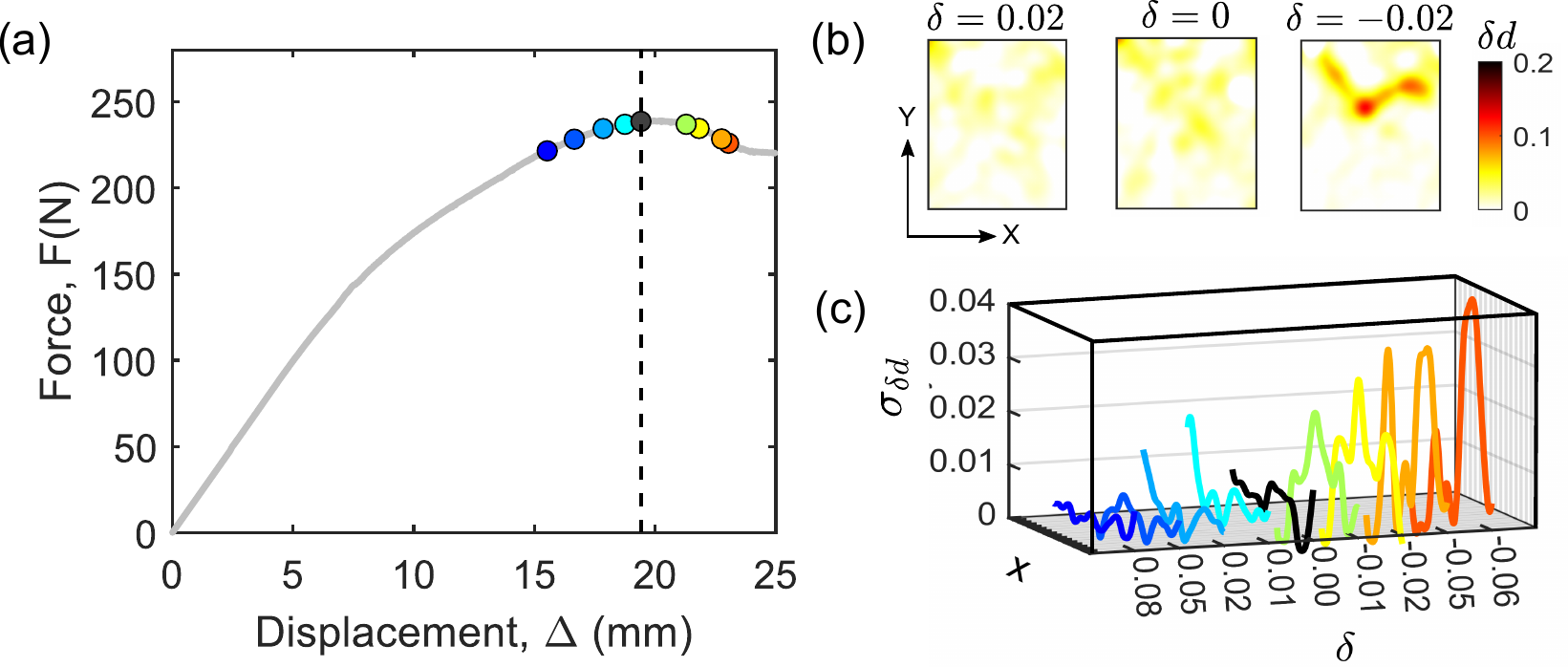}
\caption{The incremental damage field $\delta d(\vec{\mathrm{x}})$ is computed for several locations $\delta$ near peak load as indicated by solid circles on the force-displacement curve of panel (a). (b) $\delta d(\vec{\mathrm{x}})$ is represented right before ($\delta = 0.02$), at peak ($\delta = 0$) and right after ($\delta = -0.02$). Note the localization of the damage activity in the third panel suggesting that localization took place at peak load. (c) This is further confirmed by the standard deviation $\sigma_{\delta d}(x)$ of the incremental damage field that is shown to increase for $\delta <0$.}
\label{fig:fig9}
\end{figure}

We provide here the methodology employed to determine the localization threshold. The basic idea is to compute the field of incremental damage growth at different distances to peak load as shown in Fig.~\ref{fig:fig9}(a) to determine whether it is homogeneous or localized. The incremental damage field $\delta d(\vec{x}) = d(\vec{x},\delta) - d(\vec{x},\delta+\epsilon) $ where $\epsilon \ll \delta$, is shown for three different values of $\delta$ : right before peak load ($\delta = 0.02$), at peak load ($\delta = 0$) and right after ($\delta = -0.02$). We clearly see the localization of the damage growth in thin band for $\delta = -0.02$ from which we infer that the localization starts at peak load. The standard deviation $\sigma_{\delta d}(x) = \sqrt{\langle (\delta d(x) - \langle \delta (\vec{x})\rangle_\mathrm{y})^2\rangle_\mathrm{y}}$ is shown in Fig.~\ref{fig:fig9}(c) for several values of $\delta$. Here also, we see that $\sigma_{\delta d}(x)$ increases above $0.01$, suggesting that localization takes place at peak load.

\section{Numerical modeling of intermittent damage evolution and localization}
\label{app_b}To solve the damage evolution equation under quasi-static loading conditions, we adopt the procedure described in Berthier et al. \cite{berthier2021}. We consider a grid of size $L = 21$ discretized into $L^2$ elements with periodic boundary conditions. The interaction kernel is derived for the case of uni-axial compression using the method described in Dansereau \textit{et al.}~\cite{dansereau2019} and writes as
\begin{equation}
\psi(d_\circ) = \left[ \frac{E'(d_\circ)^2}{E(d_\circ)^3}\right](1 - \nu^2)\sigma_{ext}^2 \left[\frac{x^4-3y^4+6x^2y^2}{4\pi(x^2+y^2)^3}\right].
\label{kernel}
\end{equation}
\noindent Interestingly, the kernel is independent of the type of loading configuration (stress vs. strain imposed loading conditions). The amplitude of the kernel is also non-stationary, in contrast with the non-positive interaction used to describe amorphous plasticity~\cite{lin2014,lin2015}. The variations of the elastic constant with the damage level is found to follow $E_\circ = E^\circ(1 - d_\circ^2)$ with $E^\circ = 1.0~\mathrm{MPa}$. We consider a heterogeneous field of damage resistance $Y_c[d(\vec{x},t)]$ that evolves with the damage level $d(\vec{x},t)$ following the linear hardening law $Y_{c\circ} = Y_c^\circ(1 + \eta d_\circ)$, where $Y_c^\circ \simeq 1.4~\mathrm{kJ/m^3}$ and $\eta \simeq 45$ are measured experimentally. The stress (strain) is gradually increased until the damage criterion is fulfilled for one of the elements $\vec{x} = \vec{x}_0$. The damage is increased locally and the values of driving force $Y(\vec{x}_0,t)$ and resistance $Y_c(\vec{x}_0)$ are updated. The non-local redistribution of driving force given by the kernel $\psi (d_\circ)$ may then trigger a cascade of damage events. The cascade stops when all elements  are stable following which, we increase the stress (strain) again. 

The reader is invited to refer to the Supplementary Information~\cite{supp_info} for the thorough description of the local analyses of experimental precursors, the theoretical and numerical modeling of damage accumulation and the analogy with driven disordered elastic interface.

\begin{acknowledgments}
The authors gratefully acknowledge financial support from Sorbonne Universit\'e through the Emergence grant for the research project, \textit{From damage spreading to failure in quasi-brittle materials} as well as CNRS and Satt-Lutech through the tech transfer project,  \textit{Development of a technology of predictive maintenance for materials and structures under compression}.
\end{acknowledgments}


\begin{thebibliography}{80}%
\makeatletter
\providecommand \@ifxundefined [1]{%
 \@ifx{#1\undefined}
}%
\providecommand \@ifnum [1]{%
 \ifnum #1\expandafter \@firstoftwo
 \else \expandafter \@secondoftwo
 \fi
}%
\providecommand \@ifx [1]{%
 \ifx #1\expandafter \@firstoftwo
 \else \expandafter \@secondoftwo
 \fi
}%
\providecommand \natexlab [1]{#1}%
\providecommand \enquote  [1]{``#1''}%
\providecommand \bibnamefont  [1]{#1}%
\providecommand \bibfnamefont [1]{#1}%
\providecommand \citenamefont [1]{#1}%
\providecommand \href@noop [0]{\@secondoftwo}%
\providecommand \href [0]{\begingroup \@sanitize@url \@href}%
\providecommand \@href[1]{\@@startlink{#1}\@@href}%
\providecommand \@@href[1]{\endgroup#1\@@endlink}%
\providecommand \@sanitize@url [0]{\catcode `\\12\catcode `\$12\catcode
  `\&12\catcode `\#12\catcode `\^12\catcode `\_12\catcode `\%12\relax}%
\providecommand \@@startlink[1]{}%
\providecommand \@@endlink[0]{}%
\providecommand \url  [0]{\begingroup\@sanitize@url \@url }%
\providecommand \@url [1]{\endgroup\@href {#1}{\urlprefix }}%
\providecommand \urlprefix  [0]{URL }%
\providecommand \Eprint [0]{\href }%
\providecommand \doibase [0]{https://doi.org/}%
\providecommand \selectlanguage [0]{\@gobble}%
\providecommand \bibinfo  [0]{\@secondoftwo}%
\providecommand \bibfield  [0]{\@secondoftwo}%
\providecommand \translation [1]{[#1]}%
\providecommand \BibitemOpen [0]{}%
\providecommand \bibitemStop [0]{}%
\providecommand \bibitemNoStop [0]{.\EOS\space}%
\providecommand \EOS [0]{\spacefactor3000\relax}%
\providecommand \BibitemShut  [1]{\csname bibitem#1\endcsname}%
\let\auto@bib@innerbib\@empty
\bibitem [{\citenamefont {Koivisto}\ \emph {et~al.}(2016)\citenamefont
  {Koivisto}, \citenamefont {Ovaska}, \citenamefont {Miksic}, \citenamefont
  {Laurson},\ and\ \citenamefont {Alava}}]{koivisto2016}%
  \BibitemOpen
  \bibfield  {author} {\bibinfo {author} {\bibfnamefont {J.}~\bibnamefont
  {Koivisto}}, \bibinfo {author} {\bibfnamefont {M.}~\bibnamefont {Ovaska}},
  \bibinfo {author} {\bibfnamefont {A.}~\bibnamefont {Miksic}}, \bibinfo
  {author} {\bibfnamefont {L.}~\bibnamefont {Laurson}},\ and\ \bibinfo {author}
  {\bibfnamefont {M.~J.}\ \bibnamefont {Alava}},\ }\bibfield  {title} {\bibinfo
  {title} {Predicting sample lifetimes in creep fracture of heterogeneous
  materials},\ }\href@noop {} {\bibfield  {journal} {\bibinfo  {journal}
  {Physical Review E}\ }\textbf {\bibinfo {volume} {94}},\ \bibinfo {pages}
  {023002} (\bibinfo {year} {2016})}\BibitemShut {NoStop}%
\bibitem [{\citenamefont {K{\'a}d{\'a}r}\ \emph {et~al.}(2020)\citenamefont
  {K{\'a}d{\'a}r}, \citenamefont {P{\'a}l},\ and\ \citenamefont
  {Kun}}]{kadar2020}%
  \BibitemOpen
  \bibfield  {author} {\bibinfo {author} {\bibfnamefont {V.}~\bibnamefont
  {K{\'a}d{\'a}r}}, \bibinfo {author} {\bibfnamefont {G.}~\bibnamefont
  {P{\'a}l}},\ and\ \bibinfo {author} {\bibfnamefont {F.}~\bibnamefont {Kun}},\
  }\bibfield  {title} {\bibinfo {title} {Record statistics of bursts signals
  the onset of acceleration towards failure},\ }\href@noop {} {\bibfield
  {journal} {\bibinfo  {journal} {Scientific reports}\ }\textbf {\bibinfo
  {volume} {10}},\ \bibinfo {pages} {1} (\bibinfo {year} {2020})}\BibitemShut
  {NoStop}%
\bibitem [{\citenamefont {Biswas}\ \emph {et~al.}(2020)\citenamefont {Biswas},
  \citenamefont {Fernandez~Castellanos},\ and\ \citenamefont
  {Zaiser}}]{biswas2020}%
  \BibitemOpen
  \bibfield  {author} {\bibinfo {author} {\bibfnamefont {S.}~\bibnamefont
  {Biswas}}, \bibinfo {author} {\bibfnamefont {D.}~\bibnamefont
  {Fernandez~Castellanos}},\ and\ \bibinfo {author} {\bibfnamefont
  {M.}~\bibnamefont {Zaiser}},\ }\bibfield  {title} {\bibinfo {title}
  {Prediction of creep failure time using machine learning},\ }\href@noop {}
  {\bibfield  {journal} {\bibinfo  {journal} {Scientific Reports}\ }\textbf
  {\bibinfo {volume} {10}},\ \bibinfo {pages} {1} (\bibinfo {year}
  {2020})}\BibitemShut {NoStop}%
\bibitem [{\citenamefont {Debski}\ \emph {et~al.}(2021)\citenamefont {Debski},
  \citenamefont {Pradhan},\ and\ \citenamefont {Hansen}}]{debski2021}%
  \BibitemOpen
  \bibfield  {author} {\bibinfo {author} {\bibfnamefont {W.}~\bibnamefont
  {Debski}}, \bibinfo {author} {\bibfnamefont {S.}~\bibnamefont {Pradhan}},\
  and\ \bibinfo {author} {\bibfnamefont {A.}~\bibnamefont {Hansen}},\
  }\bibfield  {title} {\bibinfo {title} {Criterion for imminent failure during
  loading—discrete element method analysis},\ }\href@noop {} {\bibfield
  {journal} {\bibinfo  {journal} {Frontiers in Physics}\ }\textbf {\bibinfo
  {volume} {9}},\ \bibinfo {pages} {223} (\bibinfo {year} {2021})}\BibitemShut
  {NoStop}%
\bibitem [{\citenamefont {Ashby}\ and\ \citenamefont
  {Sammis}(1990)}]{ashby1990}%
  \BibitemOpen
  \bibfield  {author} {\bibinfo {author} {\bibfnamefont {M.}~\bibnamefont
  {Ashby}}\ and\ \bibinfo {author} {\bibfnamefont {C.}~\bibnamefont {Sammis}},\
  }\bibfield  {title} {\bibinfo {title} {The damage mechanics of brittle solids
  in compression},\ }\href@noop {} {\bibfield  {journal} {\bibinfo  {journal}
  {Pure and Applied Geophysics}\ }\textbf {\bibinfo {volume} {133}},\ \bibinfo
  {pages} {489} (\bibinfo {year} {1990})}\BibitemShut {NoStop}%
\bibitem [{\citenamefont {Lockner}\ \emph {et~al.}(1991)\citenamefont
  {Lockner}, \citenamefont {Byerlee}, \citenamefont {Kuksenko}, \citenamefont
  {Ponomarev},\ and\ \citenamefont {Sidorin}}]{lockner1991}%
  \BibitemOpen
  \bibfield  {author} {\bibinfo {author} {\bibfnamefont {D.}~\bibnamefont
  {Lockner}}, \bibinfo {author} {\bibfnamefont {J.}~\bibnamefont {Byerlee}},
  \bibinfo {author} {\bibfnamefont {V.}~\bibnamefont {Kuksenko}}, \bibinfo
  {author} {\bibfnamefont {A.}~\bibnamefont {Ponomarev}},\ and\ \bibinfo
  {author} {\bibfnamefont {A.}~\bibnamefont {Sidorin}},\ }\bibfield  {title}
  {\bibinfo {title} {Quasi-static fault growth and shear fracture energy in
  granite},\ }\href@noop {} {\bibfield  {journal} {\bibinfo  {journal}
  {Nature}\ }\textbf {\bibinfo {volume} {350}},\ \bibinfo {pages} {39}
  (\bibinfo {year} {1991})}\BibitemShut {NoStop}%
\bibitem [{\citenamefont {Kachanov}(1987)}]{kachanov1987}%
  \BibitemOpen
  \bibfield  {author} {\bibinfo {author} {\bibfnamefont {M.}~\bibnamefont
  {Kachanov}},\ }\bibfield  {title} {\bibinfo {title} {Elastic solids with many
  cracks: a simple method of analysis},\ }\href@noop {} {\bibfield  {journal}
  {\bibinfo  {journal} {International Journal of Solids and Structures}\
  }\textbf {\bibinfo {volume} {23}},\ \bibinfo {pages} {23} (\bibinfo {year}
  {1987})}\BibitemShut {NoStop}%
\bibitem [{\citenamefont {Mazars}\ and\ \citenamefont
  {Pijaudier-Cabot}(1989)}]{mazars1989}%
  \BibitemOpen
  \bibfield  {author} {\bibinfo {author} {\bibfnamefont {J.}~\bibnamefont
  {Mazars}}\ and\ \bibinfo {author} {\bibfnamefont {G.}~\bibnamefont
  {Pijaudier-Cabot}},\ }\bibfield  {title} {\bibinfo {title} {Continuum damage
  theory—application to concrete},\ }\href@noop {} {\bibfield  {journal}
  {\bibinfo  {journal} {Journal of engineering mechanics}\ }\textbf {\bibinfo
  {volume} {115}},\ \bibinfo {pages} {345} (\bibinfo {year}
  {1989})}\BibitemShut {NoStop}%
\bibitem [{\citenamefont {Kachanov}(1993)}]{kachanov1993}%
  \BibitemOpen
  \bibfield  {author} {\bibinfo {author} {\bibfnamefont {M.}~\bibnamefont
  {Kachanov}},\ }\bibfield  {title} {\bibinfo {title} {Elastic solids with many
  cracks and related problems},\ }\href@noop {} {\bibfield  {journal} {\bibinfo
   {journal} {Advances in applied mechanics}\ }\textbf {\bibinfo {volume}
  {30}},\ \bibinfo {pages} {259} (\bibinfo {year} {1993})}\BibitemShut
  {NoStop}%
\bibitem [{\citenamefont {Fortin}\ \emph {et~al.}(2006)\citenamefont {Fortin},
  \citenamefont {Stanchits}, \citenamefont {Dresen},\ and\ \citenamefont
  {Gu{\'e}guen}}]{fortin2006}%
  \BibitemOpen
  \bibfield  {author} {\bibinfo {author} {\bibfnamefont {J.}~\bibnamefont
  {Fortin}}, \bibinfo {author} {\bibfnamefont {S.}~\bibnamefont {Stanchits}},
  \bibinfo {author} {\bibfnamefont {G.}~\bibnamefont {Dresen}},\ and\ \bibinfo
  {author} {\bibfnamefont {Y.}~\bibnamefont {Gu{\'e}guen}},\ }\bibfield
  {title} {\bibinfo {title} {Acoustic emission and velocities associated with
  the formation of compaction bands in sandstone},\ }\href@noop {} {\bibfield
  {journal} {\bibinfo  {journal} {J. Geophys. Res. Solid Earth}\ }\textbf
  {\bibinfo {volume} {111}} (\bibinfo {year} {2006})}\BibitemShut {NoStop}%
\bibitem [{\citenamefont {Fortin}\ \emph {et~al.}(2009)\citenamefont {Fortin},
  \citenamefont {Stanchits}, \citenamefont {Dresen},\ and\ \citenamefont
  {Gueguen}}]{fortin2009}%
  \BibitemOpen
  \bibfield  {author} {\bibinfo {author} {\bibfnamefont {J.}~\bibnamefont
  {Fortin}}, \bibinfo {author} {\bibfnamefont {S.}~\bibnamefont {Stanchits}},
  \bibinfo {author} {\bibfnamefont {G.}~\bibnamefont {Dresen}},\ and\ \bibinfo
  {author} {\bibfnamefont {Y.}~\bibnamefont {Gueguen}},\ }\bibfield  {title}
  {\bibinfo {title} {Acoustic emissions monitoring during inelastic deformation
  of porous sandstone: comparison of three modes of deformation},\ }\href@noop
  {} {\bibfield  {journal} {\bibinfo  {journal} {Pure and Applied Geophysics}\
  }\textbf {\bibinfo {volume} {166}},\ \bibinfo {pages} {823} (\bibinfo {year}
  {2009})}\BibitemShut {NoStop}%
\bibitem [{\citenamefont {Manzato}\ \emph {et~al.}(2012)\citenamefont
  {Manzato}, \citenamefont {Shekhawat}, \citenamefont {Nukala}, \citenamefont
  {Alava}, \citenamefont {Sethna},\ and\ \citenamefont
  {Zapperi}}]{manzato2012}%
  \BibitemOpen
  \bibfield  {author} {\bibinfo {author} {\bibfnamefont {C.}~\bibnamefont
  {Manzato}}, \bibinfo {author} {\bibfnamefont {A.}~\bibnamefont {Shekhawat}},
  \bibinfo {author} {\bibfnamefont {P.~K. V.~V.}\ \bibnamefont {Nukala}},
  \bibinfo {author} {\bibfnamefont {M.~J.}\ \bibnamefont {Alava}}, \bibinfo
  {author} {\bibfnamefont {J.~P.}\ \bibnamefont {Sethna}},\ and\ \bibinfo
  {author} {\bibfnamefont {S.}~\bibnamefont {Zapperi}},\ }\bibfield  {title}
  {\bibinfo {title} {Fracture strength of disordered media: Universality,
  interactions, and tail asymptotics},\ }\href@noop {} {\bibfield  {journal}
  {\bibinfo  {journal} {Phys. Rev. Lett.}\ }\textbf {\bibinfo {volume} {108}},\
  \bibinfo {pages} {065504} (\bibinfo {year} {2012})}\BibitemShut {NoStop}%
\bibitem [{\citenamefont {Manzato}\ \emph {et~al.}(2014)\citenamefont
  {Manzato}, \citenamefont {Alava},\ and\ \citenamefont
  {Zapperi}}]{manzato2014}%
  \BibitemOpen
  \bibfield  {author} {\bibinfo {author} {\bibfnamefont {C.}~\bibnamefont
  {Manzato}}, \bibinfo {author} {\bibfnamefont {M.~J.}\ \bibnamefont {Alava}},\
  and\ \bibinfo {author} {\bibfnamefont {S.}~\bibnamefont {Zapperi}},\
  }\bibfield  {title} {\bibinfo {title} {Damage accumulation in quasibrittle
  fracture},\ }\href@noop {} {\bibfield  {journal} {\bibinfo  {journal} {Phys.
  Rev. E}\ }\textbf {\bibinfo {volume} {90}},\ \bibinfo {pages} {012408}
  (\bibinfo {year} {2014})}\BibitemShut {NoStop}%
\bibitem [{\citenamefont {Tal}\ \emph {et~al.}(2016)\citenamefont {Tal},
  \citenamefont {Evans},\ and\ \citenamefont {Mok}}]{tal2016}%
  \BibitemOpen
  \bibfield  {author} {\bibinfo {author} {\bibfnamefont {Y.}~\bibnamefont
  {Tal}}, \bibinfo {author} {\bibfnamefont {B.}~\bibnamefont {Evans}},\ and\
  \bibinfo {author} {\bibfnamefont {U.}~\bibnamefont {Mok}},\ }\bibfield
  {title} {\bibinfo {title} {Direct observations of damage during unconfined
  brittle failure of carrara marble},\ }\href@noop {} {\bibfield  {journal}
  {\bibinfo  {journal} {J. Geophys. Res. Solid Earth}\ }\textbf {\bibinfo
  {volume} {121}},\ \bibinfo {pages} {1584} (\bibinfo {year}
  {2016})}\BibitemShut {NoStop}%
\bibitem [{\citenamefont {Thilakarathna}\ \emph {et~al.}(2020)\citenamefont
  {Thilakarathna}, \citenamefont {Baduge}, \citenamefont {Mendis},
  \citenamefont {Vimonsatit},\ and\ \citenamefont {Lee}}]{thilakarathna2020}%
  \BibitemOpen
  \bibfield  {author} {\bibinfo {author} {\bibfnamefont {P.}~\bibnamefont
  {Thilakarathna}}, \bibinfo {author} {\bibfnamefont {K.~K.}\ \bibnamefont
  {Baduge}}, \bibinfo {author} {\bibfnamefont {P.}~\bibnamefont {Mendis}},
  \bibinfo {author} {\bibfnamefont {V.}~\bibnamefont {Vimonsatit}},\ and\
  \bibinfo {author} {\bibfnamefont {H.}~\bibnamefont {Lee}},\ }\bibfield
  {title} {\bibinfo {title} {Mesoscale modelling of concrete--a review of
  geometry generation, placing algorithms, constitutive relations and
  applications},\ }\href@noop {} {\bibfield  {journal} {\bibinfo  {journal}
  {Engineering Fracture Mechanics}\ }\textbf {\bibinfo {volume} {231}},\
  \bibinfo {pages} {106974} (\bibinfo {year} {2020})}\BibitemShut {NoStop}%
\bibitem [{\citenamefont {Rudnicki}\ and\ \citenamefont
  {Rice}(1975)}]{rudnicki1975}%
  \BibitemOpen
  \bibfield  {author} {\bibinfo {author} {\bibfnamefont {J.~W.}\ \bibnamefont
  {Rudnicki}}\ and\ \bibinfo {author} {\bibfnamefont {J.}~\bibnamefont
  {Rice}},\ }\bibfield  {title} {\bibinfo {title} {Conditions for the
  localization of deformation in pressure-sensitive dilatant materials},\
  }\href@noop {} {\bibfield  {journal} {\bibinfo  {journal} {Journal of the
  Mechanics and Physics of Solids}\ }\textbf {\bibinfo {volume} {23}},\
  \bibinfo {pages} {371} (\bibinfo {year} {1975})}\BibitemShut {NoStop}%
\bibitem [{\citenamefont {Olsson}(1999)}]{olsson1999}%
  \BibitemOpen
  \bibfield  {author} {\bibinfo {author} {\bibfnamefont {W.~A.}\ \bibnamefont
  {Olsson}},\ }\bibfield  {title} {\bibinfo {title} {Theoretical and
  experimental investigation of compaction bands in porous rock},\ }\href
  {https://doi.org/10.1029/1998JB900120} {\bibfield  {journal} {\bibinfo
  {journal} {J. Geophys. Res. Solid Earth}\ }\textbf {\bibinfo {volume}
  {104}},\ \bibinfo {pages} {7219} (\bibinfo {year} {1999})}\BibitemShut
  {NoStop}%
\bibitem [{\citenamefont {Lockner}(1993)}]{lockner1993}%
  \BibitemOpen
  \bibfield  {author} {\bibinfo {author} {\bibfnamefont {D.}~\bibnamefont
  {Lockner}},\ }\bibfield  {title} {\bibinfo {title} {The role of acoustic
  emission in the study of rock fracture},\ }in\ \href@noop {} {\emph {\bibinfo
  {booktitle} {International Journal of Rock Mechanics and Mining Sciences \&
  Geomechanics Abstracts}}},\ Vol.~\bibinfo {volume} {30}\ (\bibinfo
  {organization} {Elsevier},\ \bibinfo {year} {1993})\ pp.\ \bibinfo {pages}
  {883--899}\BibitemShut {NoStop}%
\bibitem [{\citenamefont {Petri}\ \emph {et~al.}(1994)\citenamefont {Petri},
  \citenamefont {Paparo}, \citenamefont {Vespignani}, \citenamefont {Alippi},\
  and\ \citenamefont {Costantini}}]{petri1994}%
  \BibitemOpen
  \bibfield  {author} {\bibinfo {author} {\bibfnamefont {A.}~\bibnamefont
  {Petri}}, \bibinfo {author} {\bibfnamefont {G.}~\bibnamefont {Paparo}},
  \bibinfo {author} {\bibfnamefont {A.}~\bibnamefont {Vespignani}}, \bibinfo
  {author} {\bibfnamefont {A.}~\bibnamefont {Alippi}},\ and\ \bibinfo {author}
  {\bibfnamefont {M.}~\bibnamefont {Costantini}},\ }\bibfield  {title}
  {\bibinfo {title} {Experimental evidence for critical dynamics in
  microfracturing processes},\ }\href@noop {} {\bibfield  {journal} {\bibinfo
  {journal} {Phys. Rev. Lett.}\ }\textbf {\bibinfo {volume} {73}},\ \bibinfo
  {pages} {3423} (\bibinfo {year} {1994})}\BibitemShut {NoStop}%
\bibitem [{\citenamefont {Deschanel}\ \emph {et~al.}(2006)\citenamefont
  {Deschanel}, \citenamefont {Vanel}, \citenamefont {Vigier}, \citenamefont
  {Godin},\ and\ \citenamefont {Ciliberto}}]{deschanel2006}%
  \BibitemOpen
  \bibfield  {author} {\bibinfo {author} {\bibfnamefont {S.}~\bibnamefont
  {Deschanel}}, \bibinfo {author} {\bibfnamefont {L.}~\bibnamefont {Vanel}},
  \bibinfo {author} {\bibfnamefont {G.}~\bibnamefont {Vigier}}, \bibinfo
  {author} {\bibfnamefont {N.}~\bibnamefont {Godin}},\ and\ \bibinfo {author}
  {\bibfnamefont {S.}~\bibnamefont {Ciliberto}},\ }\bibfield  {title} {\bibinfo
  {title} {Statistical properties of microcracking in polyurethane foams under
  tensile test, influence of temperature and density},\ }\href@noop {}
  {\bibfield  {journal} {\bibinfo  {journal} {Int. J. Frac.}\ }\textbf
  {\bibinfo {volume} {140}},\ \bibinfo {pages} {87} (\bibinfo {year}
  {2006})}\BibitemShut {NoStop}%
\bibitem [{\citenamefont {Davidsen}\ \emph {et~al.}(2007)\citenamefont
  {Davidsen}, \citenamefont {Stanchits},\ and\ \citenamefont
  {Dresen}}]{davidsen2007}%
  \BibitemOpen
  \bibfield  {author} {\bibinfo {author} {\bibfnamefont {J.}~\bibnamefont
  {Davidsen}}, \bibinfo {author} {\bibfnamefont {S.}~\bibnamefont
  {Stanchits}},\ and\ \bibinfo {author} {\bibfnamefont {G.}~\bibnamefont
  {Dresen}},\ }\bibfield  {title} {\bibinfo {title} {Scaling and universality
  in rock fracture},\ }\href@noop {} {\bibfield  {journal} {\bibinfo  {journal}
  {Phys. Rev. Lett.}\ }\textbf {\bibinfo {volume} {98}},\ \bibinfo {pages}
  {125502} (\bibinfo {year} {2007})}\BibitemShut {NoStop}%
\bibitem [{\citenamefont {Rosti}\ \emph {et~al.}(2009)\citenamefont {Rosti},
  \citenamefont {Illa}, \citenamefont {Koivisto},\ and\ \citenamefont
  {Alava}}]{rosti2009}%
  \BibitemOpen
  \bibfield  {author} {\bibinfo {author} {\bibfnamefont {J.}~\bibnamefont
  {Rosti}}, \bibinfo {author} {\bibfnamefont {X.}~\bibnamefont {Illa}},
  \bibinfo {author} {\bibfnamefont {J.}~\bibnamefont {Koivisto}},\ and\
  \bibinfo {author} {\bibfnamefont {M.~J.}\ \bibnamefont {Alava}},\ }\bibfield
  {title} {\bibinfo {title} {Crackling noise and its dynamics in fracture of
  disordered media},\ }\href@noop {} {\bibfield  {journal} {\bibinfo  {journal}
  {J. Phys. D: Appl. Phys.}\ }\textbf {\bibinfo {volume} {42}},\ \bibinfo
  {pages} {214013} (\bibinfo {year} {2009})}\BibitemShut {NoStop}%
\bibitem [{\citenamefont {Bar{\'o}}\ \emph {et~al.}(2013)\citenamefont
  {Bar{\'o}}, \citenamefont {Corral}, \citenamefont {Illa}, \citenamefont
  {Planes}, \citenamefont {Salje}, \citenamefont {Schranz}, \citenamefont
  {Soto-Parra},\ and\ \citenamefont {Vives}}]{baro2013}%
  \BibitemOpen
  \bibfield  {author} {\bibinfo {author} {\bibfnamefont {J.}~\bibnamefont
  {Bar{\'o}}}, \bibinfo {author} {\bibfnamefont {{\'A}.}~\bibnamefont
  {Corral}}, \bibinfo {author} {\bibfnamefont {X.}~\bibnamefont {Illa}},
  \bibinfo {author} {\bibfnamefont {A.}~\bibnamefont {Planes}}, \bibinfo
  {author} {\bibfnamefont {E.~K.}\ \bibnamefont {Salje}}, \bibinfo {author}
  {\bibfnamefont {W.}~\bibnamefont {Schranz}}, \bibinfo {author} {\bibfnamefont
  {D.~E.}\ \bibnamefont {Soto-Parra}},\ and\ \bibinfo {author} {\bibfnamefont
  {E.}~\bibnamefont {Vives}},\ }\bibfield  {title} {\bibinfo {title}
  {Statistical similarity between the compression of a porous material and
  earthquakes},\ }\href@noop {} {\bibfield  {journal} {\bibinfo  {journal}
  {Phys. Rev. Lett.}\ }\textbf {\bibinfo {volume} {110}},\ \bibinfo {pages}
  {088702} (\bibinfo {year} {2013})}\BibitemShut {NoStop}%
\bibitem [{\citenamefont {Castillo-Villa}\ \emph {et~al.}(2013)\citenamefont
  {Castillo-Villa}, \citenamefont {Baro}, \citenamefont {Planes}, \citenamefont
  {Salje}, \citenamefont {Sellappan}, \citenamefont {Kriven},\ and\
  \citenamefont {Vives}}]{castillo2013}%
  \BibitemOpen
  \bibfield  {author} {\bibinfo {author} {\bibfnamefont {P.}~\bibnamefont
  {Castillo-Villa}}, \bibinfo {author} {\bibfnamefont {J.}~\bibnamefont
  {Baro}}, \bibinfo {author} {\bibfnamefont {A.}~\bibnamefont {Planes}},
  \bibinfo {author} {\bibfnamefont {E.~K.~H.}\ \bibnamefont {Salje}}, \bibinfo
  {author} {\bibfnamefont {P.}~\bibnamefont {Sellappan}}, \bibinfo {author}
  {\bibfnamefont {W.~M.}\ \bibnamefont {Kriven}},\ and\ \bibinfo {author}
  {\bibfnamefont {E.}~\bibnamefont {Vives}},\ }\bibfield  {title} {\bibinfo
  {title} {Crackling noise during failure of alumina under compression: The
  effect of porosity},\ }\href@noop {} {\bibfield  {journal} {\bibinfo
  {journal} {J. Phys. Cond. Mat.}\ }\textbf {\bibinfo {volume} {25}},\ \bibinfo
  {pages} {292202} (\bibinfo {year} {2013})}\BibitemShut {NoStop}%
\bibitem [{\citenamefont {Bar{\'o}}\ \emph {et~al.}(2018)\citenamefont
  {Bar{\'o}}, \citenamefont {Dahmen}, \citenamefont {Davidsen}, \citenamefont
  {Planes}, \citenamefont {Castillo}, \citenamefont {Nataf}, \citenamefont
  {Salje},\ and\ \citenamefont {Vives}}]{baro2018}%
  \BibitemOpen
  \bibfield  {author} {\bibinfo {author} {\bibfnamefont {J.}~\bibnamefont
  {Bar{\'o}}}, \bibinfo {author} {\bibfnamefont {K.~A.}\ \bibnamefont
  {Dahmen}}, \bibinfo {author} {\bibfnamefont {J.}~\bibnamefont {Davidsen}},
  \bibinfo {author} {\bibfnamefont {A.}~\bibnamefont {Planes}}, \bibinfo
  {author} {\bibfnamefont {P.~O.}\ \bibnamefont {Castillo}}, \bibinfo {author}
  {\bibfnamefont {G.~F.}\ \bibnamefont {Nataf}}, \bibinfo {author}
  {\bibfnamefont {E.~K.}\ \bibnamefont {Salje}},\ and\ \bibinfo {author}
  {\bibfnamefont {E.}~\bibnamefont {Vives}},\ }\bibfield  {title} {\bibinfo
  {title} {Experimental evidence of accelerated seismic release without
  critical failure in acoustic emissions of compressed nanoporous materials},\
  }\href@noop {} {\bibfield  {journal} {\bibinfo  {journal} {Phys. Rev. Lett.}\
  }\textbf {\bibinfo {volume} {120}},\ \bibinfo {pages} {245501} (\bibinfo
  {year} {2018})}\BibitemShut {NoStop}%
\bibitem [{\citenamefont {Vu}\ \emph {et~al.}(2019)\citenamefont {Vu},
  \citenamefont {Amitrano}, \citenamefont {Pl\'e},\ and\ \citenamefont
  {Weiss}}]{weiss2019}%
  \BibitemOpen
  \bibfield  {author} {\bibinfo {author} {\bibfnamefont {C.-C.}\ \bibnamefont
  {Vu}}, \bibinfo {author} {\bibfnamefont {D.}~\bibnamefont {Amitrano}},
  \bibinfo {author} {\bibfnamefont {O.}~\bibnamefont {Pl\'e}},\ and\ \bibinfo
  {author} {\bibfnamefont {J.}~\bibnamefont {Weiss}},\ }\bibfield  {title}
  {\bibinfo {title} {Compressive failure as a critical transition: Experimental
  evidence and mapping onto the universality class of depinning},\ }\href
  {https://doi.org/10.1103/PhysRevLett.122.015502} {\bibfield  {journal}
  {\bibinfo  {journal} {Phys. Rev. Lett.}\ }\textbf {\bibinfo {volume} {122}},\
  \bibinfo {pages} {015502} (\bibinfo {year} {2019})}\BibitemShut {NoStop}%
\bibitem [{\citenamefont {Sornette}(1994)}]{sornette1994}%
  \BibitemOpen
  \bibfield  {author} {\bibinfo {author} {\bibfnamefont {D.}~\bibnamefont
  {Sornette}},\ }\bibfield  {title} {\bibinfo {title} {Sweeping of an
  instability: an alternative to self-organized criticality to get power laws
  without parameter tuning},\ }\href@noop {} {\bibfield  {journal} {\bibinfo
  {journal} {J. Phys. I}\ }\textbf {\bibinfo {volume} {4}},\ \bibinfo {pages}
  {209} (\bibinfo {year} {1994})}\BibitemShut {NoStop}%
\bibitem [{\citenamefont {Zapperi}\ \emph
  {et~al.}(1997{\natexlab{a}})\citenamefont {Zapperi}, \citenamefont {Ray},
  \citenamefont {Stanley},\ and\ \citenamefont {Vespignani}}]{zapperi1997}%
  \BibitemOpen
  \bibfield  {author} {\bibinfo {author} {\bibfnamefont {S.}~\bibnamefont
  {Zapperi}}, \bibinfo {author} {\bibfnamefont {P.}~\bibnamefont {Ray}},
  \bibinfo {author} {\bibfnamefont {H.~E.}\ \bibnamefont {Stanley}},\ and\
  \bibinfo {author} {\bibfnamefont {A.}~\bibnamefont {Vespignani}},\ }\bibfield
   {title} {\bibinfo {title} {First-order transition in the breakdown of
  disordered media},\ }\href@noop {} {\bibfield  {journal} {\bibinfo  {journal}
  {Phys. Rev. Lett.}\ }\textbf {\bibinfo {volume} {78}},\ \bibinfo {pages}
  {1408} (\bibinfo {year} {1997}{\natexlab{a}})}\BibitemShut {NoStop}%
\bibitem [{\citenamefont {Zapperi}\ \emph
  {et~al.}(1997{\natexlab{b}})\citenamefont {Zapperi}, \citenamefont
  {Vespignani},\ and\ \citenamefont {Stanley}}]{zapperi1997b}%
  \BibitemOpen
  \bibfield  {author} {\bibinfo {author} {\bibfnamefont {S.}~\bibnamefont
  {Zapperi}}, \bibinfo {author} {\bibfnamefont {A.}~\bibnamefont
  {Vespignani}},\ and\ \bibinfo {author} {\bibfnamefont {H.~E.}\ \bibnamefont
  {Stanley}},\ }\bibfield  {title} {\bibinfo {title} {Plasticity and avalanche
  behaviour in microfracturing phenomena},\ }\href@noop {} {\bibfield
  {journal} {\bibinfo  {journal} {Nature}\ }\textbf {\bibinfo {volume} {388}},\
  \bibinfo {pages} {658} (\bibinfo {year} {1997}{\natexlab{b}})}\BibitemShut
  {NoStop}%
\bibitem [{\citenamefont {Sornette}(2002)}]{sornette2002}%
  \BibitemOpen
  \bibfield  {author} {\bibinfo {author} {\bibfnamefont {D.}~\bibnamefont
  {Sornette}},\ }\bibfield  {title} {\bibinfo {title} {Predictability of
  catastrophic events: Material rupture, earthquakes, turbulence, financial
  crashes, and human birth},\ }\href@noop {} {\bibfield  {journal} {\bibinfo
  {journal} {Proceedings of the National Academy of Sciences}\ }\textbf
  {\bibinfo {volume} {99}},\ \bibinfo {pages} {2522} (\bibinfo {year}
  {2002})}\BibitemShut {NoStop}%
\bibitem [{\citenamefont {da~Rocha}\ and\ \citenamefont
  {Truskinovsky}(2020)}]{truskinovsky2020}%
  \BibitemOpen
  \bibfield  {author} {\bibinfo {author} {\bibfnamefont {H.~B.}\ \bibnamefont
  {da~Rocha}}\ and\ \bibinfo {author} {\bibfnamefont {L.}~\bibnamefont
  {Truskinovsky}},\ }\bibfield  {title} {\bibinfo {title} {Rigidity-controlled
  crossover: From spinodal to critical failure},\ }\href@noop {} {\bibfield
  {journal} {\bibinfo  {journal} {Physical Review Letters}\ }\textbf {\bibinfo
  {volume} {124}},\ \bibinfo {pages} {015501} (\bibinfo {year}
  {2020})}\BibitemShut {NoStop}%
\bibitem [{\citenamefont {Guarino}\ \emph {et~al.}(1998)\citenamefont
  {Guarino}, \citenamefont {Garcimartin},\ and\ \citenamefont
  {Ciliberto}}]{guarino1998}%
  \BibitemOpen
  \bibfield  {author} {\bibinfo {author} {\bibfnamefont {A.}~\bibnamefont
  {Guarino}}, \bibinfo {author} {\bibfnamefont {A.}~\bibnamefont
  {Garcimartin}},\ and\ \bibinfo {author} {\bibfnamefont {S.}~\bibnamefont
  {Ciliberto}},\ }\bibfield  {title} {\bibinfo {title} {An experimental test of
  the critical behaviour of fracture precursors},\ }\href@noop {} {\bibfield
  {journal} {\bibinfo  {journal} {Eur. Phys. J. B}\ }\textbf {\bibinfo {volume}
  {6}},\ \bibinfo {pages} {13} (\bibinfo {year} {1998})}\BibitemShut {NoStop}%
\bibitem [{\citenamefont {Girard}\ \emph {et~al.}(2010)\citenamefont {Girard},
  \citenamefont {Amitrano},\ and\ \citenamefont {Weiss}}]{girard2010}%
  \BibitemOpen
  \bibfield  {author} {\bibinfo {author} {\bibfnamefont {L.}~\bibnamefont
  {Girard}}, \bibinfo {author} {\bibfnamefont {D.}~\bibnamefont {Amitrano}},\
  and\ \bibinfo {author} {\bibfnamefont {J.}~\bibnamefont {Weiss}},\ }\bibfield
   {title} {\bibinfo {title} {Failure as a critical phenomenon in a progressive
  damage model},\ }\href@noop {} {\bibfield  {journal} {\bibinfo  {journal}
  {Journal of Statistical Mechanics: Theory and Experiment}\ }\textbf {\bibinfo
  {volume} {2010}},\ \bibinfo {pages} {P01013} (\bibinfo {year}
  {2010})}\BibitemShut {NoStop}%
\bibitem [{\citenamefont {Kun}\ \emph {et~al.}(2014)\citenamefont {Kun},
  \citenamefont {Varga}, \citenamefont {Lennartz-Sassinek},\ and\ \citenamefont
  {Main}}]{kun2014}%
  \BibitemOpen
  \bibfield  {author} {\bibinfo {author} {\bibfnamefont {F.}~\bibnamefont
  {Kun}}, \bibinfo {author} {\bibfnamefont {I.}~\bibnamefont {Varga}}, \bibinfo
  {author} {\bibfnamefont {S.}~\bibnamefont {Lennartz-Sassinek}},\ and\
  \bibinfo {author} {\bibfnamefont {I.~G.}\ \bibnamefont {Main}},\ }\bibfield
  {title} {\bibinfo {title} {Rupture cascades in a discrete element model of a
  porous sedimentary rock},\ }\href@noop {} {\bibfield  {journal} {\bibinfo
  {journal} {Phys. Rev. Lett.}\ }\textbf {\bibinfo {volume} {112}},\ \bibinfo
  {pages} {165501} (\bibinfo {year} {2014})}\BibitemShut {NoStop}%
\bibitem [{\citenamefont {Kandula}\ \emph {et~al.}(2019)\citenamefont
  {Kandula}, \citenamefont {Cordonnier}, \citenamefont {Boller}, \citenamefont
  {Weiss}, \citenamefont {Dysthe},\ and\ \citenamefont {Renard}}]{kandula2019}%
  \BibitemOpen
  \bibfield  {author} {\bibinfo {author} {\bibfnamefont {N.}~\bibnamefont
  {Kandula}}, \bibinfo {author} {\bibfnamefont {B.}~\bibnamefont {Cordonnier}},
  \bibinfo {author} {\bibfnamefont {E.}~\bibnamefont {Boller}}, \bibinfo
  {author} {\bibfnamefont {J.}~\bibnamefont {Weiss}}, \bibinfo {author}
  {\bibfnamefont {D.~K.}\ \bibnamefont {Dysthe}},\ and\ \bibinfo {author}
  {\bibfnamefont {F.}~\bibnamefont {Renard}},\ }\bibfield  {title} {\bibinfo
  {title} {Dynamics of microscale precursors during brittle compressive failure
  in carrara marble},\ }\href@noop {} {\bibfield  {journal} {\bibinfo
  {journal} {J. Geophys. Res. Solid Earth}\ }\textbf {\bibinfo {volume}
  {124}},\ \bibinfo {pages} {6121} (\bibinfo {year} {2019})}\BibitemShut
  {NoStop}%
\bibitem [{\citenamefont {Roux}\ \emph {et~al.}(1988)\citenamefont {Roux},
  \citenamefont {Hansen}, \citenamefont {Herrmann},\ and\ \citenamefont
  {Guyon}}]{roux1988}%
  \BibitemOpen
  \bibfield  {author} {\bibinfo {author} {\bibfnamefont {S.}~\bibnamefont
  {Roux}}, \bibinfo {author} {\bibfnamefont {A.}~\bibnamefont {Hansen}},
  \bibinfo {author} {\bibfnamefont {H.}~\bibnamefont {Herrmann}},\ and\
  \bibinfo {author} {\bibfnamefont {E.}~\bibnamefont {Guyon}},\ }\bibfield
  {title} {\bibinfo {title} {Rupture of heterogeneous media in the limit of
  infinite disorder},\ }\href@noop {} {\bibfield  {journal} {\bibinfo
  {journal} {Journal of statistical physics}\ }\textbf {\bibinfo {volume}
  {52}},\ \bibinfo {pages} {237} (\bibinfo {year} {1988})}\BibitemShut
  {NoStop}%
\bibitem [{\citenamefont {Delaplace}\ \emph {et~al.}(1996)\citenamefont
  {Delaplace}, \citenamefont {Pijaudier-Cabot},\ and\ \citenamefont
  {Roux}}]{delaplace1996}%
  \BibitemOpen
  \bibfield  {author} {\bibinfo {author} {\bibfnamefont {A.}~\bibnamefont
  {Delaplace}}, \bibinfo {author} {\bibfnamefont {G.}~\bibnamefont
  {Pijaudier-Cabot}},\ and\ \bibinfo {author} {\bibfnamefont {S.}~\bibnamefont
  {Roux}},\ }\bibfield  {title} {\bibinfo {title} {Progressive damage in
  discrete models and consequences on continuum modelling},\ }\href@noop {}
  {\bibfield  {journal} {\bibinfo  {journal} {Journal of the Mechanics and
  Physics of Solids}\ }\textbf {\bibinfo {volume} {44}},\ \bibinfo {pages} {99}
  (\bibinfo {year} {1996})}\BibitemShut {NoStop}%
\bibitem [{\citenamefont {Garcimartin}\ \emph {et~al.}(1997)\citenamefont
  {Garcimartin}, \citenamefont {Guarino}, \citenamefont {Bellon},\ and\
  \citenamefont {Ciliberto}}]{garcimartin1997}%
  \BibitemOpen
  \bibfield  {author} {\bibinfo {author} {\bibfnamefont {A.}~\bibnamefont
  {Garcimartin}}, \bibinfo {author} {\bibfnamefont {A.}~\bibnamefont
  {Guarino}}, \bibinfo {author} {\bibfnamefont {L.}~\bibnamefont {Bellon}},\
  and\ \bibinfo {author} {\bibfnamefont {S.}~\bibnamefont {Ciliberto}},\
  }\bibfield  {title} {\bibinfo {title} {Statistical properties of fracture
  precursors},\ }\href@noop {} {\bibfield  {journal} {\bibinfo  {journal}
  {Phys. Rev. Lett.}\ }\textbf {\bibinfo {volume} {79}},\ \bibinfo {pages}
  {3202} (\bibinfo {year} {1997})}\BibitemShut {NoStop}%
\bibitem [{\citenamefont {Moreno}\ \emph {et~al.}(2000)\citenamefont {Moreno},
  \citenamefont {G\`{o}mez},\ and\ \citenamefont {Pacheco}}]{moreno2000}%
  \BibitemOpen
  \bibfield  {author} {\bibinfo {author} {\bibfnamefont {Y.}~\bibnamefont
  {Moreno}}, \bibinfo {author} {\bibfnamefont {J.~B.}\ \bibnamefont
  {G\`{o}mez}},\ and\ \bibinfo {author} {\bibfnamefont {A.~F.}\ \bibnamefont
  {Pacheco}},\ }\bibfield  {title} {\bibinfo {title} {Fracture and second-order
  phase transitions},\ }\href@noop {} {\bibfield  {journal} {\bibinfo
  {journal} {Phys. Rev. Lett.}\ }\textbf {\bibinfo {volume} {85}},\ \bibinfo
  {pages} {2865} (\bibinfo {year} {2000})}\BibitemShut {NoStop}%
\bibitem [{\citenamefont {Weiss}\ \emph {et~al.}(2014)\citenamefont {Weiss},
  \citenamefont {Girard}, \citenamefont {Gimbert}, \citenamefont {Amitrano},\
  and\ \citenamefont {Vandembroucq}}]{weiss2014}%
  \BibitemOpen
  \bibfield  {author} {\bibinfo {author} {\bibfnamefont {J.}~\bibnamefont
  {Weiss}}, \bibinfo {author} {\bibfnamefont {L.}~\bibnamefont {Girard}},
  \bibinfo {author} {\bibfnamefont {F.}~\bibnamefont {Gimbert}}, \bibinfo
  {author} {\bibfnamefont {D.}~\bibnamefont {Amitrano}},\ and\ \bibinfo
  {author} {\bibfnamefont {D.}~\bibnamefont {Vandembroucq}},\ }\bibfield
  {title} {\bibinfo {title} {(finite) statistical size effects on compressive
  strength},\ }\href@noop {} {\bibfield  {journal} {\bibinfo  {journal}
  {Proceedings of the National Academy of Sciences}\ }\textbf {\bibinfo
  {volume} {111}},\ \bibinfo {pages} {6231} (\bibinfo {year}
  {2014})}\BibitemShut {NoStop}%
\bibitem [{\citenamefont {Renard}\ \emph {et~al.}(2018)\citenamefont {Renard},
  \citenamefont {Weiss}, \citenamefont {Mathiesen}, \citenamefont {Ben-Zion},
  \citenamefont {Kandula},\ and\ \citenamefont {Cordonnier}}]{renard2018}%
  \BibitemOpen
  \bibfield  {author} {\bibinfo {author} {\bibfnamefont {F.}~\bibnamefont
  {Renard}}, \bibinfo {author} {\bibfnamefont {J.}~\bibnamefont {Weiss}},
  \bibinfo {author} {\bibfnamefont {J.}~\bibnamefont {Mathiesen}}, \bibinfo
  {author} {\bibfnamefont {Y.}~\bibnamefont {Ben-Zion}}, \bibinfo {author}
  {\bibfnamefont {N.}~\bibnamefont {Kandula}},\ and\ \bibinfo {author}
  {\bibfnamefont {B.}~\bibnamefont {Cordonnier}},\ }\bibfield  {title}
  {\bibinfo {title} {Critical evolution of damage toward system-size failure in
  crystalline rock},\ }\href@noop {} {\bibfield  {journal} {\bibinfo  {journal}
  {J. Geophys. Res. Solid Earth}\ }\textbf {\bibinfo {volume} {123}},\ \bibinfo
  {pages} {1969} (\bibinfo {year} {2018})}\BibitemShut {NoStop}%
\bibitem [{\citenamefont {Barab\'asi}\ and\ \citenamefont
  {Stanley}(1995)}]{barabasi}%
  \BibitemOpen
  \bibfield  {author} {\bibinfo {author} {\bibfnamefont {A.~L.}\ \bibnamefont
  {Barab\'asi}}\ and\ \bibinfo {author} {\bibfnamefont {H.~E.}\ \bibnamefont
  {Stanley}},\ }\href@noop {} {\emph {\bibinfo {title} {Fractal concepts in
  surface growth}}}\ (\bibinfo  {publisher} {Cambridge University Press},\
  \bibinfo {year} {1995})\BibitemShut {NoStop}%
\bibitem [{\citenamefont {Sethna}\ \emph {et~al.}(2001)\citenamefont {Sethna},
  \citenamefont {Dahmen},\ and\ \citenamefont {Myers}}]{sethna}%
  \BibitemOpen
  \bibfield  {author} {\bibinfo {author} {\bibfnamefont {J.}~\bibnamefont
  {Sethna}}, \bibinfo {author} {\bibfnamefont {K.}~\bibnamefont {Dahmen}},\
  and\ \bibinfo {author} {\bibfnamefont {C.~R.}\ \bibnamefont {Myers}},\
  }\bibfield  {title} {\bibinfo {title} {Crackling noise},\ }\href@noop {}
  {\bibfield  {journal} {\bibinfo  {journal} {Nature}\ }\textbf {\bibinfo
  {volume} {410}},\ \bibinfo {pages} {242} (\bibinfo {year}
  {2001})}\BibitemShut {NoStop}%
\bibitem [{\citenamefont {Wiese}(2021)}]{wiese2}%
  \BibitemOpen
  \bibfield  {author} {\bibinfo {author} {\bibfnamefont {K.~J.}\ \bibnamefont
  {Wiese}},\ }\bibfield  {title} {\bibinfo {title} {Theory and experiments for
  disordered elastic manifolds, depinning, avalanches, and sandpiles},\
  }\href@noop {} {\bibfield  {journal} {\bibinfo  {journal} {arXiv preprint
  arXiv:2102.01215}\ }\textbf {\bibinfo {volume} {.}} (\bibinfo {year}
  {2021})}\BibitemShut {NoStop}%
\bibitem [{\citenamefont {Berthier}\ \emph {et~al.}(2022)\citenamefont
  {Berthier}, \citenamefont {Mayya},\ and\ \citenamefont
  {Ponson}}]{berthier2021}%
  \BibitemOpen
  \bibfield  {author} {\bibinfo {author} {\bibfnamefont {E.}~\bibnamefont
  {Berthier}}, \bibinfo {author} {\bibfnamefont {A.}~\bibnamefont {Mayya}},\
  and\ \bibinfo {author} {\bibfnamefont {L.}~\bibnamefont {Ponson}},\
  }\bibfield  {title} {\bibinfo {title} {Damage spreading in quasi-brittle
  disordered solids: Ii. what the statistics of precursors teach us about
  compressive failure},\ }\href@noop {} {\bibfield  {journal} {\bibinfo
  {journal} {Journal of the Mechanics and Physics of Solids}\ }\textbf
  {\bibinfo {volume} {162}},\ \bibinfo {pages} {104826} (\bibinfo {year}
  {2022})}\BibitemShut {NoStop}%
\bibitem [{\citenamefont {Mayya}\ \emph {et~al.}(2023)\citenamefont {Mayya},
  \citenamefont {Berthier},\ and\ \citenamefont {Ponson}}]{supp_info}%
  \BibitemOpen
  \bibfield  {author} {\bibinfo {author} {\bibfnamefont {A.}~\bibnamefont
  {Mayya}}, \bibinfo {author} {\bibfnamefont {E.}~\bibnamefont {Berthier}},\
  and\ \bibinfo {author} {\bibfnamefont {L.}~\bibnamefont {Ponson}},\
  }\href@noop {} {\bibinfo {title} {Supplementary information for ``how
  criticality meets bifurcation in compressive failure of disordered solid''}}
  (\bibinfo {year} {2023})\BibitemShut {NoStop}%
\bibitem [{\citenamefont {Poirier}\ \emph {et~al.}(1992)\citenamefont
  {Poirier}, \citenamefont {Ammi}, \citenamefont {Bideau},\ and\ \citenamefont
  {Troadec}}]{poirier1992}%
  \BibitemOpen
  \bibfield  {author} {\bibinfo {author} {\bibfnamefont {C.}~\bibnamefont
  {Poirier}}, \bibinfo {author} {\bibfnamefont {M.}~\bibnamefont {Ammi}},
  \bibinfo {author} {\bibfnamefont {D.}~\bibnamefont {Bideau}},\ and\ \bibinfo
  {author} {\bibfnamefont {J.~P.}\ \bibnamefont {Troadec}},\ }\bibfield
  {title} {\bibinfo {title} {Experimental study of the geometrical effects in
  the localization of deformation},\ }\href
  {https://doi.org/10.1103/PhysRevLett.68.216} {\bibfield  {journal} {\bibinfo
  {journal} {Phys. Rev. Lett.}\ }\textbf {\bibinfo {volume} {68}},\ \bibinfo
  {pages} {216} (\bibinfo {year} {1992})}\BibitemShut {NoStop}%
\bibitem [{\citenamefont {Karimi}\ \emph {et~al.}(2019)\citenamefont {Karimi},
  \citenamefont {Amitrano},\ and\ \citenamefont {Weiss}}]{karimi2019}%
  \BibitemOpen
  \bibfield  {author} {\bibinfo {author} {\bibfnamefont {K.}~\bibnamefont
  {Karimi}}, \bibinfo {author} {\bibfnamefont {D.}~\bibnamefont {Amitrano}},\
  and\ \bibinfo {author} {\bibfnamefont {J.}~\bibnamefont {Weiss}},\ }\bibfield
   {title} {\bibinfo {title} {From plastic flow to brittle fracture: Role of
  microscopic friction in amorphous solids},\ }\href@noop {} {\bibfield
  {journal} {\bibinfo  {journal} {Physical Review E}\ }\textbf {\bibinfo
  {volume} {100}},\ \bibinfo {pages} {012908} (\bibinfo {year}
  {2019})}\BibitemShut {NoStop}%
\bibitem [{\citenamefont {Renard}\ \emph {et~al.}(2019)\citenamefont {Renard},
  \citenamefont {McBeck}, \citenamefont {Kandula}, \citenamefont {Cordonnier},
  \citenamefont {Meakin},\ and\ \citenamefont {Ben-Zion}}]{renard2019}%
  \BibitemOpen
  \bibfield  {author} {\bibinfo {author} {\bibfnamefont {F.}~\bibnamefont
  {Renard}}, \bibinfo {author} {\bibfnamefont {J.}~\bibnamefont {McBeck}},
  \bibinfo {author} {\bibfnamefont {N.}~\bibnamefont {Kandula}}, \bibinfo
  {author} {\bibfnamefont {B.}~\bibnamefont {Cordonnier}}, \bibinfo {author}
  {\bibfnamefont {P.}~\bibnamefont {Meakin}},\ and\ \bibinfo {author}
  {\bibfnamefont {Y.}~\bibnamefont {Ben-Zion}},\ }\bibfield  {title} {\bibinfo
  {title} {Volumetric and shear processes in crystalline rock approaching
  faulting},\ }\href@noop {} {\bibfield  {journal} {\bibinfo  {journal}
  {Proceedings of the National Academy of Sciences}\ }\textbf {\bibinfo
  {volume} {116}},\ \bibinfo {pages} {16234} (\bibinfo {year}
  {2019})}\BibitemShut {NoStop}%
\bibitem [{\citenamefont {Cartwright-Taylor}\ \emph {et~al.}(2020)\citenamefont
  {Cartwright-Taylor}, \citenamefont {Main}, \citenamefont {Butler},
  \citenamefont {Fusseis}, \citenamefont {Flynn},\ and\ \citenamefont
  {King}}]{cartwright2020}%
  \BibitemOpen
  \bibfield  {author} {\bibinfo {author} {\bibfnamefont {A.}~\bibnamefont
  {Cartwright-Taylor}}, \bibinfo {author} {\bibfnamefont {I.~G.}\ \bibnamefont
  {Main}}, \bibinfo {author} {\bibfnamefont {I.~B.}\ \bibnamefont {Butler}},
  \bibinfo {author} {\bibfnamefont {F.}~\bibnamefont {Fusseis}}, \bibinfo
  {author} {\bibfnamefont {M.}~\bibnamefont {Flynn}},\ and\ \bibinfo {author}
  {\bibfnamefont {A.}~\bibnamefont {King}},\ }\bibfield  {title} {\bibinfo
  {title} {Catastrophic failure: How and when? insights from 4-d in situ x-ray
  microtomography},\ }\href@noop {} {\bibfield  {journal} {\bibinfo  {journal}
  {J. Geophys. Res. Solid Earth}\ }\textbf {\bibinfo {volume} {125}},\ \bibinfo
  {pages} {e2020JB019642} (\bibinfo {year} {2020})}\BibitemShut {NoStop}%
\bibitem [{\citenamefont {Lennartz-Sassinek}\ \emph {et~al.}(2014)\citenamefont
  {Lennartz-Sassinek}, \citenamefont {Main}, \citenamefont {Zaiser},\ and\
  \citenamefont {Graham}}]{lennartz2014}%
  \BibitemOpen
  \bibfield  {author} {\bibinfo {author} {\bibfnamefont {S.}~\bibnamefont
  {Lennartz-Sassinek}}, \bibinfo {author} {\bibfnamefont {I.}~\bibnamefont
  {Main}}, \bibinfo {author} {\bibfnamefont {M.}~\bibnamefont {Zaiser}},\ and\
  \bibinfo {author} {\bibfnamefont {C.}~\bibnamefont {Graham}},\ }\bibfield
  {title} {\bibinfo {title} {Acceleration and localization of subcritical crack
  growth in a natural composite material},\ }\href@noop {} {\bibfield
  {journal} {\bibinfo  {journal} {Phys. Rev. E}\ }\textbf {\bibinfo {volume}
  {90}},\ \bibinfo {pages} {052401} (\bibinfo {year} {2014})}\BibitemShut
  {NoStop}%
\bibitem [{\citenamefont {Davidsen}\ \emph {et~al.}(2017)\citenamefont
  {Davidsen}, \citenamefont {Kwiatek}, \citenamefont {Charalampidou},
  \citenamefont {Goebel}, \citenamefont {Stanchits}, \citenamefont {R{\"u}ck},\
  and\ \citenamefont {Dresen}}]{davidsen2017}%
  \BibitemOpen
  \bibfield  {author} {\bibinfo {author} {\bibfnamefont {J.}~\bibnamefont
  {Davidsen}}, \bibinfo {author} {\bibfnamefont {G.}~\bibnamefont {Kwiatek}},
  \bibinfo {author} {\bibfnamefont {E.-M.}\ \bibnamefont {Charalampidou}},
  \bibinfo {author} {\bibfnamefont {T.}~\bibnamefont {Goebel}}, \bibinfo
  {author} {\bibfnamefont {S.}~\bibnamefont {Stanchits}}, \bibinfo {author}
  {\bibfnamefont {M.}~\bibnamefont {R{\"u}ck}},\ and\ \bibinfo {author}
  {\bibfnamefont {G.}~\bibnamefont {Dresen}},\ }\bibfield  {title} {\bibinfo
  {title} {Triggering processes in rock fracture},\ }\href@noop {} {\bibfield
  {journal} {\bibinfo  {journal} {Phys. Rev. Lett.}\ }\textbf {\bibinfo
  {volume} {119}},\ \bibinfo {pages} {068501} (\bibinfo {year}
  {2017})}\BibitemShut {NoStop}%
\bibitem [{\citenamefont {Baud}\ \emph {et~al.}(2004)\citenamefont {Baud},
  \citenamefont {Klein},\ and\ \citenamefont {Wong}}]{baud2004}%
  \BibitemOpen
  \bibfield  {author} {\bibinfo {author} {\bibfnamefont {P.}~\bibnamefont
  {Baud}}, \bibinfo {author} {\bibfnamefont {E.}~\bibnamefont {Klein}},\ and\
  \bibinfo {author} {\bibfnamefont {T.-f.}\ \bibnamefont {Wong}},\ }\bibfield
  {title} {\bibinfo {title} {Compaction localization in porous sandstones:
  spatial evolution of damage and acoustic emission activity},\ }\href@noop {}
  {\bibfield  {journal} {\bibinfo  {journal} {Journal of Structural Geology}\
  }\textbf {\bibinfo {volume} {26}},\ \bibinfo {pages} {603} (\bibinfo {year}
  {2004})}\BibitemShut {NoStop}%
\bibitem [{\citenamefont {Lenoir}\ \emph {et~al.}(2007)\citenamefont {Lenoir},
  \citenamefont {Bornert}, \citenamefont {Desrues}, \citenamefont
  {B{\'e}suelle},\ and\ \citenamefont {Viggiani}}]{lenoir2007}%
  \BibitemOpen
  \bibfield  {author} {\bibinfo {author} {\bibfnamefont {N.}~\bibnamefont
  {Lenoir}}, \bibinfo {author} {\bibfnamefont {M.}~\bibnamefont {Bornert}},
  \bibinfo {author} {\bibfnamefont {J.}~\bibnamefont {Desrues}}, \bibinfo
  {author} {\bibfnamefont {P.}~\bibnamefont {B{\'e}suelle}},\ and\ \bibinfo
  {author} {\bibfnamefont {G.}~\bibnamefont {Viggiani}},\ }\bibfield  {title}
  {\bibinfo {title} {Volumetric digital image correlation applied to x-ray
  microtomography images from triaxial compression tests on argillaceous
  rock},\ }\href@noop {} {\bibfield  {journal} {\bibinfo  {journal} {Strain}\
  }\textbf {\bibinfo {volume} {43}},\ \bibinfo {pages} {193} (\bibinfo {year}
  {2007})}\BibitemShut {NoStop}%
\bibitem [{\citenamefont {McBeck}\ \emph {et~al.}(2020)\citenamefont {McBeck},
  \citenamefont {Aiken}, \citenamefont {Mathiesen}, \citenamefont {Ben-Zion},\
  and\ \citenamefont {Renard}}]{mcbeck2020}%
  \BibitemOpen
  \bibfield  {author} {\bibinfo {author} {\bibfnamefont {J.~A.}\ \bibnamefont
  {McBeck}}, \bibinfo {author} {\bibfnamefont {J.~M.}\ \bibnamefont {Aiken}},
  \bibinfo {author} {\bibfnamefont {J.}~\bibnamefont {Mathiesen}}, \bibinfo
  {author} {\bibfnamefont {Y.}~\bibnamefont {Ben-Zion}},\ and\ \bibinfo
  {author} {\bibfnamefont {F.}~\bibnamefont {Renard}},\ }\bibfield  {title}
  {\bibinfo {title} {Deformation precursors to catastrophic failure in rocks},\
  }\href@noop {} {\bibfield  {journal} {\bibinfo  {journal} {Geophysical
  Research Letters}\ }\textbf {\bibinfo {volume} {47}},\ \bibinfo {pages}
  {e2020GL090255} (\bibinfo {year} {2020})}\BibitemShut {NoStop}%
\bibitem [{\citenamefont {Kandula}\ \emph {et~al.}(2022)\citenamefont
  {Kandula}, \citenamefont {McBeck}, \citenamefont {Cordonnier}, \citenamefont
  {Weiss}, \citenamefont {Dysthe},\ and\ \citenamefont {Renard}}]{kandula2022}%
  \BibitemOpen
  \bibfield  {author} {\bibinfo {author} {\bibfnamefont {N.}~\bibnamefont
  {Kandula}}, \bibinfo {author} {\bibfnamefont {J.}~\bibnamefont {McBeck}},
  \bibinfo {author} {\bibfnamefont {B.}~\bibnamefont {Cordonnier}}, \bibinfo
  {author} {\bibfnamefont {J.}~\bibnamefont {Weiss}}, \bibinfo {author}
  {\bibfnamefont {D.~K.}\ \bibnamefont {Dysthe}},\ and\ \bibinfo {author}
  {\bibfnamefont {F.}~\bibnamefont {Renard}},\ }\bibfield  {title} {\bibinfo
  {title} {Synchrotron 4d x-ray imaging reveals strain localization at the
  onset of system-size failure in porous reservoir rocks},\ }\href@noop {}
  {\bibfield  {journal} {\bibinfo  {journal} {Pure and Applied Geophysics}\
  }\textbf {\bibinfo {volume} {179}},\ \bibinfo {pages} {325} (\bibinfo {year}
  {2022})}\BibitemShut {NoStop}%
\bibitem [{\citenamefont {Berthier}\ \emph {et~al.}(2017)\citenamefont
  {Berthier}, \citenamefont {D{\'e}mery},\ and\ \citenamefont
  {Ponson}}]{berthier2017}%
  \BibitemOpen
  \bibfield  {author} {\bibinfo {author} {\bibfnamefont {E.}~\bibnamefont
  {Berthier}}, \bibinfo {author} {\bibfnamefont {V.}~\bibnamefont
  {D{\'e}mery}},\ and\ \bibinfo {author} {\bibfnamefont {L.}~\bibnamefont
  {Ponson}},\ }\bibfield  {title} {\bibinfo {title} {Damage spreading in
  quasi-brittle disordered solids: I. localization and failure},\ }\href@noop
  {} {\bibfield  {journal} {\bibinfo  {journal} {Journal of the Mechanics and
  Physics of Solids}\ }\textbf {\bibinfo {volume} {102}},\ \bibinfo {pages}
  {101} (\bibinfo {year} {2017})}\BibitemShut {NoStop}%
\bibitem [{\citenamefont {Dansereau}\ \emph {et~al.}(2019)\citenamefont
  {Dansereau}, \citenamefont {D{\'e}mery}, \citenamefont {Berthier},
  \citenamefont {Weiss},\ and\ \citenamefont {Ponson}}]{dansereau2019}%
  \BibitemOpen
  \bibfield  {author} {\bibinfo {author} {\bibfnamefont {V.}~\bibnamefont
  {Dansereau}}, \bibinfo {author} {\bibfnamefont {V.}~\bibnamefont
  {D{\'e}mery}}, \bibinfo {author} {\bibfnamefont {E.}~\bibnamefont
  {Berthier}}, \bibinfo {author} {\bibfnamefont {J.}~\bibnamefont {Weiss}},\
  and\ \bibinfo {author} {\bibfnamefont {L.}~\bibnamefont {Ponson}},\
  }\bibfield  {title} {\bibinfo {title} {Collective damage growth controls
  fault orientation in quasibrittle compressive failure},\ }\href@noop {}
  {\bibfield  {journal} {\bibinfo  {journal} {Phys. Rev. Lett.}\ }\textbf
  {\bibinfo {volume} {122}},\ \bibinfo {pages} {085501} (\bibinfo {year}
  {2019})}\BibitemShut {NoStop}%
\bibitem [{\citenamefont {Hentschel}\ and\ \citenamefont
  {Procaccia}(1983)}]{hentschel1983}%
  \BibitemOpen
  \bibfield  {author} {\bibinfo {author} {\bibfnamefont {H.}~\bibnamefont
  {Hentschel}}\ and\ \bibinfo {author} {\bibfnamefont {I.}~\bibnamefont
  {Procaccia}},\ }\bibfield  {title} {\bibinfo {title} {The infinite number of
  generalized dimensions of fractals and strange attractors},\ }\href@noop {}
  {\bibfield  {journal} {\bibinfo  {journal} {Physica D: Nonlinear Phenomena}\
  }\textbf {\bibinfo {volume} {8}},\ \bibinfo {pages} {435} (\bibinfo {year}
  {1983})}\BibitemShut {NoStop}%
\bibitem [{\citenamefont {Amitrano}(2012)}]{amitrano2012}%
  \BibitemOpen
  \bibfield  {author} {\bibinfo {author} {\bibfnamefont {D.}~\bibnamefont
  {Amitrano}},\ }\bibfield  {title} {\bibinfo {title} {Variability in the
  power-law distributions of rupture events},\ }\href@noop {} {\bibfield
  {journal} {\bibinfo  {journal} {Eur. Phys. J.: Spec. Top.}\ }\textbf
  {\bibinfo {volume} {205}},\ \bibinfo {pages} {199} (\bibinfo {year}
  {2012})}\BibitemShut {NoStop}%
\bibitem [{\citenamefont {Lin}\ \emph {et~al.}(2014)\citenamefont {Lin},
  \citenamefont {Lerner}, \citenamefont {Rosso},\ and\ \citenamefont
  {Wyart}}]{lin2014}%
  \BibitemOpen
  \bibfield  {author} {\bibinfo {author} {\bibfnamefont {J.}~\bibnamefont
  {Lin}}, \bibinfo {author} {\bibfnamefont {E.}~\bibnamefont {Lerner}},
  \bibinfo {author} {\bibfnamefont {A.}~\bibnamefont {Rosso}},\ and\ \bibinfo
  {author} {\bibfnamefont {M.}~\bibnamefont {Wyart}},\ }\bibfield  {title}
  {\bibinfo {title} {Scaling description of the yielding transition in soft
  amorphous solids at zero temperature},\ }\href@noop {} {\bibfield  {journal}
  {\bibinfo  {journal} {Proceedings of the National Academy of Sciences}\
  }\textbf {\bibinfo {volume} {111}},\ \bibinfo {pages} {14382} (\bibinfo
  {year} {2014})}\BibitemShut {NoStop}%
\bibitem [{\citenamefont {Lin}\ \emph {et~al.}(2015)\citenamefont {Lin},
  \citenamefont {Gueudr{\'e}}, \citenamefont {Rosso},\ and\ \citenamefont
  {Wyart}}]{lin2015}%
  \BibitemOpen
  \bibfield  {author} {\bibinfo {author} {\bibfnamefont {J.}~\bibnamefont
  {Lin}}, \bibinfo {author} {\bibfnamefont {T.}~\bibnamefont {Gueudr{\'e}}},
  \bibinfo {author} {\bibfnamefont {A.}~\bibnamefont {Rosso}},\ and\ \bibinfo
  {author} {\bibfnamefont {M.}~\bibnamefont {Wyart}},\ }\bibfield  {title}
  {\bibinfo {title} {Criticality in the approach to failure in amorphous
  solids},\ }\href@noop {} {\bibfield  {journal} {\bibinfo  {journal} {Phys.
  Rev. Lett.}\ }\textbf {\bibinfo {volume} {115}},\ \bibinfo {pages} {168001}
  (\bibinfo {year} {2015})}\BibitemShut {NoStop}%
\bibitem [{\citenamefont {Lawn}(1993)}]{lawn}%
  \BibitemOpen
  \bibfield  {author} {\bibinfo {author} {\bibfnamefont {B.}~\bibnamefont
  {Lawn}},\ }\href@noop {} {\emph {\bibinfo {title} {Fracture of brittle
  solids}}}\ (\bibinfo  {publisher} {Cambridge University Press},\ \bibinfo
  {year} {1993})\BibitemShut {NoStop}%
\bibitem [{\citenamefont {Berthier}(2015)}]{berthier_thesis}%
  \BibitemOpen
  \bibfield  {author} {\bibinfo {author} {\bibfnamefont {E.}~\bibnamefont
  {Berthier}},\ }\emph {\bibinfo {title} {Quasi-brittle failure of
  heterogeneous materials : Damage statistics and localization}},\ \href@noop
  {} {Ph.D. thesis},\ \bibinfo  {school} {Universit\'e Pierre et Marie Curie}
  (\bibinfo {year} {2015})\BibitemShut {NoStop}%
\bibitem [{\citenamefont {Lin}\ and\ \citenamefont {Wyart}(2016)}]{lin2016}%
  \BibitemOpen
  \bibfield  {author} {\bibinfo {author} {\bibfnamefont {J.}~\bibnamefont
  {Lin}}\ and\ \bibinfo {author} {\bibfnamefont {M.}~\bibnamefont {Wyart}},\
  }\bibfield  {title} {\bibinfo {title} {Mean-field description of plastic flow
  in amorphous solids},\ }\href@noop {} {\bibfield  {journal} {\bibinfo
  {journal} {Phys. Rev. X}\ }\textbf {\bibinfo {volume} {6}},\ \bibinfo {pages}
  {011005} (\bibinfo {year} {2016})}\BibitemShut {NoStop}%
\bibitem [{\citenamefont {Liu}\ \emph {et~al.}(2016)\citenamefont {Liu},
  \citenamefont {Ferrero}, \citenamefont {Puosi}, \citenamefont {Barrat},\ and\
  \citenamefont {Martens}}]{liu2016}%
  \BibitemOpen
  \bibfield  {author} {\bibinfo {author} {\bibfnamefont {C.}~\bibnamefont
  {Liu}}, \bibinfo {author} {\bibfnamefont {E.~E.}\ \bibnamefont {Ferrero}},
  \bibinfo {author} {\bibfnamefont {F.}~\bibnamefont {Puosi}}, \bibinfo
  {author} {\bibfnamefont {J.-L.}\ \bibnamefont {Barrat}},\ and\ \bibinfo
  {author} {\bibfnamefont {K.}~\bibnamefont {Martens}},\ }\bibfield  {title}
  {\bibinfo {title} {Driving rate dependence of avalanche statistics and shapes
  at the yielding transition},\ }\href@noop {} {\bibfield  {journal} {\bibinfo
  {journal} {Phys. Rev. Lett.}\ }\textbf {\bibinfo {volume} {116}},\ \bibinfo
  {pages} {065501} (\bibinfo {year} {2016})}\BibitemShut {NoStop}%
\bibitem [{\citenamefont {Budrikis}\ \emph {et~al.}(2017)\citenamefont
  {Budrikis}, \citenamefont {Castellanos}, \citenamefont {Sandfeld},
  \citenamefont {Zaiser},\ and\ \citenamefont {Zapperi}}]{budrikis2017}%
  \BibitemOpen
  \bibfield  {author} {\bibinfo {author} {\bibfnamefont {Z.}~\bibnamefont
  {Budrikis}}, \bibinfo {author} {\bibfnamefont {D.~F.}\ \bibnamefont
  {Castellanos}}, \bibinfo {author} {\bibfnamefont {S.}~\bibnamefont
  {Sandfeld}}, \bibinfo {author} {\bibfnamefont {M.}~\bibnamefont {Zaiser}},\
  and\ \bibinfo {author} {\bibfnamefont {S.}~\bibnamefont {Zapperi}},\
  }\bibfield  {title} {\bibinfo {title} {Universal features of amorphous
  plasticity},\ }\href@noop {} {\bibfield  {journal} {\bibinfo  {journal}
  {Nature communications}\ }\textbf {\bibinfo {volume} {8}},\ \bibinfo {pages}
  {1} (\bibinfo {year} {2017})}\BibitemShut {NoStop}%
\bibitem [{\citenamefont {Ozawa}\ \emph {et~al.}(2018)\citenamefont {Ozawa},
  \citenamefont {Berthier}, \citenamefont {Biroli}, \citenamefont {Rosso},\
  and\ \citenamefont {Tarjus}}]{ozawa2018}%
  \BibitemOpen
  \bibfield  {author} {\bibinfo {author} {\bibfnamefont {M.}~\bibnamefont
  {Ozawa}}, \bibinfo {author} {\bibfnamefont {L.}~\bibnamefont {Berthier}},
  \bibinfo {author} {\bibfnamefont {G.}~\bibnamefont {Biroli}}, \bibinfo
  {author} {\bibfnamefont {A.}~\bibnamefont {Rosso}},\ and\ \bibinfo {author}
  {\bibfnamefont {G.}~\bibnamefont {Tarjus}},\ }\bibfield  {title} {\bibinfo
  {title} {Random critical point separates brittle and ductile yielding
  transitions in amorphous materials},\ }\href@noop {} {\bibfield  {journal}
  {\bibinfo  {journal} {Proceedings of the National Academy of Sciences}\
  }\textbf {\bibinfo {volume} {115}},\ \bibinfo {pages} {6656} (\bibinfo {year}
  {2018})}\BibitemShut {NoStop}%
\bibitem [{\citenamefont {Anifrani}\ \emph {et~al.}(1995)\citenamefont
  {Anifrani}, \citenamefont {Le~Floc'h}, \citenamefont {Sornette},\ and\
  \citenamefont {Souillard}}]{anifrani1995}%
  \BibitemOpen
  \bibfield  {author} {\bibinfo {author} {\bibfnamefont {J.-C.}\ \bibnamefont
  {Anifrani}}, \bibinfo {author} {\bibfnamefont {C.}~\bibnamefont {Le~Floc'h}},
  \bibinfo {author} {\bibfnamefont {D.}~\bibnamefont {Sornette}},\ and\
  \bibinfo {author} {\bibfnamefont {B.}~\bibnamefont {Souillard}},\ }\bibfield
  {title} {\bibinfo {title} {Universal log-periodic correction to
  renormalization group scaling for rupture stress prediction from acoustic
  emissions},\ }\href@noop {} {\bibfield  {journal} {\bibinfo  {journal}
  {Journal de Physique I}\ }\textbf {\bibinfo {volume} {5}},\ \bibinfo {pages}
  {631} (\bibinfo {year} {1995})}\BibitemShut {NoStop}%
\bibitem [{\citenamefont {Mayya}\ \emph {et~al.}(2020)\citenamefont {Mayya},
  \citenamefont {Berthier},\ and\ \citenamefont {Ponson}}]{Mayya}%
  \BibitemOpen
  \bibfield  {author} {\bibinfo {author} {\bibfnamefont {A.}~\bibnamefont
  {Mayya}}, \bibinfo {author} {\bibfnamefont {E.}~\bibnamefont {Berthier}},\
  and\ \bibinfo {author} {\bibfnamefont {L.}~\bibnamefont {Ponson}},\
  }\href@noop {} {\bibinfo {title} {Proc\'ed\'e et dispositif d’analyse
  d’une structure. french patent application fr2002824}} (\bibinfo {year}
  {2020})\BibitemShut {NoStop}%
\bibitem [{\citenamefont {Nataf}\ \emph {et~al.}(2014)\citenamefont {Nataf},
  \citenamefont {Castillo-Villa}, \citenamefont {Bar{\'o}}, \citenamefont
  {Illa}, \citenamefont {Vives}, \citenamefont {Planes},\ and\ \citenamefont
  {Salje}}]{nataf2014b}%
  \BibitemOpen
  \bibfield  {author} {\bibinfo {author} {\bibfnamefont {G.~F.}\ \bibnamefont
  {Nataf}}, \bibinfo {author} {\bibfnamefont {P.~O.}\ \bibnamefont
  {Castillo-Villa}}, \bibinfo {author} {\bibfnamefont {J.}~\bibnamefont
  {Bar{\'o}}}, \bibinfo {author} {\bibfnamefont {X.}~\bibnamefont {Illa}},
  \bibinfo {author} {\bibfnamefont {E.}~\bibnamefont {Vives}}, \bibinfo
  {author} {\bibfnamefont {A.}~\bibnamefont {Planes}},\ and\ \bibinfo {author}
  {\bibfnamefont {E.~K.}\ \bibnamefont {Salje}},\ }\bibfield  {title} {\bibinfo
  {title} {Avalanches in compressed porous si o 2-based materials},\
  }\href@noop {} {\bibfield  {journal} {\bibinfo  {journal} {Phys. Rev. E}\
  }\textbf {\bibinfo {volume} {90}},\ \bibinfo {pages} {022405} (\bibinfo
  {year} {2014})}\BibitemShut {NoStop}%
\bibitem [{\citenamefont {Davidsen}\ \emph {et~al.}(2021)\citenamefont
  {Davidsen}, \citenamefont {Goebel}, \citenamefont {Kwiatek}, \citenamefont
  {Stanchits}, \citenamefont {Bar{\'o}},\ and\ \citenamefont
  {Dresen}}]{davidsen2021}%
  \BibitemOpen
  \bibfield  {author} {\bibinfo {author} {\bibfnamefont {J.}~\bibnamefont
  {Davidsen}}, \bibinfo {author} {\bibfnamefont {T.}~\bibnamefont {Goebel}},
  \bibinfo {author} {\bibfnamefont {G.}~\bibnamefont {Kwiatek}}, \bibinfo
  {author} {\bibfnamefont {S.}~\bibnamefont {Stanchits}}, \bibinfo {author}
  {\bibfnamefont {J.}~\bibnamefont {Bar{\'o}}},\ and\ \bibinfo {author}
  {\bibfnamefont {G.}~\bibnamefont {Dresen}},\ }\bibfield  {title} {\bibinfo
  {title} {What controls the presence and characteristics of aftershocks in
  rock fracture in the lab?},\ }\href@noop {} {\bibfield  {journal} {\bibinfo
  {journal} {Journal of Geophysical Research: Solid Earth}\ }\textbf {\bibinfo
  {volume} {126}},\ \bibinfo {pages} {e2021JB022539} (\bibinfo {year}
  {2021})}\BibitemShut {NoStop}%
\bibitem [{\citenamefont {Xu}\ \emph {et~al.}(2019)\citenamefont {Xu},
  \citenamefont {Borrego}, \citenamefont {Planes}, \citenamefont {Ding},\ and\
  \citenamefont {Vives}}]{xu2019}%
  \BibitemOpen
  \bibfield  {author} {\bibinfo {author} {\bibfnamefont {Y.}~\bibnamefont
  {Xu}}, \bibinfo {author} {\bibfnamefont {A.~G.}\ \bibnamefont {Borrego}},
  \bibinfo {author} {\bibfnamefont {A.}~\bibnamefont {Planes}}, \bibinfo
  {author} {\bibfnamefont {X.}~\bibnamefont {Ding}},\ and\ \bibinfo {author}
  {\bibfnamefont {E.}~\bibnamefont {Vives}},\ }\bibfield  {title} {\bibinfo
  {title} {Criticality in failure under compression: Acoustic emission study of
  coal and charcoal with different microstructures},\ }\href@noop {} {\bibfield
   {journal} {\bibinfo  {journal} {Phys. Rev. E}\ }\textbf {\bibinfo {volume}
  {99}},\ \bibinfo {pages} {033001} (\bibinfo {year} {2019})}\BibitemShut
  {NoStop}%
\bibitem [{\citenamefont {Zreihan}\ \emph {et~al.}(2019)\citenamefont
  {Zreihan}, \citenamefont {Faran}, \citenamefont {Vives}, \citenamefont
  {Planes},\ and\ \citenamefont {Shilo}}]{zreihan2019}%
  \BibitemOpen
  \bibfield  {author} {\bibinfo {author} {\bibfnamefont {N.}~\bibnamefont
  {Zreihan}}, \bibinfo {author} {\bibfnamefont {E.}~\bibnamefont {Faran}},
  \bibinfo {author} {\bibfnamefont {E.}~\bibnamefont {Vives}}, \bibinfo
  {author} {\bibfnamefont {A.}~\bibnamefont {Planes}},\ and\ \bibinfo {author}
  {\bibfnamefont {D.}~\bibnamefont {Shilo}},\ }\bibfield  {title} {\bibinfo
  {title} {Relations between stress drops and acoustic emission measured during
  mechanical loading},\ }\href@noop {} {\bibfield  {journal} {\bibinfo
  {journal} {Phys. Rev. Mater.}\ }\textbf {\bibinfo {volume} {3}},\ \bibinfo
  {pages} {043603} (\bibinfo {year} {2019})}\BibitemShut {NoStop}%
\bibitem [{\citenamefont {Parisi}\ \emph {et~al.}(2017)\citenamefont {Parisi},
  \citenamefont {Procaccia}, \citenamefont {Rainone},\ and\ \citenamefont
  {Singh}}]{parisi2017}%
  \BibitemOpen
  \bibfield  {author} {\bibinfo {author} {\bibfnamefont {G.}~\bibnamefont
  {Parisi}}, \bibinfo {author} {\bibfnamefont {I.}~\bibnamefont {Procaccia}},
  \bibinfo {author} {\bibfnamefont {C.}~\bibnamefont {Rainone}},\ and\ \bibinfo
  {author} {\bibfnamefont {M.}~\bibnamefont {Singh}},\ }\bibfield  {title}
  {\bibinfo {title} {Shear bands as manifestation of a criticality in yielding
  amorphous solids},\ }\href@noop {} {\bibfield  {journal} {\bibinfo  {journal}
  {Proceedings of the National Academy of Sciences}\ }\textbf {\bibinfo
  {volume} {114}},\ \bibinfo {pages} {5577} (\bibinfo {year}
  {2017})}\BibitemShut {NoStop}%
\bibitem [{\citenamefont {Schindelin}\ \emph {et~al.}(2012)\citenamefont
  {Schindelin}, \citenamefont {Arganda-Carreras}, \citenamefont {Frise},
  \citenamefont {Kaynig}, \citenamefont {Longair}, \citenamefont {Pietzsch},
  \citenamefont {Preibisch}, \citenamefont {Rueden}, \citenamefont {Saalfeld},
  \citenamefont {Schmid} \emph {et~al.}}]{fiji}%
  \BibitemOpen
  \bibfield  {author} {\bibinfo {author} {\bibfnamefont {J.}~\bibnamefont
  {Schindelin}}, \bibinfo {author} {\bibfnamefont {I.}~\bibnamefont
  {Arganda-Carreras}}, \bibinfo {author} {\bibfnamefont {E.}~\bibnamefont
  {Frise}}, \bibinfo {author} {\bibfnamefont {V.}~\bibnamefont {Kaynig}},
  \bibinfo {author} {\bibfnamefont {M.}~\bibnamefont {Longair}}, \bibinfo
  {author} {\bibfnamefont {T.}~\bibnamefont {Pietzsch}}, \bibinfo {author}
  {\bibfnamefont {S.}~\bibnamefont {Preibisch}}, \bibinfo {author}
  {\bibfnamefont {C.}~\bibnamefont {Rueden}}, \bibinfo {author} {\bibfnamefont
  {S.}~\bibnamefont {Saalfeld}}, \bibinfo {author} {\bibfnamefont
  {B.}~\bibnamefont {Schmid}}, \emph {et~al.},\ }\bibfield  {title} {\bibinfo
  {title} {Fiji: an open-source platform for biological-image analysis},\
  }\href@noop {} {\bibfield  {journal} {\bibinfo  {journal} {Nature methods}\
  }\textbf {\bibinfo {volume} {9}},\ \bibinfo {pages} {676} (\bibinfo {year}
  {2012})}\BibitemShut {NoStop}%
\bibitem [{\citenamefont {Lebyodkin}\ \emph {et~al.}(2013)\citenamefont
  {Lebyodkin}, \citenamefont {Shashkov}, \citenamefont {Lebedkina},
  \citenamefont {Mathis}, \citenamefont {Dobron},\ and\ \citenamefont
  {Chmelik}}]{lebyodkin2013}%
  \BibitemOpen
  \bibfield  {author} {\bibinfo {author} {\bibfnamefont {M.}~\bibnamefont
  {Lebyodkin}}, \bibinfo {author} {\bibfnamefont {I.}~\bibnamefont {Shashkov}},
  \bibinfo {author} {\bibfnamefont {T.}~\bibnamefont {Lebedkina}}, \bibinfo
  {author} {\bibfnamefont {K.}~\bibnamefont {Mathis}}, \bibinfo {author}
  {\bibfnamefont {P.}~\bibnamefont {Dobron}},\ and\ \bibinfo {author}
  {\bibfnamefont {F.}~\bibnamefont {Chmelik}},\ }\bibfield  {title} {\bibinfo
  {title} {Role of superposition of dislocation avalanches in the statistics of
  acoustic emission during plastic deformation},\ }\href@noop {} {\bibfield
  {journal} {\bibinfo  {journal} {Physical Review E}\ }\textbf {\bibinfo
  {volume} {88}},\ \bibinfo {pages} {042402} (\bibinfo {year}
  {2013})}\BibitemShut {NoStop}%
\bibitem [{\citenamefont {Houdoux}\ \emph {et~al.}(2018)\citenamefont
  {Houdoux}, \citenamefont {Nguyen}, \citenamefont {Amon},\ and\ \citenamefont
  {Crassous}}]{houdoux2018}%
  \BibitemOpen
  \bibfield  {author} {\bibinfo {author} {\bibfnamefont {D.}~\bibnamefont
  {Houdoux}}, \bibinfo {author} {\bibfnamefont {T.~B.}\ \bibnamefont {Nguyen}},
  \bibinfo {author} {\bibfnamefont {A.}~\bibnamefont {Amon}},\ and\ \bibinfo
  {author} {\bibfnamefont {J.}~\bibnamefont {Crassous}},\ }\bibfield  {title}
  {\bibinfo {title} {Plastic flow and localization in an amorphous material:
  experimental interpretation of the fluidity},\ }\href@noop {} {\bibfield
  {journal} {\bibinfo  {journal} {Physical Review E}\ }\textbf {\bibinfo
  {volume} {98}},\ \bibinfo {pages} {022905} (\bibinfo {year}
  {2018})}\BibitemShut {NoStop}%
\bibitem [{\citenamefont {Houdoux}\ \emph {et~al.}(2021)\citenamefont
  {Houdoux}, \citenamefont {Amon}, \citenamefont {Marsan}, \citenamefont
  {Weiss},\ and\ \citenamefont {Crassous}}]{houdoux2021}%
  \BibitemOpen
  \bibfield  {author} {\bibinfo {author} {\bibfnamefont {D.}~\bibnamefont
  {Houdoux}}, \bibinfo {author} {\bibfnamefont {A.}~\bibnamefont {Amon}},
  \bibinfo {author} {\bibfnamefont {D.}~\bibnamefont {Marsan}}, \bibinfo
  {author} {\bibfnamefont {J.}~\bibnamefont {Weiss}},\ and\ \bibinfo {author}
  {\bibfnamefont {J.}~\bibnamefont {Crassous}},\ }\bibfield  {title} {\bibinfo
  {title} {Micro-slips in an experimental granular shear band replicate the
  spatiotemporal characteristics of natural earthquakes},\ }\href@noop {}
  {\bibfield  {journal} {\bibinfo  {journal} {Communications Earth \&
  Environment}\ }\textbf {\bibinfo {volume} {2}},\ \bibinfo {pages} {90}
  (\bibinfo {year} {2021})}\BibitemShut {NoStop}%
\bibitem [{\citenamefont {Glasser}\ and\ \citenamefont
  {Goldhirsch}(2001)}]{glasser2001}%
  \BibitemOpen
  \bibfield  {author} {\bibinfo {author} {\bibfnamefont {B.}~\bibnamefont
  {Glasser}}\ and\ \bibinfo {author} {\bibfnamefont {I.}~\bibnamefont
  {Goldhirsch}},\ }\bibfield  {title} {\bibinfo {title} {Scale dependence,
  correlations, and fluctuations of stresses in rapid granular flows},\
  }\href@noop {} {\bibfield  {journal} {\bibinfo  {journal} {Physics of
  Fluids}\ }\textbf {\bibinfo {volume} {13}},\ \bibinfo {pages} {407} (\bibinfo
  {year} {2001})}\BibitemShut {NoStop}%
\end{thebibliography}

\begin{thebibliography}{17}%
\makeatletter
\providecommand \@ifxundefined [1]{%
 \@ifx{#1\undefined}
}%
\providecommand \@ifnum [1]{%
 \ifnum #1\expandafter \@firstoftwo
 \else \expandafter \@secondoftwo
 \fi
}%
\providecommand \@ifx [1]{%
 \ifx #1\expandafter \@firstoftwo
 \else \expandafter \@secondoftwo
 \fi
}%
\providecommand \natexlab [1]{#1}%
\providecommand \enquote  [1]{``#1''}%
\providecommand \bibnamefont  [1]{#1}%
\providecommand \bibfnamefont [1]{#1}%
\providecommand \citenamefont [1]{#1}%
\providecommand \href@noop [0]{\@secondoftwo}%
\providecommand \href [0]{\begingroup \@sanitize@url \@href}%
\providecommand \@href[1]{\@@startlink{#1}\@@href}%
\providecommand \@@href[1]{\endgroup#1\@@endlink}%
\providecommand \@sanitize@url [0]{\catcode `\\12\catcode `\$12\catcode
  `\&12\catcode `\#12\catcode `\^12\catcode `\_12\catcode `\%12\relax}%
\providecommand \@@startlink[1]{}%
\providecommand \@@endlink[0]{}%
\providecommand \url  [0]{\begingroup\@sanitize@url \@url }%
\providecommand \@url [1]{\endgroup\@href {#1}{\urlprefix }}%
\providecommand \urlprefix  [0]{URL }%
\providecommand \Eprint [0]{\href }%
\providecommand \doibase [0]{https://doi.org/}%
\providecommand \selectlanguage [0]{\@gobble}%
\providecommand \bibinfo  [0]{\@secondoftwo}%
\providecommand \bibfield  [0]{\@secondoftwo}%
\providecommand \translation [1]{[#1]}%
\providecommand \BibitemOpen [0]{}%
\providecommand \bibitemStop [0]{}%
\providecommand \bibitemNoStop [0]{.\EOS\space}%
\providecommand \EOS [0]{\spacefactor3000\relax}%
\providecommand \BibitemShut  [1]{\csname bibitem#1\endcsname}%
\let\auto@bib@innerbib\@empty
\bibitem [{\citenamefont {Schindelin}\ \emph {et~al.}(2012)\citenamefont
  {Schindelin}, \citenamefont {Arganda-Carreras}, \citenamefont {Frise},
  \citenamefont {Kaynig}, \citenamefont {Longair}, \citenamefont {Pietzsch},
  \citenamefont {Preibisch}, \citenamefont {Rueden}, \citenamefont {Saalfeld},
  \citenamefont {Schmid} \emph {et~al.}}]{fiji}%
  \BibitemOpen
  \bibfield  {author} {\bibinfo {author} {\bibfnamefont {J.}~\bibnamefont
  {Schindelin}}, \bibinfo {author} {\bibfnamefont {I.}~\bibnamefont
  {Arganda-Carreras}}, \bibinfo {author} {\bibfnamefont {E.}~\bibnamefont
  {Frise}}, \bibinfo {author} {\bibfnamefont {V.}~\bibnamefont {Kaynig}},
  \bibinfo {author} {\bibfnamefont {M.}~\bibnamefont {Longair}}, \bibinfo
  {author} {\bibfnamefont {T.}~\bibnamefont {Pietzsch}}, \bibinfo {author}
  {\bibfnamefont {S.}~\bibnamefont {Preibisch}}, \bibinfo {author}
  {\bibfnamefont {C.}~\bibnamefont {Rueden}}, \bibinfo {author} {\bibfnamefont
  {S.}~\bibnamefont {Saalfeld}}, \bibinfo {author} {\bibfnamefont
  {B.}~\bibnamefont {Schmid}}, \emph {et~al.},\ }\bibfield  {title} {\bibinfo
  {title} {Fiji: an open-source platform for biological-image analysis},\
  }\href@noop {} {\bibfield  {journal} {\bibinfo  {journal} {Nature methods}\
  }\textbf {\bibinfo {volume} {9}},\ \bibinfo {pages} {676} (\bibinfo {year}
  {2012})}\BibitemShut {NoStop}%
\bibitem [{\citenamefont {Glasser}\ and\ \citenamefont
  {Goldhirsch}(2001)}]{glasser2001scale}%
  \BibitemOpen
  \bibfield  {author} {\bibinfo {author} {\bibfnamefont {B.}~\bibnamefont
  {Glasser}}\ and\ \bibinfo {author} {\bibfnamefont {I.}~\bibnamefont
  {Goldhirsch}},\ }\bibfield  {title} {\bibinfo {title} {Scale dependence,
  correlations, and fluctuations of stresses in rapid granular flows},\
  }\href@noop {} {\bibfield  {journal} {\bibinfo  {journal} {Physics of
  Fluids}\ }\textbf {\bibinfo {volume} {13}},\ \bibinfo {pages} {407} (\bibinfo
  {year} {2001})}\BibitemShut {NoStop}%
\bibitem [{\citenamefont {Berthier}\ \emph {et~al.}(2017)\citenamefont
  {Berthier}, \citenamefont {D{\'e}mery},\ and\ \citenamefont
  {Ponson}}]{SI_berthier2017}%
  \BibitemOpen
  \bibfield  {author} {\bibinfo {author} {\bibfnamefont {E.}~\bibnamefont
  {Berthier}}, \bibinfo {author} {\bibfnamefont {V.}~\bibnamefont
  {D{\'e}mery}},\ and\ \bibinfo {author} {\bibfnamefont {L.}~\bibnamefont
  {Ponson}},\ }\bibfield  {title} {\bibinfo {title} {Damage spreading in
  quasi-brittle disordered solids: I. localization and failure},\ }\href@noop
  {} {\bibfield  {journal} {\bibinfo  {journal} {Journal of the Mechanics and
  Physics of Solids}\ }\textbf {\bibinfo {volume} {102}},\ \bibinfo {pages}
  {101} (\bibinfo {year} {2017})}\BibitemShut {NoStop}%
\bibitem [{\citenamefont {Dansereau}\ \emph {et~al.}(2019)\citenamefont
  {Dansereau}, \citenamefont {D{\'e}mery}, \citenamefont {Berthier},
  \citenamefont {Weiss},\ and\ \citenamefont {Ponson}}]{SI_dansereau2019}%
  \BibitemOpen
  \bibfield  {author} {\bibinfo {author} {\bibfnamefont {V.}~\bibnamefont
  {Dansereau}}, \bibinfo {author} {\bibfnamefont {V.}~\bibnamefont
  {D{\'e}mery}}, \bibinfo {author} {\bibfnamefont {E.}~\bibnamefont
  {Berthier}}, \bibinfo {author} {\bibfnamefont {J.}~\bibnamefont {Weiss}},\
  and\ \bibinfo {author} {\bibfnamefont {L.}~\bibnamefont {Ponson}},\
  }\bibfield  {title} {\bibinfo {title} {Collective damage growth controls
  fault orientation in quasibrittle compressive failure},\ }\href@noop {}
  {\bibfield  {journal} {\bibinfo  {journal} {Physical review letters}\
  }\textbf {\bibinfo {volume} {122}},\ \bibinfo {pages} {085501} (\bibinfo
  {year} {2019})}\BibitemShut {NoStop}%
\bibitem [{\citenamefont {Lemaitre}(1992)}]{Lemaitre}%
  \BibitemOpen
  \bibfield  {author} {\bibinfo {author} {\bibfnamefont {J.}~\bibnamefont
  {Lemaitre}},\ }\href@noop {} {\emph {\bibinfo {title} {A course on damage
  mechanics}}},\ Amsterdam\ (\bibinfo  {publisher} {Springer Verlag},\ \bibinfo
  {year} {1992})\BibitemShut {NoStop}%
\bibitem [{\citenamefont {Girard}\ \emph {et~al.}(2010)\citenamefont {Girard},
  \citenamefont {Amitrano},\ and\ \citenamefont {Weiss}}]{SI_girard2010}%
  \BibitemOpen
  \bibfield  {author} {\bibinfo {author} {\bibfnamefont {L.}~\bibnamefont
  {Girard}}, \bibinfo {author} {\bibfnamefont {D.}~\bibnamefont {Amitrano}},\
  and\ \bibinfo {author} {\bibfnamefont {J.}~\bibnamefont {Weiss}},\ }\bibfield
   {title} {\bibinfo {title} {Failure as a critical phenomenon in a progressive
  damage model},\ }\href@noop {} {\bibfield  {journal} {\bibinfo  {journal}
  {Journal of Statistical Mechanics: Theory and Experiment}\ }\textbf {\bibinfo
  {volume} {2010}},\ \bibinfo {pages} {P01013} (\bibinfo {year}
  {2010})}\BibitemShut {NoStop}%
\bibitem [{\citenamefont {Thilakarathna}\ \emph {et~al.}(2020)\citenamefont
  {Thilakarathna}, \citenamefont {Baduge}, \citenamefont {Mendis},
  \citenamefont {Vimonsatit},\ and\ \citenamefont {Lee}}]{SI_thilakarathna2020}%
  \BibitemOpen
  \bibfield  {author} {\bibinfo {author} {\bibfnamefont {P.}~\bibnamefont
  {Thilakarathna}}, \bibinfo {author} {\bibfnamefont {K.~K.}\ \bibnamefont
  {Baduge}}, \bibinfo {author} {\bibfnamefont {P.}~\bibnamefont {Mendis}},
  \bibinfo {author} {\bibfnamefont {V.}~\bibnamefont {Vimonsatit}},\ and\
  \bibinfo {author} {\bibfnamefont {H.}~\bibnamefont {Lee}},\ }\bibfield
  {title} {\bibinfo {title} {Mesoscale modelling of concrete--a review of
  geometry generation, placing algorithms, constitutive relations and
  applications},\ }\href@noop {} {\bibfield  {journal} {\bibinfo  {journal}
  {Engineering Fracture Mechanics}\ }\textbf {\bibinfo {volume} {231}},\
  \bibinfo {pages} {106974} (\bibinfo {year} {2020})}\BibitemShut {NoStop}%
\bibitem [{\citenamefont {Kachanov}(1993)}]{SI_kachanov1993}%
  \BibitemOpen
  \bibfield  {author} {\bibinfo {author} {\bibfnamefont {M.}~\bibnamefont
  {Kachanov}},\ }\bibfield  {title} {\bibinfo {title} {Elastic solids with many
  cracks and related problems},\ }\href@noop {} {\bibfield  {journal} {\bibinfo
   {journal} {Advances in applied mechanics}\ }\textbf {\bibinfo {volume}
  {30}},\ \bibinfo {pages} {259} (\bibinfo {year} {1993})}\BibitemShut
  {NoStop}%
\bibitem [{\citenamefont {Berthier}\ \emph {et~al.}(2022)\citenamefont
  {Berthier}, \citenamefont {Mayya},\ and\ \citenamefont
  {Ponson}}]{SI_berthier2021}%
  \BibitemOpen
  \bibfield  {author} {\bibinfo {author} {\bibfnamefont {E.}~\bibnamefont
  {Berthier}}, \bibinfo {author} {\bibfnamefont {A.}~\bibnamefont {Mayya}},\
  and\ \bibinfo {author} {\bibfnamefont {L.}~\bibnamefont {Ponson}},\
  }\bibfield  {title} {\bibinfo {title} {Damage spreading in quasi-brittle
  disordered solids: Ii. what the statistics of precursors teach us about
  compressive failure},\ }\href@noop {} {\bibfield  {journal} {\bibinfo
  {journal} {Journal of the Mechanics and Physics of Solids}\ }\textbf
  {\bibinfo {volume} {162}},\ \bibinfo {pages} {104826} (\bibinfo {year}
  {2022})}\BibitemShut {NoStop}%
\bibitem [{\citenamefont {Amitrano}(2012)}]{SI_amitrano2012}%
  \BibitemOpen
  \bibfield  {author} {\bibinfo {author} {\bibfnamefont {D.}~\bibnamefont
  {Amitrano}},\ }\bibfield  {title} {\bibinfo {title} {Variability in the
  power-law distributions of rupture events},\ }\href@noop {} {\bibfield
  {journal} {\bibinfo  {journal} {The European Physical Journal Special
  Topics}\ }\textbf {\bibinfo {volume} {205}},\ \bibinfo {pages} {199}
  (\bibinfo {year} {2012})}\BibitemShut {NoStop}%
\bibitem [{\citenamefont {Lin}\ \emph {et~al.}(2014)\citenamefont {Lin},
  \citenamefont {Lerner}, \citenamefont {Rosso},\ and\ \citenamefont
  {Wyart}}]{SI_lin2014}%
  \BibitemOpen
  \bibfield  {author} {\bibinfo {author} {\bibfnamefont {J.}~\bibnamefont
  {Lin}}, \bibinfo {author} {\bibfnamefont {E.}~\bibnamefont {Lerner}},
  \bibinfo {author} {\bibfnamefont {A.}~\bibnamefont {Rosso}},\ and\ \bibinfo
  {author} {\bibfnamefont {M.}~\bibnamefont {Wyart}},\ }\bibfield  {title}
  {\bibinfo {title} {Scaling description of the yielding transition in soft
  amorphous solids at zero temperature},\ }\href@noop {} {\bibfield  {journal}
  {\bibinfo  {journal} {Proceedings of the National Academy of Sciences}\
  }\textbf {\bibinfo {volume} {111}},\ \bibinfo {pages} {14382} (\bibinfo
  {year} {2014})}\BibitemShut {NoStop}%
\bibitem [{\citenamefont {Lin}\ \emph {et~al.}(2015)\citenamefont {Lin},
  \citenamefont {Gueudr{\'e}}, \citenamefont {Rosso},\ and\ \citenamefont
  {Wyart}}]{SI_lin2015}%
  \BibitemOpen
  \bibfield  {author} {\bibinfo {author} {\bibfnamefont {J.}~\bibnamefont
  {Lin}}, \bibinfo {author} {\bibfnamefont {T.}~\bibnamefont {Gueudr{\'e}}},
  \bibinfo {author} {\bibfnamefont {A.}~\bibnamefont {Rosso}},\ and\ \bibinfo
  {author} {\bibfnamefont {M.}~\bibnamefont {Wyart}},\ }\bibfield  {title}
  {\bibinfo {title} {Criticality in the approach to failure in amorphous
  solids},\ }\href@noop {} {\bibfield  {journal} {\bibinfo  {journal} {Physical
  review letters}\ }\textbf {\bibinfo {volume} {115}},\ \bibinfo {pages}
  {168001} (\bibinfo {year} {2015})}\BibitemShut {NoStop}%
\bibitem [{\citenamefont {Lin}\ and\ \citenamefont {Wyart}(2016)}]{SI_lin2016}%
  \BibitemOpen
  \bibfield  {author} {\bibinfo {author} {\bibfnamefont {J.}~\bibnamefont
  {Lin}}\ and\ \bibinfo {author} {\bibfnamefont {M.}~\bibnamefont {Wyart}},\
  }\bibfield  {title} {\bibinfo {title} {Mean-field description of plastic flow
  in amorphous solids},\ }\href@noop {} {\bibfield  {journal} {\bibinfo
  {journal} {Physical review X}\ }\textbf {\bibinfo {volume} {6}},\ \bibinfo
  {pages} {011005} (\bibinfo {year} {2016})}\BibitemShut {NoStop}%
\bibitem [{\citenamefont {Liu}\ \emph {et~al.}(2016)\citenamefont {Liu},
  \citenamefont {Ferrero}, \citenamefont {Puosi}, \citenamefont {Barrat},\ and\
  \citenamefont {Martens}}]{SI_liu2016}%
  \BibitemOpen
  \bibfield  {author} {\bibinfo {author} {\bibfnamefont {C.}~\bibnamefont
  {Liu}}, \bibinfo {author} {\bibfnamefont {E.~E.}\ \bibnamefont {Ferrero}},
  \bibinfo {author} {\bibfnamefont {F.}~\bibnamefont {Puosi}}, \bibinfo
  {author} {\bibfnamefont {J.-L.}\ \bibnamefont {Barrat}},\ and\ \bibinfo
  {author} {\bibfnamefont {K.}~\bibnamefont {Martens}},\ }\bibfield  {title}
  {\bibinfo {title} {Driving rate dependence of avalanche statistics and shapes
  at the yielding transition},\ }\href@noop {} {\bibfield  {journal} {\bibinfo
  {journal} {Physical review letters}\ }\textbf {\bibinfo {volume} {116}},\
  \bibinfo {pages} {065501} (\bibinfo {year} {2016})}\BibitemShut {NoStop}%
\bibitem [{\citenamefont {Ozawa}\ \emph {et~al.}(2018)\citenamefont {Ozawa},
  \citenamefont {Berthier}, \citenamefont {Biroli}, \citenamefont {Rosso},\
  and\ \citenamefont {Tarjus}}]{SI_ozawa2018}%
  \BibitemOpen
  \bibfield  {author} {\bibinfo {author} {\bibfnamefont {M.}~\bibnamefont
  {Ozawa}}, \bibinfo {author} {\bibfnamefont {L.}~\bibnamefont {Berthier}},
  \bibinfo {author} {\bibfnamefont {G.}~\bibnamefont {Biroli}}, \bibinfo
  {author} {\bibfnamefont {A.}~\bibnamefont {Rosso}},\ and\ \bibinfo {author}
  {\bibfnamefont {G.}~\bibnamefont {Tarjus}},\ }\bibfield  {title} {\bibinfo
  {title} {Random critical point separates brittle and ductile yielding
  transitions in amorphous materials},\ }\href@noop {} {\bibfield  {journal}
  {\bibinfo  {journal} {Proceedings of the National Academy of Sciences}\
  }\textbf {\bibinfo {volume} {115}},\ \bibinfo {pages} {6656} (\bibinfo {year}
  {2018})}\BibitemShut {NoStop}%
\bibitem [{\citenamefont {Vu}\ \emph {et~al.}(2019)\citenamefont {Vu},
  \citenamefont {Amitrano}, \citenamefont {Pl\'{e}},\ and\ \citenamefont
  {Weiss}}]{SI_weiss2019}%
  \BibitemOpen
  \bibfield  {author} {\bibinfo {author} {\bibfnamefont {C.~C.}\ \bibnamefont
  {Vu}}, \bibinfo {author} {\bibfnamefont {D.}~\bibnamefont {Amitrano}},
  \bibinfo {author} {\bibfnamefont {O.}~\bibnamefont {Pl\'{e}}},\ and\ \bibinfo
  {author} {\bibfnamefont {J.}~\bibnamefont {Weiss}},\ }\bibfield  {title}
  {\bibinfo {title} {Compressive failure as a critical transition: Experimental
  evidence and mapping onto the universality class of depinning},\ }\href@noop
  {} {\bibfield  {journal} {\bibinfo  {journal} {Phys. Rev. Lett.}\ }\textbf
  {\bibinfo {volume} {122}},\ \bibinfo {pages} {015502} (\bibinfo {year}
  {2019})}\BibitemShut {NoStop}%
\bibitem [{\citenamefont {Kandula}\ \emph {et~al.}(2019)\citenamefont
  {Kandula}, \citenamefont {Cordonnier}, \citenamefont {Boller}, \citenamefont
  {Weiss}, \citenamefont {Dysthe},\ and\ \citenamefont
  {Renard}}]{SI_kandula2019}%
  \BibitemOpen
  \bibfield  {author} {\bibinfo {author} {\bibfnamefont {N.}~\bibnamefont
  {Kandula}}, \bibinfo {author} {\bibfnamefont {B.}~\bibnamefont {Cordonnier}},
  \bibinfo {author} {\bibfnamefont {E.}~\bibnamefont {Boller}}, \bibinfo
  {author} {\bibfnamefont {J.}~\bibnamefont {Weiss}}, \bibinfo {author}
  {\bibfnamefont {D.~K.}\ \bibnamefont {Dysthe}},\ and\ \bibinfo {author}
  {\bibfnamefont {F.}~\bibnamefont {Renard}},\ }\bibfield  {title} {\bibinfo
  {title} {Dynamics of microscale precursors during brittle compressive failure
  in carrara marble},\ }\href@noop {} {\bibfield  {journal} {\bibinfo
  {journal} {Journal of Geophysical Research: Solid Earth}\ }\textbf {\bibinfo
  {volume} {124}},\ \bibinfo {pages} {6121} (\bibinfo {year}
  {2019})}\BibitemShut {NoStop}%
\end{thebibliography}
%

\newpage
\onecolumngrid
\appendix
\renewcommand\appendixname{SI Appendix}
\setcounter{figure}{0}  
\setcounter{table}{0} 

\section*{Supplementary Information}

\begin{table*}[h]
\caption{List of notations used in the manuscript and supplementary information}
\begin{tabular}{ll}
$L, H,b$ & : Length, height and thickness of the specimen\\
$F$ & : Force experienced by the specimen as recorded at the macroscopic scale\\
$F_\mathrm{el}, F_\mathrm{c}$ & : Elastic limit and peak force experienced by the specimen \\
$\Delta$ & : Applied displacement by the loading machine during the experiment \\
$E$&: Elastic modulus of the specimen \\
$\nu$&: Poisson's ratio of the specimen \\
$\sigma_{\mathrm{ext}}$&: Nominal stress experienced by the specimen  during the experiment \\
$\epsilon_{\mathrm{ext}}$&: Nominal strain experienced by the specimen  solid during the experiment \\

$W$, $\Delta W$ &: Input work and incremental input work by the loading machine \\
$E_{el}$, $\Delta E_{el}$ & : Total elastic energy and incremental elastic energy in the specimen corresponding to input  work  \\
 $E_{d}$, $\Delta E_{d}$ & : Total energy dissipated and incremental dissipation during damage growth in the specimen\\
  ${}_{\mathrm{global}}, {}_{\mathrm{local}}$ &: Subscripts specifying the mode of analysis - macroscopic response (global) and at the cells (local) \\
\multirow{2}{*}{$\Delta_{\mathrm{ini}}$,  $\Delta_{\mathrm{end}}$}   &: Values of displacement corresponding to the start  and end of the damage cascade\\
&~~in equivalent force control scenario \\
 $\delta$ & : Normalized  distance to failure\\
$\rho$ &: Dissipation energy density of the material element \\
$\rho_{\mathrm{th}}$ &: Thresholded value of  the dissipation energy density of the material element \\

 $S$&: Size of the damage cascade \\
$\xi$&: Spatial extent of the damage cascade \\
$T$&: Duration of the damage cascade \\
$A$ &: Size of the load drops \\
$dN_S/dt$&: Activity rate of the cascades \\
$dN_A/dt$&: Activity rate of the load drops \\
$\tau_w , \tau_F$&: Waiting time between the cascades in terms of time and force\\
$\Delta d^\ast$&: Characteristic damage value of the cascades  \\
$d_{\mathrm{f}}$&: Fractal dimension \\
$z$&: Dynamic exponent \\
$\theta$&: Exponent  characterizing the marginal stability   \\
$\beta$ &: Exponent characterizing the decay of distribution of cascade size, $S$ close to failure \\
$\beta_{tot}$ &: Exponent characterizing the decay of the stress-integrated distribution of cascade size, $S$ \\
$\alpha$ & : Exponent characterizing the divergence of size of precursors on the approach to failure \\ 

$d_\circ$&: Average damage level in the specimen \\
$\delta d(\vec{x})$ &: Perturbations to the mean damage level whose average over the field is zero \\
$d(\vec{x},t)$ &: Local damage level at any time $t$ during the experiment\\
$\Delta d(\vec{x},t)$ &: Incremental damage from $d_\circ$ at site $\vec{x}$ during the experiment \\
 $Y[\vec{x},d(\vec{x},t),X_0]$&: Local damage driving force determined for a fixed external driving parameter(stress or strain), $X_0$.  \\
 $\psi(d_\circ)$ &: Non-local interaction kernel describing the redistribution of elastic energy  \\
 $Y_c[\vec{x},d(\vec{x},t)]$&: Local damage resistance corresponding to the threshold for damage growth  \\
  $\eta$&: Damage hardening parameter characterizing the variation of damage resistance with damage level  \\
$y_c[\vec{x},d(\vec{x},t)]$ &: Fluctuations in the field of damage resistance visited by the damage like elastic interface\\ 
$\delta Y(\vec{x})$ &: Distance to local failure, i.e., incremental damage in line with energy based damage criterion \\
$\mathcal{K}$ &: Stiffness of springs driving the elastic interface describing the global stability of damage evolution \\
$v_\mathrm{m}$ &: Velocity of the rigid plate that is connected to the pseudo-interface \\
$\delta d_\circ$&: Incremental damage that occurs whenever the damage criterion is satisfied \\
$ \langle \, \rangle $ &: mean of the enclosed quantity \\
\end{tabular}
\label{tab:tab0}
\end{table*}
\newpage

\section{Experimental precursors from an equivalent force control scenario} 
The macroscopic response of the specimen under displacement controlled loading conditions comprises intermittently occurring load drops during which damage grows, followed by elastic reloading phases. Here, we construct an equivalent force control scenario (cyan curve in Fig.~\ref{fig:local}(a)) from the displacement control experiment (blue curve). As a result, damage evolution is interpreted as a micro-instability manifesting as a jump of displacement from $\Delta_\mathrm{ini}$ to $\Delta_\mathrm{end}$ at constant force, wherein distinct damage events (force drops outlined in red) are separated by silent times corresponding to elastic re-loading. The macroscopic response is thus reconstructed as a sequence of force plateaus followed by elastic reloading.
\subsection{Precursors size at the global scale} 
For the equivalent force control scenario, the incremental work done by the external force during silent damage periods  (for $\Delta\leq \Delta_\mathrm{ini}$ and $\Delta_\mathrm{end}\leq\Delta$) is converted into elastic energy, $dW = dE_\mathrm{el} = \frac{1}{2}F \Delta$. During the equivalent force control cascade (for $\Delta_\mathrm{ini} \leq \Delta \leq \Delta_\mathrm{end}$), only a  part of the incremental work $dW = F (\Delta_{\mathrm{end}} - \Delta_{\mathrm{ini}})$ is stored in the material as elastic energy. The remaining part $\Delta E_\mathrm{d} = \Delta W - \Delta E_\mathrm{el} =  \Delta W/2$ is dissipated through damage. This quantity defines the avalanche size $S_\mathrm{global}$ at the global (macroscopic) scale. 

\subsection{Precursors duration and damage events within a cascade} 
\begin{figure}
\centering
\includegraphics[width=0.95\textwidth]{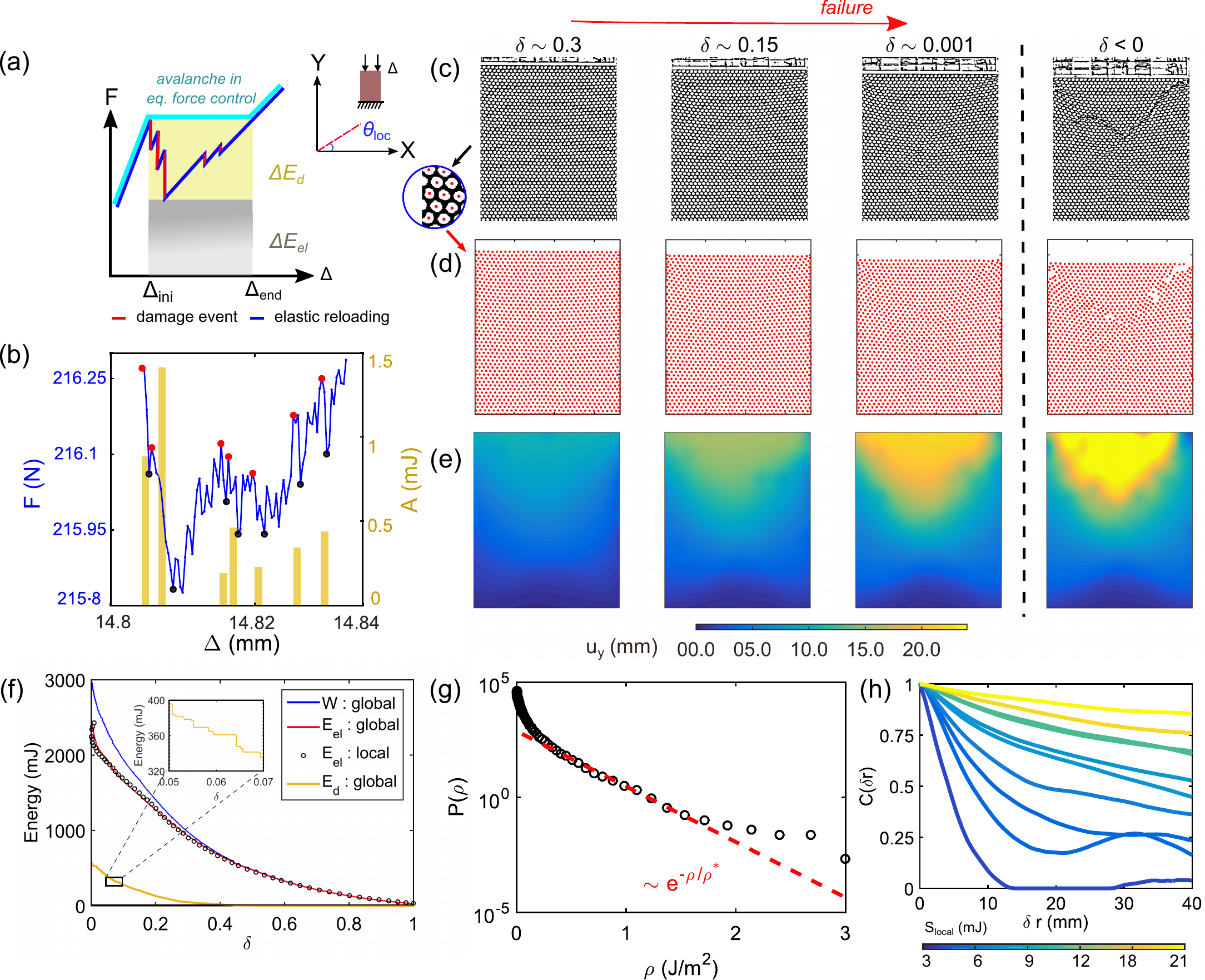}
\caption{
(a) Schematic of a damage precursor (or avalanche) in an equivalent force control scenario during the compression experiment in displacement control. A schematic representation of the localization angle $\theta_{\mathrm{loc}}$. (b) Global analysis of a precursor as measured in our experiments: The damage events constituting the avalanche are revealed by sudden load drops, identified by the red and black circles, denoting the beginning and the end of the event, respectively. The vertical bars provide the size $A$ of the damage event that is defined from the energy dissipated during the event. (c) Binary format of the snapshots of the specimen taken at different distances from localization. (d) The images are post-treated to identify the position  of the cell centers that are tracked during the experiment. (e) The displacement of the cell centers with respect to their initial position are coarse-grained to compute the displacement field $u_{y}(\vec{x})$  at various distances to localization. (f) Evolution of the different energies during the experiment as obtained from our global and local analyses. The work $W$ of the loading machine is converted into elastic energy $E_\mathrm{el}$ (computed both at the local and global scale) and dissipated energy $E_\mathrm{d}$. The inset highlights the intermittent evolution of the dissipated energy, a feature reminiscent of the avalanche dynamics of the damage field. (g) Distribution of increments $\rho$ of local dissipation energy density observed during an avalanche \--- the distribution is computed here from the increments extracted from all the avalanches of a single experiment. It follows an exponential decay $P(\rho) \propto \mathrm{e}^{-\rho/\rho^\star}$, where the exponential cut-off $\rho^\star \simeq 0.18~\mathrm{J/m}^2$ is used to threshold the dissipation energy density maps of single avalanches. 
(h) Auto-correlation functions $C(\delta r)$ of the thresholded maps of dissipation energy density $\rho_{th}$ of $10$ precursors of different sizes. The spatial extent $\xi$ of the precursors is obtained from the fit of the correlation function by an exponential decay $C(\delta r) \propto \mathrm{-e}^{\delta r/\xi}$. Note that larger precursors have smaller slopes and hence larger spatial extent.
}
\label{fig:local}
\end{figure}

The duration of the force plateau is dependent on the cross-head velocity of the loading machine. Thus, it cannot be used to define the intrinsic duration $T$ of a precursor. To determine the duration of a cascade of damage events, we focus on the load drops observed during the force-plateau  (Fig.~\ref{fig:local}(b)). The intermittent load drops are reminiscent of the individual events within a cascade of damage growth during a force controlled experiment.  Taking $\tau$ as the characteristic duration for these damage events, the duration of a cascade writes as $T = N\tau$, where $N$ is the number of load drops during the cascade. Following this framework, we consider the magnitude of the load drops, denoted by $A$, as the size of the damage events during the cascade. Only the load drops of magnitude greater than $0.1~\mathrm{N}$ are considered as individual events. As damage events manifest as highly correlated spatio-temporal clusters, load drops are expected to be macroscopic manifestation of the clusters of correlated individual damage events observed at the local scale, see Fig.~2(b) in the main article.
\subsection{Precursors size at the local scale} 
The continuous image acquisition during the experiment allows for the tracking of both the location and the deformation of individual cells,  which are then used to analyze the spatio-temporal structure of damage cascades at the local scale. The images are recorded at a rate of 10 frames per second  such that each frame corresponds to a duration comparable to the duration of the smallest force plateaus. First, we obtain the binary formats of the images processed using the open source Fiji software \cite{fiji}, Fig.~\ref{fig:local}(c). The center of the cells and their relative circularity are then  tracked to extract the local displacement field and local damage field, respectively. The coordinates of the cell centers recorded at different distances to failure during a typical experiment is shown in Fig.~\ref{fig:local}(d). Using a coarse-graining technique~\cite{glasser2001scale} based on mass conservation, we obtain the coarse-grained displacement field $u_x (x,y)~\mathrm{and}~u_y(x,y)$ as well as the  damage field $d(x,y)$ of an equivalent disordered continuum. Fig.~\ref{fig:local}(e) depicts the field $u_y(x,y)$ during a typical experiment. Taking the gradient of the displacement field, we obtain the strain fields $\epsilon_{xx}(\vec{x}) = \frac{du_x}{dx},~ \epsilon_{yy}(\vec{x}) = \frac{du_y}{dy}~\mathrm{and}~ \epsilon_{xy}(\vec{x}) = \frac{1}{2}\left(\frac{du_x}{dy} + \frac{du_x}{dy}\right)$. The Poisson ratio $\nu$ is obtained from a linear fit of $\epsilon_{xx}(\vec{x})$ vs. $\epsilon_{yy}(\vec{x})$ obtained over time and space. To track the damage evolution at the local scale, we define the relative circularity of the cells as our internal damage variable. It turns out  that the damage level does not affect the Poisson ratio ($\nu = 0.26$) that remains nearly constant all along the compression test. However, the Young's modulus ($E$), inferred from the global specimen stiffness $\frac{\sigma_{yy}}{\epsilon_{yy}} = \frac{E}{(1  - \nu^2)}$, decays with the average damage level $d_\circ$. In the following, we consider a damage dependent elasticity (elasto-damageability) to describe the constitutive response of the cellular solid. We use the relation $E \sim \mathcal{O}(d_\circ^2)$ inferred from the average elastic modulus of the specimen $E$ and the average damage level $d_\circ$ to determine the field $E_{\mathrm{local}}(\vec{x},t)$ of local Young's moduli from the damage field $d(\vec{x},t)$ where $\vec{x} \rightarrow (x,y)$. 
The field of elastic energy stored in the specimen per unit volume is finally obtained from the relation,
\begin{equation}\label{eq:Eel}
    E_\mathrm{el}^\mathrm{local}(\vec{x},t) = \frac{E_\mathrm{local}(\vec{x},t)}{2(1 - \nu^2)} \left[\epsilon_\mathrm{xx}^2(\vec{x},t) + \epsilon_\mathrm{yy}^2(\vec{x},t) + 2 \, \nu \epsilon_\mathrm{xx}(\vec{x},t) \epsilon_\mathrm{yy}(\vec{x},t)  + 2 (1 - \nu)\epsilon_\mathrm{xy}^2(\vec{x},t) \right].
\end{equation}
\noindent $\epsilon_\mathrm{xx}(\vec{x},t), \epsilon_\mathrm{yy}(\vec{x},t)~\mathrm{and}~\epsilon_\mathrm{xy}(\vec{x},t)$ are the local values of normal strains perpendicular and parallel to the direction of loading and shear strains, respectively. At the macroscopic scale,  the damage cascades are evidenced from load drops at near constant displacements. At the local scale, they manifest as a local decrease in stored elastic energy of the damaging element. In practice, we consider the material elements for which the elastic energy decreases  as damaging elements and compute the local increment of dissipation energy as $\delta E_\mathrm{d}^\mathrm{local}(\vec{x}) = -\delta E_\mathrm{el}^\mathrm{local}(\vec{x})$. The evolution of the  different terms contributing significantly to the total energy of the isolated system composed of the specimen and the loading machine, namely the input work, the elastic energy and the dissipated energy as computed from the global analysis is given in Fig.~\ref{fig:local}(f). Our measurement of the total elastic energy from the local strains and elastic moduli are shown to be in good agreement with this global analysis. As damage is accompanied by a decrease in local elastic energy, we obtain the volumetric rate of dissipated energy as
\begin{equation}
    \rho(\vec{x},t_i) = b[E_\mathrm{el}^\mathrm{local}(\vec{x},t_i) - E_\mathrm{el}^\mathrm{local}(\vec{x},t_{i+1})].
\end{equation}
\noindent where $b$ is the specimen thickness and the time steps $t_i$ correspond to the displacements $\Delta(t_i) \in \left [\Delta_\mathrm{ini} , \Delta_\mathrm{end} \right]$. Combining all the increments of dissipated energy belonging to the same damage cascade provides the field $\rho(\vec{x}) = \sum_{i} \rho(\vec{x},t_i) $ of energy dissipated during a given precursor. At the specimen level, we obtain the precursor size byintegration over the whole specimen
\begin{equation}
S_\mathrm{local} =  \iint\rho(x,y) \, \mathrm{d}x \mathrm{d}y
\label{sloc}
\end{equation}
\noindent The distribution of $\rho(\vec{x})$ obtained from all cascades follows an exponential decay $P(\rho) \propto \mathrm{e}^{-\rho / \rho^\ast}$ with $\rho^\ast \simeq 0.18~\mathrm{J/m^2}$ (Fig.~\ref{fig:local}(g)). The value of $\rho^\ast$ is used as a threshold to obtain binary formats of the dissipation density maps $\rho_{\mathrm{th}}(\vec{x})$ of precursors (inset of Fig. 2(a) in the main article). 
\subsection{Spatial extent of the precursors}
To determine the spatial extent $\xi$ of the precursors, we compute the 2D auto-correlation of the thresholded dissipation density maps $\rho_\mathrm{th}(\vec{x})$. The correlation function $C(\delta r) = \langle \rho_\mathrm{th}(\vec{x}).\rho_\mathrm{th}(\vec{x} + \delta\vec{x}) \rangle_{\vec{x}, |\delta \vec{x}| = \delta r}/\langle \rho_\mathrm{th}(\vec{x})^2 \rangle_{\vec{x}}$ is shown in Fig.~\ref{fig:local}(h) for precursors of different sizes $S_\mathrm{local}$. The function decreases faster for smaller avalanche size, pointing out a smaller spatial extent. The correlation function can be fitted by an exponential decay $C(\delta r) \propto \mathrm{e}^{-\delta r/\xi}$, defining the correlation length $\xi$. This quantity is plotted as a function of the precursors' size and duration in Fig.~2(c) and (d) of the main article.

\section{Determination of field of the damage resistance}
\subsection{Retrospective determination from the local damage driving force} 
\begin{figure}[t]
\centering
\includegraphics[width=0.85\textwidth]{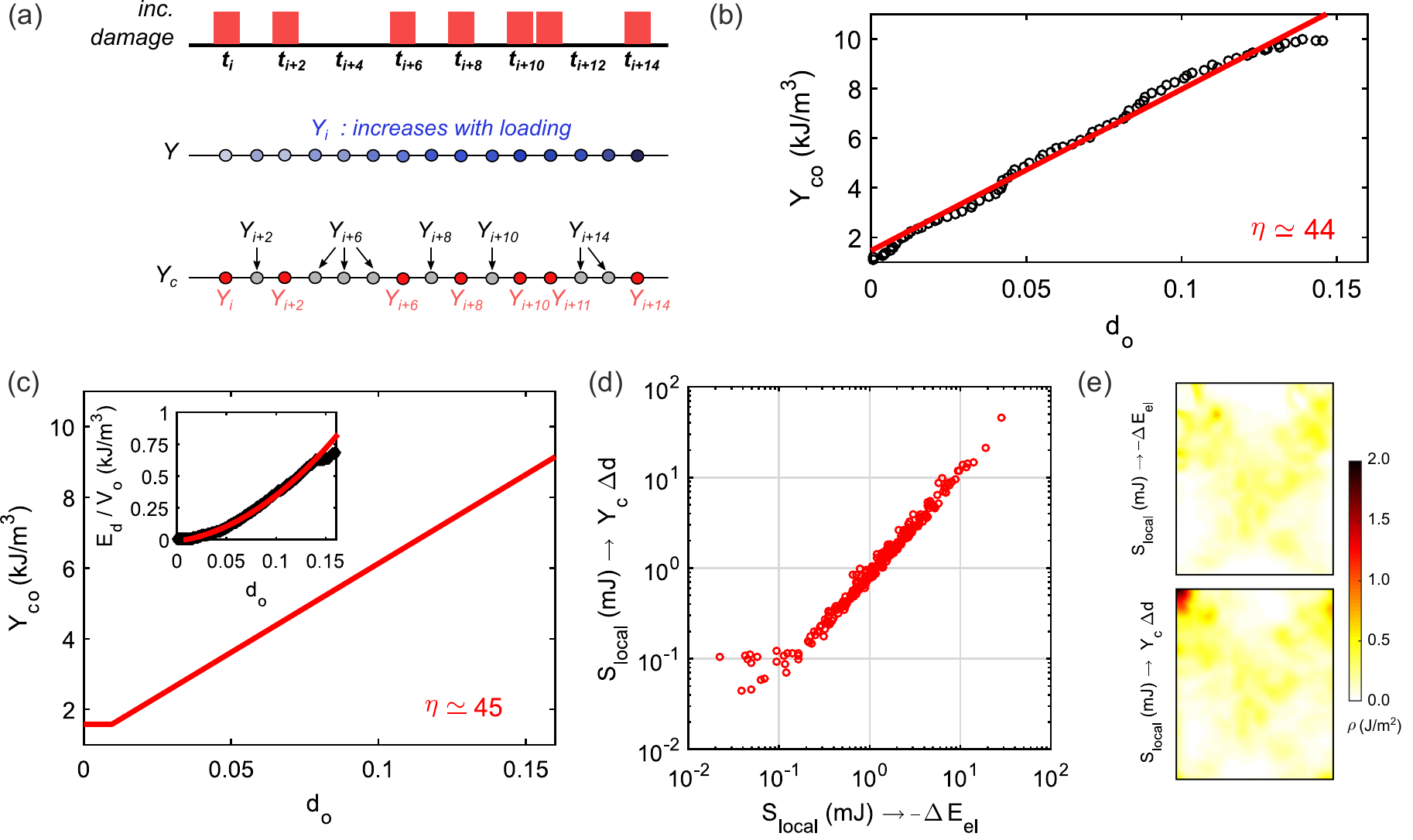}
\caption{(a) Schematic of the retrospective assignment procedure used to determine the local damage resistance $Y_\mathrm{c}[\vec{x},d(\vec{x},t)]$ from the field of damage driving force $Y[\vec{x},d(\vec{x},t),t]$. (b) Variations of the average damage resistance $Y_\mathrm{c \circ}$ with respect to the damage level $d_\circ$. The increase of the damage resistance can be described by a linear hardening $Y_\mathrm{c \circ}(d_\circ) = Y_\mathrm{c}^\circ (1 + \eta \, d_\circ)$ with a hardening coefficient $\eta \simeq 44$. (c) Variations of the average damage resistance $Y_{c\circ}$ with the average damage level $d_\circ$ inferred from the variations with respect to $d-\circ$ of the accumulated dissipated energy at the specimen level that can be described (see inset) by a quadratic function (in red). (c)  Comparison of the precursor size $S_{\mathrm{local}}$ obtained our two independent methods: from the variations of elastic energy in the abscissa and from the local increase of the damage variable and the damage resistance $Y_{c\circ}(d_\circ)$ in the ordinates. (e) Dissipated energy density maps of a typical precursor as obtained from the two methods.}
\label{fig:e_balance}
\end{figure}

One of the most appealing feature of our experiments is the full-field measurement of the damage field and its evolution over time. This can be harnessed to access the field of local damage resistance $Y_\mathrm{c}(\vec{x})$, a quantity that has been hardly measured in the literature, even at the specimen scale. The proposed methodology consists in using the energy based damage criterion
\begin{equation}
\left \{
\begin{array}{ll} 
Y[\vec{x},d(\vec{x},t)] &< Y_c[\vec{x},d(\vec{x},t)]~~~~\Rightarrow~~ \delta d(\vec{x},t) = 0 \\
Y[\vec{x},d(\vec{x},t)] &= Y_c[\vec{x},d(\vec{x},t)]~~~~\Rightarrow~~ \delta d(\vec{x},t) > 0
\end{array}
\right .
\label{dmg_crit}
\end{equation}
\noindent that, as shown in the following, provides retrospectively the local value of the damage resistance $Y_c[\vec{x},d(\vec{x},t)]$ each time a damage event takes place. First, we compute the field  of damage driving force by differentiating Eq.~\ref{eq:Eel} with respect to the damage parameter
\begin{equation}
    Y(\vec{x},t) = -\frac{E'_{\mathrm{local}}(\vec{x},t)}{2(1 - \nu^2)}\left[\epsilon_{xx}^2(\vec{\mathrm{x}},t) + \epsilon_{yy}^2(\vec{\mathrm{x}},t) + 2 \, \nu \epsilon_{xx}(\vec{\mathrm{x}},t) \epsilon_{yy}(\vec{\mathrm{x}},t)  + 2 (1 - \nu)\epsilon_{xy}^2(\vec{\mathrm{x}},t) \right].
\end{equation}
\noindent where the derivative to the local elastic modulus $E'_{\mathrm{local}} \simeq \frac{dE}{dd_\circ}(d(\vec{x},t))$. Here, we assume that the variations of the average macroscopic elastic modulus with the average damage level $d_\circ$ hold at the local scale too.  As schematically illustrated in Fig.~\ref{fig:e_balance}(a), the value of the local damage driving force $Y(\vec{x},t)$ is assigned to the damage resistance $Y_\mathrm{c}(\vec{x},d(\vec{x},t))$ each time a damage event takes place. The damage driving force $Y(\vec{x},t_i)$ at the onset of damage growth $t_i$ is also assigned to the damage resistance $Y_\mathrm{c}(\vec{x},d(\vec{x},\underset{\widetilde{}}{t})) = Y(\vec{x},t_i)$ during the whole sequence of elastic reloading preceding $t_i$, i.e. for $\underset{\widetilde{}}{t} \in [t_{i-k},t_{i-1}...t_i]$ where $t_{i-k}$ is the time at which the previous damage event in the same material element $\vec{x}$ took place. Now considering the average damage resistance $Y_\mathrm{c\circ}(d_\circ) = \langle Y_\mathrm{c}(\vec{x},d(\vec{x},t)) \rangle_{d(\vec{x},t) = d_\circ}$ at some given damage level $d_\circ$, we see in Fig.~\ref{fig:e_balance}(b) that, after an initial transient regime during which damage evolution is dominated by randomly distributed small localized events, $Y_\mathrm{c\circ}(d_\circ)$ {\it increases} with damage. This hardening can be described by a linear relation $Y_\mathrm{c \circ}(d_\circ) = Y_\mathrm{c}^\circ (1 + \eta \, d_\circ)$ where $Y_\mathrm{c}^\circ \simeq 1.4~\mathrm{kJ/m}^3$ and $\eta \simeq 44$.
\subsection{Alternative methodology:  determination from the accumulated dissipated energy}
As an alternative to the above retrospective determination of $Y_c$ at the local scale, the average value of damage resistance can be estimated from the variation of accumulated dissipated energy using
\begin{equation}
 Y_{c\circ}(d_\circ)= \frac{1}{V_\circ}\frac{dE_d}{dd_\circ}
\end{equation}
\noindent where $V_\circ$ is the initial volume of the cellular solid and $E_d$ is the accumulated dissipated energy obtained from the global balance of energy (see inset of Fig.~\ref{fig:e_balance}(c)). Assuming a linear variation  $Y_{c\circ} \propto  \eta d_\circ$, the accumulated dissipated energy is fitted by a quadratic function providing $Y_{c}^\circ = 1.1\mathrm{kJ/m^3~\mathrm{and}~\eta \simeq 45}$, both values that are close to the one inferred from the previous method. Using this linear relation, we then determine the field of damage resistance $Y_{c}(\vec{x},t) = Y_{c\circ}(d(\vec{x},t))$ from the damage field. The energy dissipated at each time step is then computed as $\rho(\vec{x},t_i) = \mathrm{max}(Y_c(\vec{x},t_i)\delta d(\vec{x}, t_i),0)$, where $\delta d(\vec{x},t_i)$ is the incremental damage growth in $\vec{x}$. The dissipated energy density $\rho(\vec{x})$ during one precursor and the precursor size $S_{local}$ follow from \eqref{sloc}. A comparison  with the first method is presented in Fig.~\ref{fig:e_balance}(d) that shows a good agreement. The maps of local dissipated energy density of a typical cascade as obtained from the two approaches are compared in Fig.~\ref{fig:e_balance}(e). The  discrepancies in the maps of the dissipation energy density are localized at the periphery of the specimen where the rather large damage may be attributed to frictional effects with the walls. Thus, we observe a good agreement between both methods, thus validating the assumption of local energy balance made previously. 

\section{Theoretical modeling of the evolution of the damage field}
We now detail the damage model used to describe our experiments. It intends to account for the co-action of material disorder and long-range elastic interactions in the aftermath of a damage event, in the spirit of the physics-based non-local damage models proposed in Berthier \textit{et al.}~\cite{SI_berthier2017} and Dansereau \textit{et al.}~\cite{SI_dansereau2019}. In particular, we provide the redistribution kernel characterizing the long-range interactions derived here for the particular case of a 2D specimen compressed between two fixed lateral walls. The first step, detailed in sections \ref{S3A} and \ref{S3B} for force and displacement imposed conditions, respectively consists in calculating the distribution of damage driving force resulting from a heterogeneous distribution of damage (see Eqs.~(\ref{eq13}) and (\ref{eq19}) that we write in a generic form valid for both loading conditions in \eqref{eq21}). We then extend our calculations to the case of a monotonically increasing loading amplitude in section~\ref{S3C}. In section~\ref{S3D} we derive the damage evolution equation from the damage criterion of \eqref{dmg_crit}. We  then explain in section~\ref{S3E} how the evolution equation of the damage field belongs to the theoretical framework of driven disordered elastic interfaces. The stability of the damage spreading process is studied in section~\ref{S3F}. The  localization threshold is shown to coincide with peak load. Lastly, we show in section~\ref{S3G} that the rate of damage growth diverges on approaching localization.

We start by considering that the level of damage in each material element located in $\vec{x}$ is $ d(\vec{x}) \geq 0 $, where $d = 0$ corresponds to the intact initial material. The level of damage is assumed to affect the elastic modulus $E[d(\vec{x})]$ of the cellular solid. Its Poisson's ratio $\nu$ is assumed to be independent of the damage level, as supported by our experimental observations (Sec.~S1C). The constitutive behavior of the cellular solid is then the one of an elasto-damageable solid under plane stress conditions
\begin{equation}
\centering
\boldsymbol{\epsilon} = \frac{(1+\nu)\boldsymbol{\sigma}- \nu\mathrm{Tr}(\boldsymbol{\sigma})} {E}.
\label{eq6}
\end{equation}
Due to the lateral confinement $(\epsilon_{xx} = 0)$, the in-plane stress components follow: $\sigma_{yy} =  \epsilon_{yy}E/(1 - \nu^2)$ along the loading direction and $\sigma_{xx} = \nu\sigma_{yy}$ along the lateral direction. In absence of out-of-plane stress ($\sigma_{zz} = 0$), Hooke's law predicts $\epsilon_{zz}=-\epsilon_{yy}\nu/(1 - \nu)$. The elastic energy per unit volume of the material, $w(\boldsymbol{\epsilon},\boldsymbol{\sigma}) = (1/2)~\boldsymbol{\epsilon} \colon \boldsymbol{\sigma}$, can then be expressed  as
\begin{equation}
\centering
w(\boldsymbol{\epsilon},\boldsymbol{\sigma})  = \frac{(1+\nu)\mathrm{Tr}(\boldsymbol{\sigma}^2)- \nu {\mathrm{Tr}(\boldsymbol{\sigma})}^2}{2E}
\label{eq7}
\end{equation}
\subsection{Damage driving force under force control conditions}\label{S3A} For a homogeneous damage level $d_\circ$, the elastic energy release rate that drives damage growth follows
\begin{equation}
Y(d_\circ,\boldsymbol{\sigma}) = \left.\frac{\partial w}{\partial d_\circ}\right|_{\boldsymbol{\sigma}} = \frac{\overline{\mu}'(d_\circ)\mathrm{Tr}(\boldsymbol{\sigma}^2)}{4} + \frac{{\nu E'(d_\circ)\mathrm{Tr}(\boldsymbol{\sigma})}^2}{2E(d_\circ)^2}
\label{eq8}
\end{equation}
where $\mu = E/2(1 + \nu) $ is the Lam{\'e} constant and  $\overline{\mu} = 1/\mu$. The prime denotes the derivative with respect to the damage variable $d_\circ$. To examine the effect of material disorder on the damage field and in particular on the stress redistributions following individual damage events, we introduce weak variations in the damage field $d(\vec{x}) = d_\circ + \delta d(\vec{x})$ where $d_\circ$ is the average damage level and  $\delta d(\vec{x}) \ll d_\circ$ are its perturbations. The heterogeneities in the damage field result from the spatial variations of the elastic modulus. Consequently, the stress field is heterogeneous and writes as $\boldsymbol{\sigma}(\vec{x}) = \boldsymbol{\sigma_\circ} + \delta \boldsymbol{\sigma}(\vec{x})$ where 
\begin{equation}
\boldsymbol{\sigma_\circ} = \sigma_{\mathrm{ext}}\begin{bmatrix} \nu & 0 & 0\\ 0&1 & 0 \\ 0 & 0 & 0  \end{bmatrix}~\mathrm{and}~\delta \boldsymbol{\sigma}(\vec{x}) = \begin{bmatrix} \delta\sigma_{xx}(\vec{x}) & \delta\sigma_{xy}(\vec{x})& 0\\ \delta\sigma_{yx}(\vec{x}) & \delta\sigma_{yy}(\vec{x}) & 0 \\ 0& 0& 0 \end{bmatrix}.
\end{equation}
Here, $\sigma_{\mathrm{ext}} = F/bL$ is the average stress imposed by the loading machine to the upper surface $bL$ of the specimen. As a consequence, the damage driving force provided in ~\eqref{eq8} is also heterogeneous and can be decomposed as $Y [d(\vec{x})] = Y_\circ(d_\circ) + \delta Y [d(\vec{x})]$ where
\begin{equation}
\left \{
\begin{array}{ll} 
Y_\circ(d_\circ,\boldsymbol{\sigma_\circ}) &=  \frac{\overline{\mu}'(d_\circ)\mathrm{Tr}(\boldsymbol{\sigma_\circ}^2)}{4} + \frac{{\nu E'(d_\circ)\mathrm{Tr}(\boldsymbol{\sigma_\circ})}^2}{2E(d_\circ)^2} \\
\delta Y[d(\vec{x}),\boldsymbol{\sigma}_\circ] &= \Bigg[\frac{\overline{\mu}''(d_\circ)\mathrm{Tr}(\boldsymbol{\sigma_\circ}^2)}{4} - \frac{{\nu \mathrm{Tr}(\boldsymbol{\sigma_\circ})}^2}{2}\left(\frac{2E'(d_\circ)^2}{E(d_\circ)^3} - \frac{E''(d_\circ)}{E(d_\circ)^2} \right)\Bigg] \delta d(\vec{x})  +
\Bigg[\frac{\overline{\mu}'(d_\circ)\mathrm{Tr}(\boldsymbol{\sigma_\circ}\delta\boldsymbol{\sigma}(\vec{x}))}{2} + \frac{\nu E'(d_\circ)\mathrm{Tr}(\boldsymbol{\sigma_\circ}) \mathrm{Tr}(\delta\boldsymbol{\sigma}(\vec{x}))}{E(d_\circ)^2}\Bigg].
\end{array}
\right.
\label{eq9}
\end{equation}
The first term in the expression of $\delta Y [d(\vec{x})]$ is a local term, namely its value in $\vec{x}$ depends only on the damage level in $\vec{x}$. It can be obtained by considering a homogeneous damage level $d_\circ$ and varying $Y[d_\circ = d(\vec{x})]$ with respect to it as follows
\begin{equation}
    \centering
  \left.\frac{\partial Y_\circ}{\partial d_\circ}\right|_{\sigma_\circ} = \frac{(1 - \nu^2)}{2}\sigma_{ext}^2 \left[\frac{2E'(d_\circ)^2}{E(d_\circ)^3} - \frac{E''(d_\circ)}{E(d_\circ)^2} \right].
  \label{eq10}
\end{equation}
To calculate the non-local contributions to $\delta Y [d(\vec{x})]$ (i.e., the second term in \eqref{eq9}) we first need to evaluate the stress perturbations $\delta \boldsymbol{\sigma}(\vec{x})$ emerging from the  spatial variations of elastic modulus, as performed in Dansereau \textit{et al.}~\citep{SI_dansereau2019}. We then use a perturbative approach in Fourier space and such calculations that gives
\begin{equation}
\delta\boldsymbol{\tilde\sigma}(q_x,q_y) = \frac{\overline{\mu}'(d_\circ)}{\overline{\mu}(d_\circ)}\left( \mathbf{O}\cdot \boldsymbol{\sigma_\circ}\cdot\mathbf{O} - \nu \left[ \left(1 - \mathbf{O}\right) \colon \boldsymbol{\sigma_\circ}\right] \mathbf{O}\right) \delta \tilde{d}(q_x,q_y),
\label{eq11}
\end{equation}
where $\mathbf{O}$ stands for the Oseen tensor, $\mathbf{O} = 1- \mathbf{Q}$ with the tensor $\mathbf{Q_{ij}} = q_iq_j/|\vec{q}|^2$. $\delta\boldsymbol{\tilde\sigma}(q_x,q_y)$ and $\delta \tilde{d}(q_x,q_y)$ are the 2D Fourier transform  of the stress field perturbations and the damage field perturbations, respectively. Substituting \eqref{eq11} into \eqref{eq9}, we obtain the non-local contributions to the damage driving force that writes as $\tilde{\psi}  \delta \tilde{d}$ in Fourier space, that defines the interaction kernel $\tilde{\psi}(q_x,q_y)$. In real space and in polar coordinates, it writes as $\psi (r,\theta)\sim g(\theta)/r^{2}$, where $r$ is the distance from the damage event at the origin of the stress redistribution and $g(\theta)$ is an angular function with $\theta = \arctan(y/x)$. As expected, the elastic interactions decay as a power law of the distance $r$, resulting in long-range interactions between material elements during damage spreading. The interaction kernel, derived here for the case of uni-axial compression with lateral confinement follows
\begin{equation}
\centering
\psi(d_\circ) = \left[ \frac{E'(d_\circ)^2}{E(d_\circ)^3}\right](1 - \nu^2)\sigma_{ext}^2 \left[\frac{x^4-3y^4+6x^2y^2}{4\pi(x^2+y^2)^3}\right] \rightarrow ||\psi|| \left[\frac{\cos^4{\theta}-3\sin^4{\theta}+6\cos^2{\theta}\sin^2{\theta}}{4\pi r^6}\right]
\label{eq12}
\end{equation}
in cartesian coordinates. The pre-factor $||\psi(d_\circ)|| =  \frac{E'(d_\circ)^2}{E(d_\circ)^3}(1 - \nu^2)\sigma_{\mathrm{ext}}^2$  is independent of $\vec{x}$. In Fourier space, the interaction kernel follows $\tilde{\psi}(\vec{q}) \sim \frac{q_x}{||\vec{q}||} \sim -cos^4(\omega)$ where $\omega = \arctan(\frac{q_x}{q_y})$. We finally obtain the spatial distribution of damage driving force 
\begin{equation}
\centering
Y[d(\vec{x}),\boldsymbol{\sigma_\circ}] = Y(d_\circ,\boldsymbol{\sigma_\circ}) + \left.\frac{\partial Y_\circ}{\partial d_\circ}\right|_{\boldsymbol{\sigma_\circ}} \delta d(\vec{x})  + \psi(d_\circ) \ast \delta d(\vec{x}).
\label{eq13}
\end{equation}
Interestingly, the kernel $\psi(d_\circ)$ that describes here the distribution of damage driving force in a heterogeneously damaged specimen can also be interpreted as describing the driving force redistribution following an individual damage event. Considering the damage perturbations $\delta d(\vec{x}) = \delta (\vec{x} - \vec{x_0}) \delta d_\circ$ corresponding to an individual event taking place in $\vec{x_0}$ where $\delta(u)$ is the Dirac function and $\delta d_\circ$ the damage event amplitude, we obtain the spatial distribution of incremental driving force $\delta Y(\vec{x}) = \delta d_\circ \psi(\vec{x} - \vec{x_0})$ resulting from the damage event.  It turns out that the material regions located next to $\vec{x}_\circ$ perpendicularly to the loading axis are reloaded while the material regions located below and above the damaged element are screened as shown in Fig.3((a)).

\subsection{Damage driving force under displacement control conditions} \label{S3B}The driving force for damage growth computed in \eqref{eq8} for a homogeneous distribution of damage level $d_\circ$  is now expressed as a function of the strain field $\boldsymbol{\epsilon}$ 
\begin{equation}
Y(d_\circ,\boldsymbol{\epsilon}) = -\mu'(d_\circ)\mathrm{Tr}(\boldsymbol{\epsilon}^2) -  \frac{\lambda'(d_\circ){\mathrm{Tr}(\boldsymbol{\epsilon})}^2}{2},
\label{eq14}
\end{equation}
where $\mu = E/2(1 + \nu) $ and $\lambda = E \nu/(1-2\nu)(1+\nu)$ are the Lam{\'e} constants. It is important to note that the expression of $Y$ does not depend on the type of loading conditions, i.e., Eqs.~(\ref{eq14}) and (\ref{eq8}) are equivalent. In presence of weak variations of the damage field $d(\vec{x}) = d_\circ + \delta d(\vec{x})$, the strain field writes as the sum of two contributions $\boldsymbol{\epsilon}= \boldsymbol{\epsilon_\circ} + \delta \boldsymbol{\epsilon}$ where

\begin{equation}
\boldsymbol{\epsilon_\circ} = \epsilon_{ext}\begin{bmatrix} 0 & 0 & 0\\ 0&1 & 0 \\ 0 & 0 & -\frac{\nu}{1-\nu}  \end{bmatrix}~\mathrm{and}~\delta \boldsymbol{\epsilon}(\vec{x}) = \begin{bmatrix} \delta\epsilon_{xx}(\vec{x}) & \delta\epsilon_{xy}(\vec{x})& 0\\ \delta\epsilon_{yx}(\vec{x}) & \delta\epsilon_{yy}(\vec{x}) & 0 \\ 0& 0& 0 \end{bmatrix}.
\end{equation}

The field of damage driving force, $Y [d(\vec{x}),\boldsymbol{\epsilon_\circ}] = Y_\circ(d_\circ,\boldsymbol{\epsilon_\circ}) + \delta Y [d(\vec{x}),\boldsymbol{\epsilon_\circ}]$, including its perturbations $\delta Y [d(\vec{x}),\boldsymbol{\epsilon_\circ}]$ follows
\begin{equation}
\left \{
\begin{array}{ll} 
Y_\circ(d_\circ,\boldsymbol{\epsilon_\circ}) &= -\mu'(d_\circ)\mathrm{Tr}(\boldsymbol{\epsilon_\circ}^2) -  \frac{\lambda'(d_\circ){\mathrm{Tr}(\boldsymbol{\epsilon_\circ})}^2}{2}, \\
\delta Y[d(\vec{x}),\boldsymbol{\epsilon_\circ}] &= -\Bigg[\mu''(d_\circ)\mathrm{Tr}(\boldsymbol{\epsilon_\circ}^2) + \frac{{\lambda''(d_\circ) \mathrm{Tr}(\boldsymbol{\epsilon_\circ})}^2}{2} \Bigg] \delta d(\vec{x}) -
 \Bigg[ 2\mu'(d_\circ)\mathrm{Tr}(\boldsymbol{\epsilon_\circ}\delta\boldsymbol{\epsilon}(\vec{x})) + \lambda' (d_\circ) E'\mathrm{Tr}(\boldsymbol{\epsilon_\circ}) \mathrm{Tr}(\delta\boldsymbol{\epsilon}(\vec{x})) \Bigg].
 \end{array}
 \right.
\label{eq15}
\end{equation}
$\delta \boldsymbol{\epsilon}(\vec{x})$ follows from ~\eqref{eq6} 
\begin{equation}
\centering
\delta \boldsymbol{\epsilon}(\vec{x}) = \left[ \frac{(1 + \nu) \delta\boldsymbol{\sigma}(\vec{x}) - \nu \mathrm{Tr}(\delta \boldsymbol{\sigma}(\vec{x}))}{E(d_\circ)}\right] - \left[\frac{E'(d_\circ)}{E(d_\circ)^2}\right] \left[ (1 + \nu) \boldsymbol{\sigma_\circ} - \nu \mathrm{Tr}(\boldsymbol{\sigma_\circ}) \right].
\label{eq16}
\end{equation}
The first term in bracket in~\eqref{eq15} provides the local contribution to the driving force perturbations
\begin{equation}
\left.\frac{\partial Y_\circ}{\partial d_\circ}\right|_{\epsilon_\circ} = - \left[\frac{E''(d_\circ)}{E(d_\circ)^2}\right]\frac{(1 - \nu^2)}{2} \sigma_{\mathrm{ext}}^2.
\label{eq17}
\end{equation}
Using ~\eqref{eq16}, the second term in brackets in~\eqref{eq15} simplifies as $
\left[\frac{2E'(d_\circ)^2}{E(d_\circ)^3}\right]\frac{(1 - \nu^2)}{2}\sigma_{\mathrm{ext}}^2 \delta d(\vec{x}) +\psi(d_\circ) \ast \delta d(\vec{x}) $
where $\psi(d_\circ)$ is the interaction kernel given in~\eqref{eq7}. Thus, the field of damage driving force under displacement control conditions writes as 
\begin{equation}
\centering
Y[d(\vec{x}),\boldsymbol{\epsilon_\circ}] = Y(d_\circ, \boldsymbol{\epsilon_\circ}) + \left.\frac{\partial Y_\circ}{\partial d_\circ}\right|_{\epsilon_\circ} \delta d(\vec{x})  + \left( \left[\frac{2E'(d_\circ)^2}{E(d_\circ)^3}\right]\frac{(1 - \nu^2)}{2}\sigma_{\mathrm{ext}}^2\right)\delta d(\vec{x}) +\psi(d_\circ) \ast \delta d(\vec{x}).
\label{eq19}
\end{equation}
Comparing \eqref{eq19} with \eqref{eq13} and noticing that the spatial distribution of damage driving force is independent of the type of loading conditions, we obtain
\begin{equation}
\begin{split}
    \centering
    \left.\frac{\partial Y_\circ}{\partial d_\circ}\right|_{\boldsymbol{\sigma_\circ}} &= \left.\frac{\partial Y_\circ}{\partial d_\circ}\right|_{\boldsymbol{\epsilon_\circ}} + \left[\frac{2E'(d_\circ)^2}{E(d_\circ)^3}\right]\frac{(1 - \nu^2)}{2}\sigma_{\mathrm{ext}}^2,\\
     &= \left.\frac{\partial Y_\circ}{\partial d_\circ}\right|_{\boldsymbol{\epsilon_\circ}} + ||\psi(d_\circ)||.
    \label{eq20}
    \end{split}
\end{equation}

\noindent Finally we can generalize the damage driving force corresponding to a perturbed damage field $d(\vec{x}) = d_\circ + \delta d(\vec{x})$ using 
\begin{equation}
\centering
Y[d(\vec{x})] = Y_\circ(d_\circ) +  \Psi(d_\circ) \ast \delta d(\vec{x}), 
\label{eq21}
\end{equation}
\noindent where the generalized interaction kernel follows $\Psi(d_\circ)  = \left.\frac{\partial Y_\circ}{\partial d_\circ}\right|_{\sigma_\circ} \delta(\vec{x}) + \psi(d_\circ)$. The first term provides the local contribution to the damage driving force while the second term corresponds to the non-local part controlled by the interaction kernel $\psi$.

\subsection{Damage evolution under increasing loading amplitude}\label{S3C} We now examine the 
evolution of the damage field $d(\vec{x},t)$ in response to an external driving $X_\circ(t)$ (which can be either force or displacement) that increases linearly with time: $X_\circ = X_\circ(0) + v_{\mathrm{ext}}t$. The damage field can be expressed as $ d(\vec{x},t) = d_\circ(0) + \Delta d(\vec{x},t)$, where the incremental damage field $ \Delta d(\vec{x},t) \ll d_\circ(0)$. Therefore, the reference (homogeneous) damage level corresponding to the driving $X_\circ(t)$ is $d_\circ(t) = d_\circ(0) + \langle \Delta d \rangle_{\vec{\mathrm{x}}}(t) $ and the corresponding damage field perturbations writes as $\delta d(\vec{x},t) = \Delta d(\vec{x},t) - \langle\Delta d\rangle_{\vec{\mathrm{x}}} $. We now use the previous result of \eqref{eq21} to obtain the damage driving force  
\begin{equation}
Y[d(\vec{x},t),X_\circ(t)] = Y_\circ[d_\circ(t), X_\circ(t)] + \Psi(d_\circ)  \ast \left[\Delta d(\vec{x},t) - \langle\Delta d\rangle_{\vec{\mathrm{x}}} \right].
\end{equation}

\noindent Note that this linearization remains valid over a short period of time $t \ll X_\circ/v_{\mathrm{ext}}$ where $v_{\mathrm{ext}}$ is the velocity of the external driving  imposed by loading machine. By expanding the first term and using the expression of the kernel $\Psi$, one obtains
\begin{equation}
\begin{split}
    Y[d(\vec{x},t),t] &= Y[d_\circ(0), X_\circ(0)] +  \left.\frac{\partial Y_\circ}{\partial X_\circ}\right|_{d_\circ(0)}v_{\mathrm{ext}}t + \left.\frac{\partial Y_\circ}{\partial d_\circ}\right|_{X_\circ(0)} \langle\Delta d\rangle_{\vec{\mathrm{x}}}~~~+ \\
    &~~~\left.\frac{\partial Y_\circ}{\partial d_\circ}\right|_{\boldsymbol{\sigma_\circ}(0)}\left[\Delta d(\vec{x},t) - \langle\Delta d\rangle_{\vec{\mathrm{x}}} \right] + \psi[d_\circ(0)] \ast \left[\Delta d(\vec{x},t) - \langle\Delta d\rangle_{\vec{\mathrm{x}}} \right]
    \end{split}
\end{equation}
\noindent where only first order terms proportional to $\Delta d$ have been kept. For the force control case $(X_\circ \rightarrow \boldsymbol{\sigma_\circ})$, the damage driving force is given by

\begin{equation}
    Y[d(\vec{x},t),t] = Y[d_\circ(0), \boldsymbol{\sigma_\circ}(0)] + \left.\frac{\partial Y_\circ}{\partial \sigma_\circ}\right|_{d_\circ(0)}v_{\mathrm{ext}}t + \left.\frac{\partial Y_\circ}{\partial d_\circ}\right|_{\boldsymbol{\sigma_\circ(0)}} \Delta d(\vec{x},t) + \psi[d_\circ(0)] \ast \left[\Delta d(\vec{x},t) - \langle\Delta d\rangle_{\vec{\mathrm{x}}} \right].
\label{eq24}    
\end{equation}

\noindent Similarly, under displacement control conditions $(X_\circ \rightarrow \boldsymbol{\epsilon_\circ})$ the damage driving force writes as

\begin{equation}
    Y[d(\vec{x},t),t] = Y[d_\circ(0), \boldsymbol{\epsilon_\circ}(0)] + \left.\frac{\partial Y_\circ}{\partial \epsilon_\circ}\right|_{d_\circ(0)}v_{\mathrm{ext}}t + \left.\frac{\partial Y_\circ}{\partial d_\circ}\right|_{\boldsymbol{\sigma_\circ}(0)} \Delta d(\vec{x},t) + \psi[d_\circ(0)] \ast \left[\Delta d(\vec{x},t) - \langle\Delta d\rangle_{\vec{\mathrm{x}}} \right] - ||\psi[d_\circ(0)]||\langle\Delta d\rangle_{\vec{\mathrm{x}}} .
\label{eq25}    
\end{equation}
\noindent where the \eqref{eq20}, $\left.\frac{\partial Y_\circ}{\partial d_\circ}\right|_{\boldsymbol{\epsilon_\circ}(0)} = \left.\frac{\partial Y_\circ}{\partial d_\circ}\right|_{\boldsymbol{\sigma_\circ}(0)} - ||\psi[d_\circ(0)]||$ has been used. The damage driving force under force and displacement control conditions are thus similar, up to the last term in \eqref{eq25}. This term acts as a mean-field restoring force. Its effect on the stability of the damage growth process is discussed in the following.

\subsection{Generalized damage driving force} \label{S3D}
We now consider the resistance to damage growth $Y_c[d(\vec{x},t)]$. Its linearized expression follows 
\begin{equation}
Y_c[d(\vec{x},t)] = Y_c^\circ[d_\circ(0)] +  \left.\frac{dY_{c\circ}}{dd_\circ}\right|_{d_\circ(0)} \Delta d(\vec{x},t) + y_c[\vec{x},d(\vec{x},t)]
\end{equation}
\noindent where $\frac{dY_{c\circ}}{dd_\circ} = Y_{c}^\circ\eta$ with $Y_{c}^\circ$ the damage resistance of the intact material, $\eta $ the  hardening parameter and $y_c[\vec{x},d(\vec{x},t)] = Y_{c}[d(\vec{x},t)] - Y_{c\circ}(d_\circ)$, the heterogeneous contribution to the field of damage resistance. It is practical to introduce the generalized damage driving force
\begin{equation}
    \mathcal{F}[d(\vec{x},t),t] = Y[d(\vec{x},t),t] - Y_\mathrm{c}[d(\vec{x},t)].
\end{equation}

\noindent In particular, under force control conditions, the expression of the generalized driving force can be written, following \eqref{eq24}, as 

\begin{equation}
    \mathcal{F}[d(\vec{x},t),t] = \left.\frac{\partial Y_\circ}{\partial \boldsymbol{\sigma_\circ}}\right|_{d_\circ(0)} v_{\mathrm{ext}}t +  \left.\frac{\partial (Y_\circ - Y_{c\circ})}{\partial d_\circ}\right|_{\boldsymbol{\sigma_\circ}(0)} \Delta d(\vec{x},t) +  \psi[d_\circ(0)] \ast \left[\Delta d(\vec{x},t) - \langle\Delta d \rangle_{\vec{\mathrm{x}}} \right] - y_c[\vec{x},d(\vec{x},t)].
    \label{eq26}
\end{equation}
\noindent The generalized driving force under displacement control conditions follows

\begin{equation}
    \mathcal{F}[d(\vec{x},t),t] = \left. \frac{\partial Y_\circ}{\partial \boldsymbol{\epsilon_\circ}}\right|_{d_\circ(0)} v_{\mathrm{ext}} t +  \left.\frac{\partial (Y_\circ - Y_{c\circ})}{\partial d_\circ}\right|_{\boldsymbol{\sigma_\circ}(0)} \Delta d(\vec{x},t) +  \psi[d_\circ(0)] \ast \left[\Delta d(\vec{x},t) - \langle\Delta d \rangle _{\vec{\mathrm{x}}}\right] - y_c[\vec{x},d(\vec{x},t)]- || \psi[d_\circ(0)]|| \langle \Delta d\rangle_{\vec{\mathrm{x}}}
    \label{eq27}
\end{equation}

\begin{figure}
\centering
\includegraphics[width=0.66\textwidth]{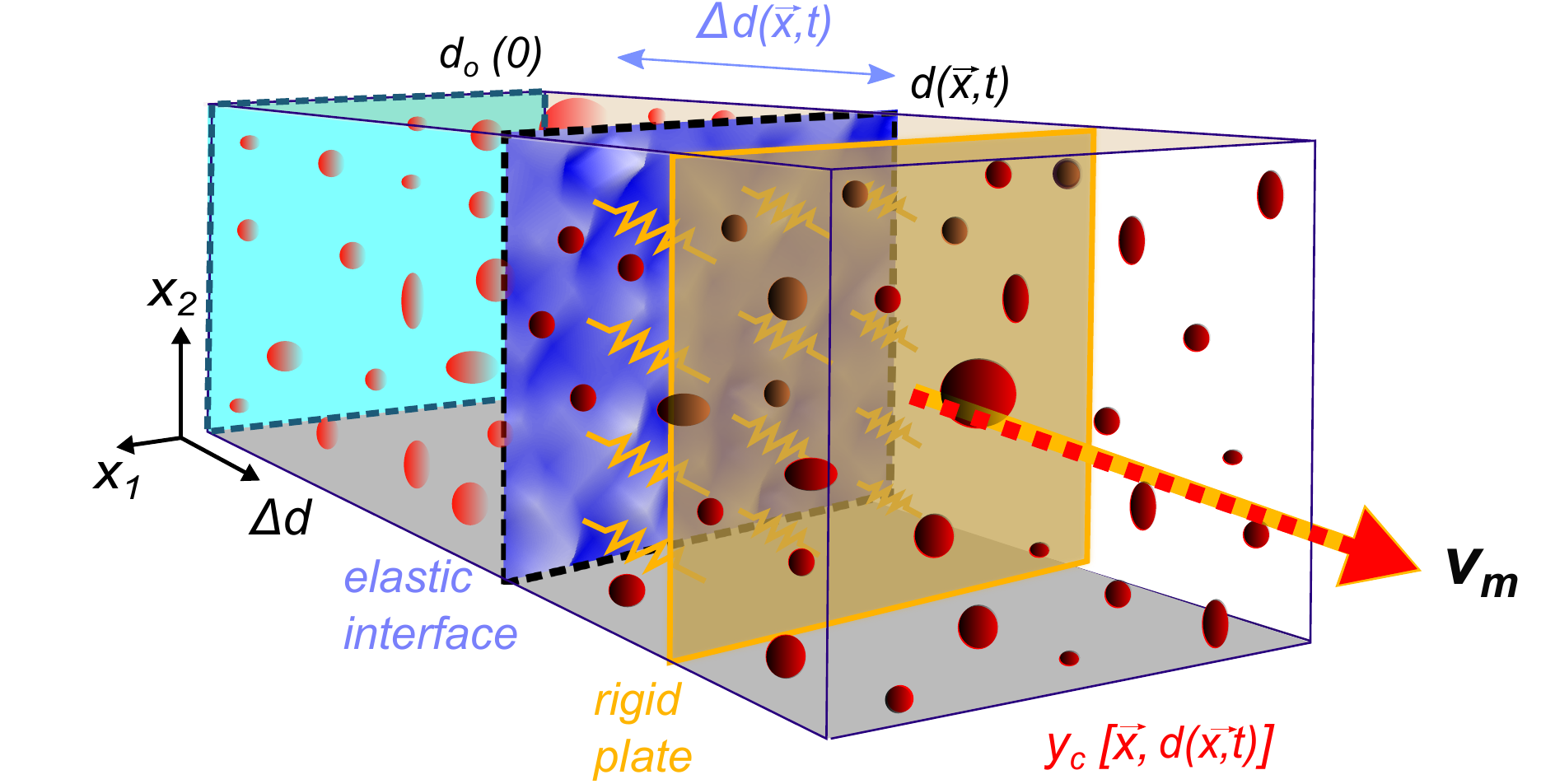}
\caption{Schematic representation of the damage level as an elastic interface (in blue) driven through a disordered field of damage resistance $y_c [\vec{x},d(\vec{x},t)]$. The red blobs are representative of the regions of relatively higher value of damage resistance. The interface is initially flat $d_\circ(0)$. The rigid plate (in yellow) is driven at the speed $v_\mathrm{m}$ and pulls on the interface with linear springs of stiffness $\mathcal{K}$. Consequently the interface also moves with an average velocity $v_\mathrm{m}$. The interplay between the disorder and the interface elasticity roughens the interface $d(\vec{x},t)$ at time $t$ and leads to an avalanche-like dynamics as it progresses through the disorder field of damage resistance. }
\label{fig:kernel}
\end{figure}
\subsection{Analogy to driven disordered elastic interfaces} \label{S3E}The framework of driven disordered elastic interfaces describes the intermittent response of an elastic manifold to a continuous external drive as it propagates through a disordered field of resistance. Following a local depinning event, the co-action of disorder and elasticity may generate a cascade of depinning events. This phenomenon yields robust scaling laws relating the characteristic features of the cascades all together - size, duration and spatial extent. Here, we seek to recast the evolution equation of the damage field under force controlled conditions to the one of a driven pseudo-interface. By considering an over-damped dynamics $\dot{\Delta d} \propto \mathcal{F}$, where $\dot{\Delta d}$ is the local damage growth rate, and rearranging the terms of \eqref{eq26}, we obtain
\begin{equation}
   \dot{\Delta d}(\vec{x},t) \propto \mathcal{K}\left[ v_\mathrm{m}t - \Delta d(\vec{x},t)\right] +  \psi[d_\circ(0)] \ast \left[\Delta d(\vec{x},t) - \langle\Delta d \rangle_{\vec{\mathrm{x}}} \right] - y_c[\vec{x},d(\vec{x},t)],
    \label{eq28}
\end{equation}
\begin{equation}
\centering
\mathrm{where~}\mathcal{K}[\boldsymbol{\sigma_\circ}(0)] = \left.\frac{\partial (Y_{c\circ} - Y_\circ)}{\partial d_\circ}\right|_{\boldsymbol{\sigma_\circ}(0)}  \mathrm {~~and~~}v_\mathrm{m}[\boldsymbol{\sigma_\circ}(0)] = \frac{\partial Y_\circ/\partial \boldsymbol{\sigma_\circ}}{\mathcal{K}[\boldsymbol{\sigma_\circ}(0)]}v_{\mathrm{ext}}. 
\label{eq29}
\end{equation}

\noindent Equation (\ref{eq28}) describes a 2D elastic interface driven through a heterogeneous field of damage resistance $y_c[\vec{x},d(\vec{x},t)]$. The interface is driven at an average speed $v_\mathrm{m}$ by Hookean springs of stiffness $\mathcal{K}$ connecting the interface to a rigid plate moving at the  speed $v_\mathrm{m}$, as schematically illustrated in Fig.~\ref{fig:kernel}.  The competition between the disorder in the field $y_c[\vec{x},d(\vec{x},t)]$ of damage resistance and the interface elasticity controls the roughness $\delta d(\vec{x},t)$ of the interface. Following each damage event,  multiple regions of the interface may move forward as a result of  the redistribution of the local driving force along the interface, leading to cascades of damage events. Notably, the disorder $y_c[\vec{x},d(\vec{x},t)]$ is a function of both the location $\vec{x}$ and the damage field $d(\vec{x},t)$, so it takes the value of damage resistance actually visited by the interface. This leads to the strongly non-linear response of the interface to the smoothly varying drive observed in our experiments.
Under displacement control conditions, the damage evolution follows a similar equation
\begin{equation}
    \dot{\Delta d}(\vec{x},t) \propto \mathcal{K}\left[ v_\mathrm{m}t - \Delta d(\vec{x},t)\right] +  \psi[d_\circ(0)] \ast \left[\Delta d(\vec{x},t) - \langle\Delta d \rangle_{\vec{\mathrm{x}}} \right] - y_c[\vec{x},d(\vec{x},t)] - || \psi[d_\circ(0)]|| \langle \Delta d\rangle_{\vec{\mathrm{x}}}
    \label{eq30}
\end{equation}
\begin{equation}
\centering
\mathrm{where~}\mathcal{K}[\boldsymbol{\sigma_\circ}(0)] = \left.\frac{\partial (Y_{c\circ} - Y_\circ)}{\partial d_\circ}\right|_{\boldsymbol{\sigma_\circ}(0)}  \mathrm {~~and~~}v_\mathrm{m}[\boldsymbol{\epsilon_\circ}(0)] = \frac{\partial Y_\circ/\partial \boldsymbol{\sigma_\circ}}{\mathcal{K} + ||\psi||} v_{\mathrm{ext}} 
\end{equation}
\noindent where the relation $\frac{\partial \boldsymbol{\sigma_\circ}}{\partial \boldsymbol{\epsilon_\circ}} =   \frac{E}{(1 - \nu^2)}\frac{\mathcal{K}}{\mathcal{K} + ||\psi||}$ has been used. The comparison of the  Eqs. (\ref{eq29}) and (\ref{eq30}) shows that the mean damage growth rate $v_\mathrm{m}$ normalized by the loading rate $v_\mathrm{ext}$ (that provides the speed of the rigid plate pulling on the interface)  is slower under displacement control conditions. In addition, an extra term $\sim - \langle \Delta d \rangle_{\vec{\mathrm{x}}}$ that contributes to stiffen the interface (and so to stabilize it), applies.

\subsection{Onset of damage localization}\label{S3F} We use our theoretical framework to investigate the stability of the damage growth process. Consider a homogeneous system, i.e.,  $y_c[\vec{x},d(\vec{x},t)] = 0$ and a harmonic positive perturbation of the damage field $d(\vec{x}) = d_\circ + \Delta d_\circ(1 + \cos{(\vec{q_\circ}\cdot\vec{x})})$ of amplitude $\Delta d_\circ$. The corresponding damage driving force $\mathcal{F}$ derived in Eqs. (\ref{eq26}) and (\ref{eq27}) writes as
\begin{equation}
\begin{split}
        \mathcal{F}(\vec{q_\circ}\cdot\vec{x}) &=  \mathcal{K}(d_\circ) \left[v_\mathrm{m}t - \Delta d_\circ\right] \delta(\vec{x}) +  \left[\psi(d_\circ)  - \mathcal{K}(d_\circ) \delta(\vec{x}) \right]\ast \Delta d_\circ\cos{(\vec{q_\circ}\cdot\vec{x})} ~~~~~~~~~~~~~~~~~~~~~~\mathrm{under~force~control,} \\
                \mathcal{F}(\vec{q_\circ}\cdot\vec{x}) &=  \mathcal{K}(d_\circ) \left[v_\mathrm{m}t - \Delta d_\circ\right] \delta(\vec{x}) +  \left[\psi(d_\circ)  - \mathcal{K}(d_\circ)\delta(\vec{x}) \right]\ast \Delta d_\circ\cos{(\vec{q_\circ}\cdot\vec{x})} - || \psi(d_\circ)||\Delta d_\circ ~~\mathrm{under~ displacement~control.} 
    \end{split}
    \label{eq31}
\end{equation}

\noindent 
 
Assuming a homogeneous perturbation $d(\vec{x}) = d_\circ + \Delta d_\circ$, the convolution product vanishes and the stability criterion of the damage evolution reduces to
\begin{equation}
\left \{
\begin{array}{ll} 
\mathcal{K} > 0 & ~~~~~~~~~~~~~~~~~~~\mathrm{under~force~control~conditions,} \\
\mathcal{K} + ||\psi|| &> 0~~~~~~~~~~~~~~\mathrm{under~ displacement~control~conditions.} 
\end{array}
\right.
\label{eq32}
\end{equation}

These conditions provide the homogeneous damage level at which the damage evolution is unstable. Under force control conditions, it corresponds to peak load where $\mathcal{K} = \left. \frac{\partial(Y_\circ - Y_{c\circ})}{\partial d_\circ}\right|_{\boldsymbol{\sigma_\circ}}$ changes sign. Under displacement control, the stability criterion predicts that damage growth is more  stable as $\mathcal{K} + ||\psi(d_\circ)|| > \mathcal{K}$. In particular, the stability condition is ensured even beyond peak load. Equation (\ref{eq31}) also provides the perturbations in damage driving force resulting from periodic perturbations $\Delta d_\circ\cos{(\vec{q_\circ}\cdot\vec{x})}$ of the damage field. The emergence of the localization band is inferred from the criterion \cite{SI_dansereau2019}.
\begin{equation}
    \tilde{\psi}(\vec{q}_\circ, d_\circ) -  \mathcal{K}(d_\circ) = 0,
    \label{eq33}
\end{equation}

\noindent Indeed, \eqref{eq33} ensures that if damage grows in regions of high damage level, at $q_\circ \cdot \vec{x} = 0, \pm2\pi,\pm4\pi,\ldots\ $, the driving force would also grow in these regions, resulting in the unstable growth of damage. As $\tilde{\psi}(\vec{q}_\circ, d_\circ) \sim -\cos^4(\omega_\circ)$ where $\omega_\circ = \arctan(q_{y\circ} / q_{x\circ}) $(see \eqref{eq12}), the interaction kernel in Fourier space $\tilde{\psi}$ is always negative and is exactly equal to zero for the polar angle $\omega_c = \pi/2$. As  a result, the localization criterion reduces to
\begin{equation}
\mathcal{K}(d_\circ) = 0
    \label{eq34}
\end{equation}

which coincides with the criterion for peak load. The inclination  $\theta_{loc} = \pi/2 - \omega_c$ of the localization band is given by the most unstable mode (the one that first reaches the localization criterion described by \eqref{eq33}), namely the one that maximizes $\tilde{\psi}(\vec{q}_\circ)$ and first satisfying the condition $\tilde{\psi}(\vec{q}_\circ) = 0$. The prediction of the emergence of an horizontal localization band ($\theta_{loc} = 0$), perpendicular to the main loading axis, is compatible with our experimental observations (Fig.~\ref{fig:local}C right panel and Video S1). 

These results call for a few comments. First, under force control conditions, the localization threshold coincides with peak load. In other words, the heterogeneous mode of instability (i.e., the growth of the localization band within the specimen) is activated simultaneously with the homogeneous one (i.e., the unstable growth of the average damage level $d_\circ$). This is not the case under displacement control conditions as the extra term $-||\psi(d_\circ)||\Delta d_{\vec{\mathrm{x}}}$ in the evolution equation (\ref{eq27}) delays the homogeneous instability (that takes place after peak load) without affecting the threshold of the heterogeneous instability. The experimental observation of an increase of the precursory activity close to peak load for both loading conditions then supports that this phenomenon takes place at the approach of an instability, irrespective of its homogeneous or heterogeneous nature. In the following section, we detail the connection between the divergence of the precursor size and the presence of an instability at peak load.

\subsection{Divergence of precursors} \label{S3G}We study the variations of damage driving force on approaching peak load where damage localization occurs. Its linear expansion around the critical damage level $d_c$ at peak load (and localization threshold) $X_c$ follows
\begin{equation}
    \mathcal{F}_\circ(d_\circ, X_\circ) = \mathcal{F}_\circ(d_{c},X_c) + \frac{\partial\mathcal{F}_\circ}{\partial d_\circ}(d_\circ - d_{c}) + \frac{\partial\mathcal{F}_\circ}{\partial X_\circ}(X_\circ - X_{c}) + \frac{1}{2}\frac{\partial^2 \mathcal{F}_\circ}{\partial d_\circ^2}(d_\circ - d_{c})^2 + \frac{1}{2}\frac{\partial^2\mathcal{F}_\circ}{\partial X_\circ^2}(X_\circ - X_{c})^2.
    \label{eq35}
\end{equation}
\noindent  Equilibrium ensures that both $\mathcal{F}_\circ(d_\circ,X_\circ) = 0~\mathrm{and}~\mathcal{F}_\circ(d_{c},X_c) = 0$. At peak load, we also have $\left.\frac{\partial\mathcal{F}_\circ}{\partial d_\circ}\right|_{X_c} = 0 \Rightarrow \mathcal{K}(d_c,X_c) = 0$. As a result, for $d_\circ < d_{c}$, \eqref{eq35} simplifies as

\begin{equation}
    d_\circ = d_{c} -  A_\circ \sqrt{X_{c} - X_\circ} \Rightarrow  v_\mathrm{m} \propto 1 / \sqrt{(X_{c} - X_\circ}.
    \label{eq36}
\end{equation}
\noindent where $A_\circ$ is a positive constant and $v_\mathrm{m} = \dot{d_\circ}$  is the average damage growth rate. Note that the second order term $\mathcal{O}(X_\circ - X_{c})^2$ has been neglected. Reminding that  $(X_{c} - X_\circ)$ is proportional to the distance to failure  $\delta$, we obtain $dE_\mathrm{d}/dt \propto \dot{d_\circ} \propto 1/\sqrt{\delta}$. Coming back to the evolution of the precursors, the dissipation rate $dE_\mathrm{d}/dt$ is given by the product of the precursor size with the precursor rate $dE_\mathrm{d}/dt = \langle S \rangle \, dN_\mathrm{S}/dt $. As we observe a constant activity rate, we thus expect the size $S $  of precursors to diverge as
\begin{equation}
    \langle S \rangle \propto 1/ \sqrt{\delta},
\end{equation}
\noindent a prediction consistent with the experimental observations.

\section{Numerical modeling  of the evolution of the damage field} 
We solve the damage evolution equations (\ref{eq28}) and (\ref{eq30}) numerically to obtain accurate predictions of the exponents involved in  the various scaling relations characterizing precursory cascades observed experimentally. An additional motivation of our numerical resolution of the damage evolution equations is  to validate the  assumptions  that the statistical features of the specimen's intermittent response under force control conditions can be approximated from analyzing the damage cascades in an equivalent scenario constructed from displacement control experimental data. We consider an isotropic elasto-damageable specimen of size $L \times L$ discretized into $L^2$ elements with periodic boundary conditions. The degradation of the material elastic modulus with damage is described by a polynomial function $E_\circ = E^\circ(1 - d_\circ^2)$ with $E^\circ = 1 \mathrm{MPa}$, in line with the non-linear degradation of stiffness observed experimentally. This differs from the linear approximation made in most damage mechanics models \cite{Lemaitre}. Note however that similar non-linear descriptions have been adopted to describe the experiments \citep{SI_girard2010, SI_thilakarathna2020}. We remind that this continuum description of the damage induced softening through the damage dependent modulus $E_\circ(d_\circ)$ applies at a scale larger than the characteristic size of the microscopic dissipative mechanisms (e.g., microcracks) \citep{SI_kachanov1993}. This continuous description also allows for modeling the non-local elastic interactions between material elements through the interaction kernel defined in \eqref{eq21}, computed in \eqref{eq11} and represented in Fig.3(a) of the main article. To set the damage resistance, we use the experimentally measured hardening behavior  $Y_{c\circ}(d_\circ) = Y_c^\circ(1 + \eta d_\circ)$  and consider $Y_c^\circ = 1.4~\mathrm{kJ/m^3}$  and $\eta = 40$. The field of $Y_c$ has an initial Gaussian disorder $N(0, 0.05)$ accounting for $y_c$ in the theoretical formulation. The value of incremental damage $\delta d_\circ$ to be added whenever the damage criterion is fulfilled is taken as $0.001$. Using the above parameters, we simulate twenty realizations each for force and displacement imposed conditions and analyze the intermittent damage evolution preceding localization.

\subsection{Force control case} We adopt the numerical procedure employed by Berthier et al.~\citep{SI_berthier2021} to simulate damage evolution \eqref{eq28} under quasi-static loading conditions. We increase the stress gradually to ensure that damage grows by an increment $\delta d_\circ$ in one material element at the location $\vec{x} = \vec{x}_0$ so that $d(\vec{x}_0) \rightarrow d(\vec{x}_0) + \delta d_\circ$. Hence, there is no explicit time defined in the simulations. The value of $Y(x_0)$ at the location $\vec{x}_0$ is increased by an increment $Y'\delta d_\circ$ and $Y_c(\vec{x}_0)$ is increased by an increment $(Y_{c}'\delta d_\circ + y_c(x_0))$ where $y_c(x_0) \ll Y_{c}'\delta d_\circ$. As $Y' < Y_c'$, owing to the stability of the damage growth process, no additional damage event takes place at $\vec{x}_0$. However, the non-local redistribution of damage driving force, given by $\delta d_\circ\psi(\sigma_0) (\vec{x} - \vec{x}_0)$, may trigger another damage event elsewhere and ultimately a damage cascade. Once the cascade is over, the stress is increased again. The initial disorder in the damage resistance ensures that cascades  are small at the early stages of the damage process, i.e., redistributions do not trigger large cascades. On approaching failure, the damage cascades' size increases and their corresponding spatial extent reaches the system size at localization. As the redistributions span the entire system, multiple clusters may nucleate during a cascade. The intermittent damage evolution is thus formulated as arising from the co-action of the disorder and long-range interactions. 

\begin{figure}
\centering
\includegraphics[width=\textwidth]{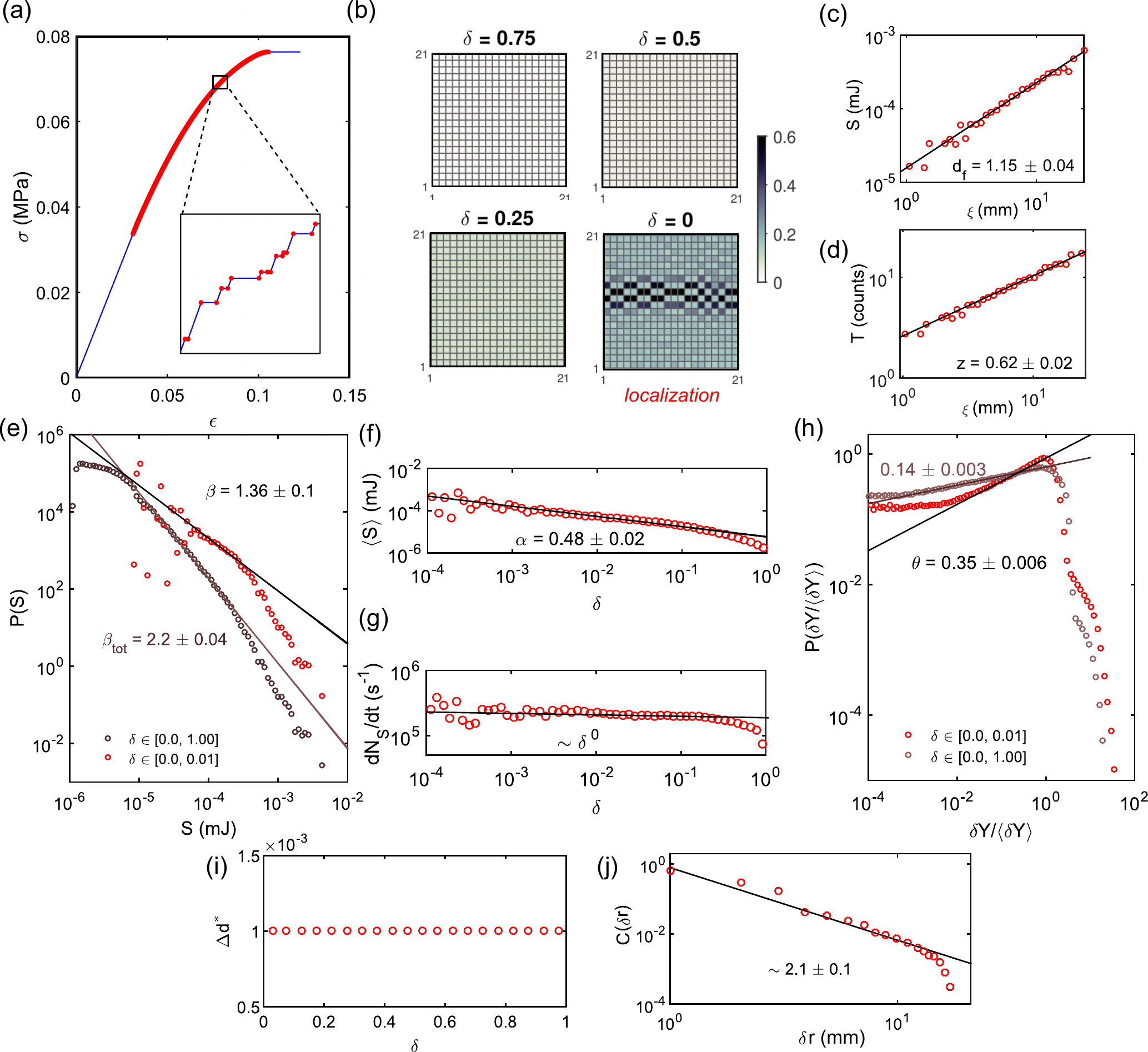}
\caption{(a) Typical stress-strain response obtained during the simulations of damage spreading under force control conditions show jumps of displacement at constant force corresponding to damage growth. Damage localization takes place at peak load. After it, damage evolution becomes unstable. (b) Snapshot of the damage field at various distances from failure. Damage grows rather homogeneously until localization. Variation of (c) cascade size $S$ and (d) duration, $T$ with the spatial extent $\xi$ are related by robust scaling exponents, thus defining the fractal dimension $d_\mathrm{f}$ and dynamic exponent $z$.  (e) The distribution of cascade sizes $S$ computed at different ranges of $\delta$ on approaching failure reveals the non-stationary nature of damage evolution, i.e., two different scaling exponents. The average size of cascades is shown to diverge as a power-law on approaching failure (f) Their activity rate remains constant, (g). (h) Distribution of distance to local failure, $\delta Y$ defining the exponent $\theta > 0$ reminiscent of the non-positive interaction kernel. (i) Variation of the characteristic damage increment of an avalanche with distance to failure $\delta$. (j) Spatial correlation of incremental damage field along X-axis displaying a decay $\sim 1/r^2$ reminiscent of the interaction kernel $\psi \sim 1/r^2$.}
\label{fig:force_control}
\end{figure}

\paragraph*{Precursors to damage localization: }
The computed stress-strain response during damage evolution is presented in Fig.~\ref{fig:force_control}(a) and consists of  a sequence of force plateaus followed by elastic loading (inset of Fig.~\ref{fig:force_control}(a)). The evolution of damage at various distances to failure for a typical numerical experiment is presented in Fig.~\ref{fig:force_control}(b) manifesting a homogeneous damage evolution until localization. The localization band depicted in the final panel is in good agreement with the theoretical predictions $\theta_{loc} = 0$. Each cycle of redistribution, irrespective of the number of elements being damaged, is assumed to constitute a single damage event of energy $A$ as incremental damage in elements are assumed to occur simultaneously. The energy of the cascade is therefore $S = \sum A$. Further, the number of redistribution cycles is taken as the duration of the cascade ($T$) and the spatial extent of the largest cluster is taken as the characteristic length of the cascades ($\xi$). The cascades' energy and duration are shown to scale with the characteristic length $\xi$ with exponents $d_\mathrm{f} \simeq 1.15$ and $z \simeq 0.62$, see Figs.~\ref{fig:force_control}(c) and \ref{fig:force_control}(d). The sizes of the cascades are distributed as a power law,  Fig.~\ref{fig:force_control}(e), with $\beta \simeq 1.36$. Considering all cascades from the beginning of the compression experiment, we measure a larger exponent $\beta_{\mathrm{tot}} \simeq 2.2$. The relation between both exponents is derived by taking  into account the non-stationary nature of damage evolution. The relation $\beta  < \beta_{\mathrm{tot}}$ results from the divergence of the exponential cut-off  $S^\ast$ to the distribution of avalanche sizes ~\cite{SI_amitrano2012}
\begin{equation}
    P_{\delta}(S) = C_oS^{ -\beta}\mathrm{exp}\left(\frac{-S}{S^\ast}\right).
\end{equation}
\noindent where $C_o$ is a normalization constant and $S^\ast \sim \delta^{-\alpha/(2 - \beta)}$. The last relation is inferred from the divergence $\langle S \rangle \sim \delta^{-\alpha}$ of the average cascade size $\langle S \rangle$ and its relation with $S^\ast$ through 
\begin{equation}
    \langle S \rangle = \int_{0}^{\infty}S P_\delta(S) dS \rightarrow \langle S \rangle \sim {S^\ast}^{(2 - \beta)}.
\end{equation}
\noindent The sizes' distribution of all the cascades of an experiment, until failure i.e., for $\delta \in [0, 1]$  is
\begin{equation}
    P(S) = \frac{C_o}{N_{\mathrm{tot}}}\int_{0}^{1}\frac{dN_S}{dt}(\delta)P_\delta(S)d\delta,
\end{equation}
\noindent where $\frac{dN_S}{dt}(\delta) \sim \delta^{-\epsilon}$ is the rate of cascades at some distance to failure $\delta$ and is a constant, i.e, $\epsilon \approx 0$. Using the relation between $\langle S \rangle$ and $S^\ast$ and the scaling of $\langle S \rangle$ with $\delta$, we have
\begin{equation}
    P(S) = \frac{C_o}{N_{\mathrm{tot}}}\int_{0}^{1}\delta^{ -\epsilon}S^{ -\beta} \mathrm{exp}\left(S\delta^{\frac{\alpha}{(2 - \beta)}}\right)d\delta.
\end{equation}
\begin{equation}
\Rightarrow P(S) \sim S^{ -\beta_{\mathrm{tot}}} \sim S^{ -\beta - (\epsilon + 1)(2 - \beta)/(\alpha)} \rightarrow \beta_{\mathrm{tot}} = \beta + \frac{2 - \beta}{\alpha}
\end{equation}
This relation is consistent with the exponents measured in our simulations.

\paragraph*{Variation of the size and the activity rate of damage cascades on approaching damage localization: }
The variation of average cascade size $\langle S \rangle$ with the distance to failure is shown in Fig.~\ref{fig:force_control}(f). We measure the exponent $\alpha \simeq 0.48$. The activity rate of cascades, $dN_S/dt$, is found independent of the distance to failure (Fig.~\ref{fig:force_control}(g)) similar to our experimental observations. 

\paragraph*{Implications of a non-positive interaction kernel: } The non-positive nature of the elastic interactions provide interesting additional traits to the critical features of the damage field. In particular, as only a fraction of material elements are loaded while the others are shielded after each damage event, the (marginal) stability of the elements is altered after each redistribution cycle. Here, we explore the stability numerically by introducing the distance to local failure $\delta Y(\vec{x}) = Y_c(\vec{x}) - Y(\vec{x})$. An additional non-trivial exponent $\theta > 0$ has been shown to characterize the distribution $P(\delta Y)$ of the distance to local failure in case of yielding of amorphous materials \cite{SI_lin2014,SI_lin2015,SI_lin2016}. A similar description also holds for elasto-damage solids. The exponent characterizing $P(\delta Y)$  is measured in Fig.~\ref{fig:force_control}(h). The exponent $\theta$ takes a small value $\simeq 0.14$ and, unlike other critical exponents $\beta$, $d_f$ and $z$, increases on approaching failure \cite{SI_lin2015,SI_lin2016} to a value $\theta \simeq 0.35$. This increase might be due to the increasing amplitude of the pre-factor of the interaction kernel $|| \psi||$. Another interesting manifestation of the non-positive elastic interactions is the constant depth of the cascades, i.e, the interface never progresses locally by more than one elementary step $\delta d_\circ$ during a cascade. The variation of the characteristic depth $\Delta d^\ast$ with the distance to failure is found to be constant in Fig.~\ref{fig:force_control}(i), implying that all material elements damage once during a cascade. 

We also validate numerically our method of deducing the nature of elastic interactions from the incremental damage field during a cascade that we used experimentally to characterize the elastic interactions. Figure~\ref{fig:force_control}(j) shows the variations along the X-axis of the 2D auto-correlation function $C(\vec{\delta r}) = \langle \delta d(\vec{r}) \cdot \delta d(\vec{r} + \vec{\delta r}) \rangle_\mathrm{\vec{r}}$ of the incremental damage field averaged over several avalanches. The correlations are found to decay as $\sim 1/\delta r^2$. Overall, we obtained a detailed statistical description of precursory activity for the force control scenario from our numerical model. In addition, it captures quantitatively the main features observed experimentally. In the next section, we explore numerically the damage field evolution under displacement control conditions and validate the protocol employed experimentally for studying avalanches under equivalent force control conditions.

\subsection{Equivalent force control case} We now simulate damage spreading under displacement control conditions. We  follow the same numerical scheme as mentioned previously for the force control case. In addition,  we include the term $-|| \psi || \langle \Delta d\rangle_{\vec{\mathrm{x}}}$ to the local driving force $Y$, in line with the damage evolution equation (\ref{eq30}). Consequently, under these loading conditions the damage growth manifests as  load drops at fixed displacement. 

\begin{figure}
\centering
\includegraphics[width=\textwidth]{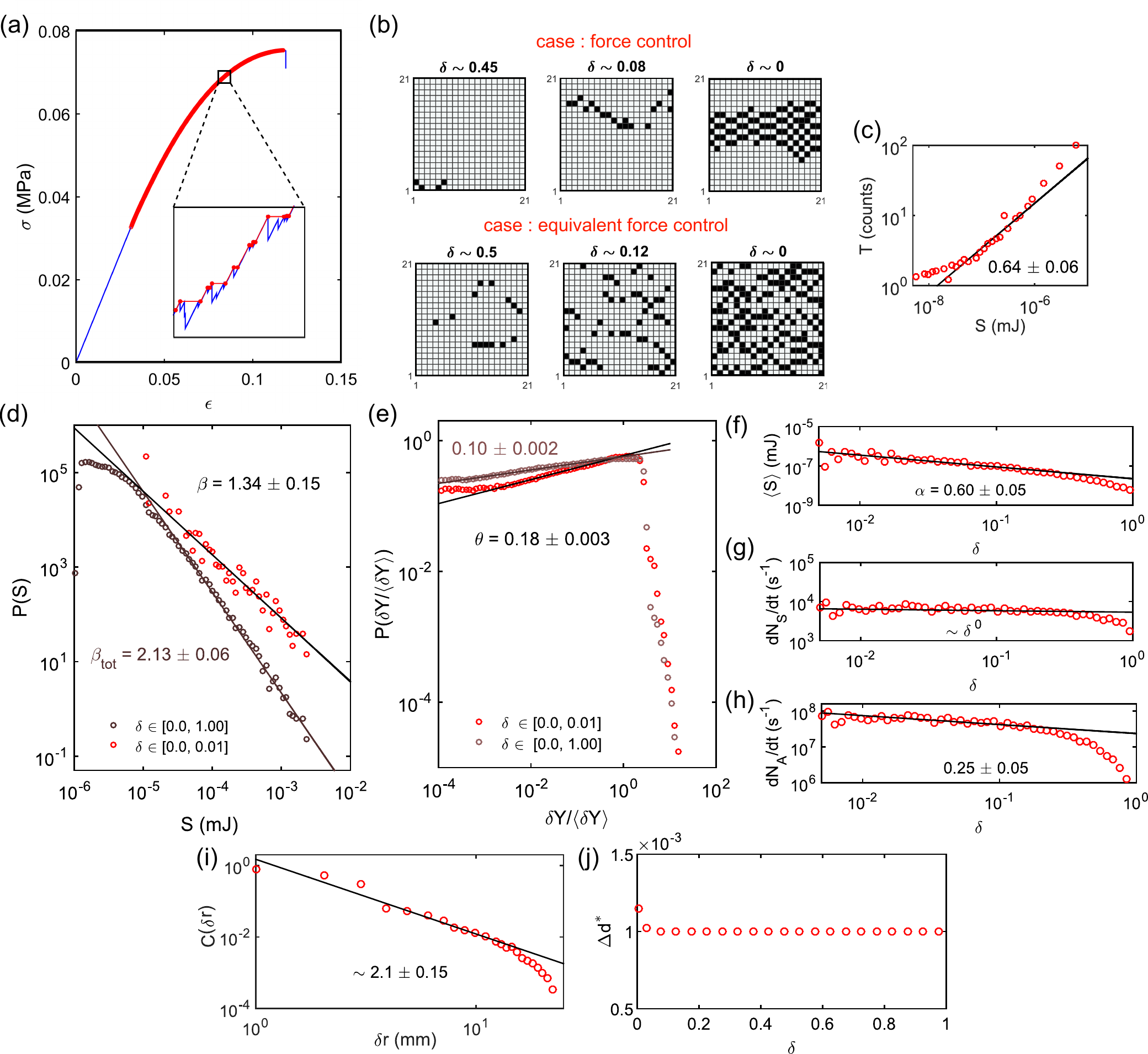}
\caption{(a) Typical stress-strain response obtained during the simulations of damage spreading in displacement control conditions show load drops at constant displacement corresponding to damage growth. We construct an equivalent force control scenario from load drops similar to the experiments to obtain damage cascades. (b) Snapshot of the incremental damage field during a damage cascade at various distances from failure in force  control condition is compared to the equivalent scenario. Note the cascades in equivalent force control case has more clusters and is spatially distributed. (c) Scaling of duration $T$ with cascade size, $S$. (d) Size distribution of cascade size at different ranges of $\delta$. (e) Size distribution of distance to local failure, $\delta Y$, in the vicinity of failure, $\delta \in [0, 0.01]$ and  $\delta \in [0.25, 0.5]$.  Variation of (f) average cascade size, (g) activity rate of cascades and (h) event activity rate (load drops) with distance to failure, $\delta$. (i) Variation of 2D correlation of incremental damage field along X-axis with distance. (j) Variation of the characteristic damage with distance to failure $\delta$.}
\label{fig:disp_control}
\end{figure}

Using the methodology proposed in the main manuscript for the analysis of experimental data, we obtain an equivalent force control scenario from the displacement control simulations (inset of Fig.~\ref{fig:disp_control}(a)). A comparison of the spatial structure of damage cascades in the equivalent scenario with cascades during force control condition are presented in Fig.~\ref{fig:disp_control}(b). They show remarkable differences while their statistics are similar as we will see in the following.  Notably, cascades in the equivalent force control conditions appear as two dimensional, in qualitative agreement with the experimental observations, see inset of Fig. 2(a). We then determine the exponent $z/d_\mathrm{f}$ describing the scaling relation between cascade size $S$ and its duration $T$ $ \simeq 0.64$, see Fig.~\ref{fig:disp_control}(c). The distribution of the cascade sizes is characterized by power-law relations with exponents $\beta \simeq 1.35$ and $\beta_{\mathrm{tot}} \simeq 2.2$  as shown in Fig.~\ref{fig:disp_control}(d). Similar to the force control case, the exponents characterizing the distribution of distance to local failure is non-zero. Interestingly, the value $\theta \simeq 0.18$ observed for the present case is much smaller than the force control case as shown in Fig.~\ref{fig:disp_control}(e). This is perhaps due to the presence of the stabilizing term $\sim -\langle \Delta d \rangle_{\vec{\mathrm{x}}}$ in displacement  loading condition that impedes the spatial organization of cascades. Nevertheless and most importantly, the divergence of the average cascade size and the constant activity rate on approaching failure are similar to the features observed under force control, as shown in Fig.~\ref{fig:disp_control}(f) and Fig.~\ref{fig:disp_control}(g), respectively. This highlights  the universal nature of the intermittent material response approaching bifurcation at peak load. The divergence of the rate of dissipation is therefore $dE_d/dt \propto 1/\sqrt{\delta}$. In terms of load drops, the event activity rate is shown to vary with distance to failure in Fig.~\ref{fig:disp_control}(h). Lastly, we observe that the radial decay of the 2D correlation function of incremental damage field taken along the X-axis as well as the depth of cascades during the displacement control simulations resemble the results obtained under force control, see Fig.~\ref{fig:disp_control}(i) and Fig.~\ref{fig:disp_control}(j). Thus, we show here that using an equivalent force control scenario results in cascades that are statistically similar to that obtained  under force control conditions. The differences in exponents are rather minimal except in case of $\theta$ which characterizes the local stability of damage field between precursors instead of the nature of precursors themselves. Also, the activity rate of cascades in both scenarios were a constant.  Such considerations thus allow for an improved understanding of the spatio-temporal structure of damage cascades in our experiments which otherwise in true force controlled conditions would be challenging for observation.

\newpage

\section{Comparison of exponents with literature \\ and prediction for the case of 3D disordered solids} As few models incorporating long-range interactions exist for compressive failure \cite{SI_girard2010}, we compare in Table \ref{tab:tab1} the exponents measured in our work with predictions from  elasto-plastic models of yielding of amorphous solids as they also feature a non-positive long-range redistribution kernel. In these studies \cite{SI_lin2014,SI_liu2016,SI_ozawa2018}, the kernel is of the shear-type: the driving force is reloaded for the elements along the diagonals and unloaded for those along the horizontal and vertical axes. Our observations are shown to be in good agreement with the theoretical predictions for the 2D case, except for the exponent $\theta$. The difference arises from the different kernels considered theoretically that corresponds to shear and the one involved in our experiments that corresponds to compression. This also explains why the exponents observed in our experiments match well with predictions drawn from our simulations, which consider a compression kernel. We believe that the exponents predicted for the 3D case constitute realistic predictions for precursory cascades during compressive failure of disordered materials. These predicted values might be significantly different from the exponents measured experimentally ~\cite{SI_weiss2019,SI_kandula2019}. We believe that the procedure employed to analyze their experiments does not capture the full damage cascades, but instead the individual clusters composing it.  Note also that the limited spatial resolution (studies using acoustic emission \cite{SI_weiss2019}) or temporal resolution (studies based on X-ray tomography \cite{SI_kandula2019}) might hinder the complete  characterization of damage cascades from individual damage events in these studies. 
\begin{table}
\centering
\caption{Comparison of exponents with numerical and experimental data from literature.}
\begin{tabular}{llccl|ll}
\hline
   & &\multicolumn{2}{c}{present work} & \multicolumn{2}{c}{literature : numerical}  & \\ \hline
   & expression &experiment & numerical & case - 2D & case - 3D & literature (exp.)\\  
\hline 
$d_\mathrm{f}$ &  $S\propto  \xi^{d_\mathrm{f}}$  & 1.07 $\pm$~0.07  & 1.15 & 1.15${}^{a}$, 1.10${}^{b}$, 0.90${}^{c}$& 1.50${}^{b}$, 1.38${}^{c}$ &2.0${}^{e}$, 2.1${}^{f}$ \\
$z$ & $T \propto  \xi^{z}$    & 0.53 $\pm$~0.11  & 0.62   &  0.57${}^{b}$, 0.57${}^{c}$ & 0.65${}^{b}$, 0.82${}^{c}$ &  1.0${}^{e}$ \\
$\theta$& $P(\delta Y) \propto  \delta Y^\theta$& 0.24 $\pm$~0.03  & 0.35~(0.18) & 0.57${}^{b}$, 0.52${}^{c}$,  & 0.35${}^{b}$, 0.37${}^{c}$, 0.2${}^{d}$&- \\
$\beta$& $P(S) \propto  S^{-\beta}$  & 1.30 $\pm$~0.11  & 1.36~(1.34) & 1.8${}^{a}$, 1.35${}^{b}$, 1.28${}^{c}$ & 1.45${}^{b}$, 1.25${}^{c}$, 1.2${}^{d}$ &1.4${}^{e}$,  2.1${}^{f}$ \\
$\alpha$ & $S \propto  \delta^{-\alpha}$ & 0.57 $\pm$~0.04  & 0.48~(0.60) & 0.4${}^{a}$ &- &1.3${}^{e}$,  2.6${}^{f}$\\
$z/d_\mathrm{f}$ & $T \propto  S^{z/d_\mathrm{f}}$  & 0.49 $\pm$~0.14  & 0.53~(0.64) & 0.51${}^{b}$, 0.63${}^{c}$&0.43${}^{b}$, 0.58${}^{c}$ &  0.5${}^{e}$ \\
$\beta_{\mathrm{tot}}$& $\beta_{\mathrm{tot}} = \beta + \frac{2-\beta}{\alpha}$  & 2.32 $\pm$~0.18 & 2.2~(2.13) & 2.7${}^{a}$, 2.65${}^{b}$, 2.72${}^{c}$& 2.55${}^{b}$, 2.62${}^{c}$, 2.8${}^{d}$&1.75${}^{e}$,  2.5${}^{f}$ \\
\hline \\
\end{tabular}
\\
The values in brackets in fourth column are obtained for the case of equivalent force control from displacement control simulations. Exponents of ${}^{a}$Girard \textit{et al.} \cite{SI_girard2010} and ${}^{d}$Ozawa \textit{et al.} \cite{SI_ozawa2018} are for cascades preceding the critical stress while exponents in ${}^{b}$Lin \textit{et al.}\cite{SI_lin2014} and ${}^{c}$Liu \textit{et al.} \cite{SI_liu2016} were obtained for stationary cases of yielding in amorphous materials. ${}^{e}$Vu \textit{et al.} \cite{SI_weiss2019} examine the failure of concrete samples using acoustic emission and ${}^{f}$Kandula \textit{et al.} \cite{SI_kandula2019} process the X-ray tomography image stack of  Carrara marble specimens.
\label{tab:tab1}
\end{table}


\section{Movies}
The two movies available online describe the spatio-temporal structure of the damage cascades measured during compressive failure. The first movie provides the global and local damage evolution as well as the raw images as obtained from the camera. The second video highlights the highly correlated clusters constituting a damage cascade as given in Fig.2(b) of the main article. 

\begin{itemize}
\item {\textbf{Video-1.mp4} : (A) Typical force-displacement response during compression  experiments. A red tracer marks the damage cascades on the curve. (B) Images of the deformation of the cells recorded by our high speed camera. Note that the localization band is barely visible to the naked eye at peak load but becomes prominent afterwards as progressive collapse of the cells in the band takes place. (C) The coordinates of the cells from the particle tracking (marked as $+$) superposed over the map of dissipation energy density for various damage cascades. (D) Variations of cascade sizes $S$ with distance to failure $\delta$, updated with each new cascade whose size is displayed at the top right corner}
\item{\textbf{Video-2.mp4} : (A) Spatial structure of the damage clusters as deciphered from a single time step. (B) The augmented spatio-temporal organization of the various clusters constituting the damage cascade shown in middle panel of Fig.~2(a)}
\end{itemize}

\end{document}